\documentclass[preprint]{aastex}
\usepackage{apjfonts}
\usepackage{epsfig}
\usepackage{graphics}
\usepackage{natbib,graphicx}

\newcommand{\rhostar}{\ensuremath{\rho_\star}}

\newcommand{\gcmc}{\ensuremath{\rm g\,cm^{-3}}}
\newcommand{\logg}{\ensuremath{\log{g}}}

\begin{document}
\title{Secure TTV Mass Measurements: \\ Ten Kepler Exoplanets between 3 and 8 M$_{\oplus}$ with Diverse Densities and Incident Fluxes}
\author{Daniel Jontof-Hutter\altaffilmark{1}, Eric B. Ford\altaffilmark{1}, Jason F. Rowe\altaffilmark{2}, Jack J. Lissauer\altaffilmark{3}, Daniel C. Fabrycky\altaffilmark{4}, Christa Van Laerhoven\altaffilmark{5}, Eric Agol\altaffilmark{6}, Katherine M. Deck\altaffilmark{7}, Tomer Holczer\altaffilmark{8}, and Tsevi Mazeh\altaffilmark{8}}
\email{dxj14@psu.edu}
\altaffiltext{1}{Department of Astronomy, Pennsylvania State University, University Park, PA 16802, USA}
\altaffiltext{2}{D\'epartement de Physique, Universit\'e de Montr\'eal, Montr\'eal, QC  H3T 1J4, Canada,}
\altaffiltext{3}{Space Science and Astrobiology Division, MS 245-3, NASA Ames Research Center, Moffett Field, CA 94035, USA}
\altaffiltext{4}{Department of Astronomy and Astrophysics, University of Chicago, 5640 South Ellis Avenue, Chicago, IL 60637, USA}
\altaffiltext{5}{Canadian Institute for Theoretical Astrophysics, 60 St. George Street, Toronto, ON M5S 3H8, Canada}
\altaffiltext{6}{Department of Astronomy, University of Washington, Seattle, WA 98195, USA}
\altaffiltext{7}{Division of Geological and Planetary Sciences, California Institute of Technology, Pasadena, CA, USA}
\altaffiltext{8}{School of Physics and Astronomy, Raymond and Beverly Sackler Faculty of Exact Sciences, Tel Aviv University, Tel Aviv 69978, Israel}

\begin{abstract}
We infer dynamical masses in eight multi-planet systems using transit times measured from \textit{Kepler}'s complete dataset, including short-cadence data where available. Of the eighteen dynamical masses that we infer, ten pass multiple tests for robustness. These are in systems: Kepler-26 (KOI-250), Kepler-29 (KOI-738), Kepler-60 (KOI-2086), Kepler-105 (KOI-115), and Kepler-307 (KOI-1576). Kepler-105 c has a radius of 1.3 R$_{\oplus}$ and a density consistent with an Earth-like composition. 

Strong TTV signals were detected from additional planets, but their inferred masses were sensitive to outliers or consistent solutions could not be found with independently-measured transit times, including planets orbiting: Kepler-49 (KOI-248), Kepler-57 (KOI-1270), Kepler-105 (KOI-115) and Kepler-177 (KOI-523). Nonetheless, strong upper limits on the mass of Kepler-177 c imply an extremely low density $\sim 0.1$ g cm$^{-3}$.

In most cases, individual orbital eccentricities were poorly constrained due to degeneracies in TTV inversion. For five planet pairs in our sample, strong secular interactions imply a moderate-to-high likelihood of apsidal alignment over a wide range of possible eccentricities. We also find solutions for the three planets known to orbit Kepler-60 in a Laplace-like resonance chain. However, non-librating solutions also match the transit-timing data. 

For six systems, we calculate more precise stellar parameters than previously known, enabling useful constraints on planetary densities where we have secure mass measurements. Placing these exoplanets on the mass-radius diagram, we find a wide range of densities is observed among sub-Neptune mass planets and that the range in observed densities is anti-correlated with incident flux. 
\end{abstract}
\section{Introduction}
With the rich \textit{Kepler} dataset, much progress has been made recently in measuring precise radii and masses of sub-Neptune mass exoplanets. Both measurements are essential for characterizing planetary bulk densities and modeling bulk compositions. The \textit{Kepler} mission  has substantially increased the number of characterized sub-Neptune mass planets. The sub-Neptune regime in mass is of particular interest because it includes the transition from rocky planets to worlds that retain deep atmospheres. 

Planetary radii relative to the host star are directly measured from the depth of transit, and precision in absolute planetary radii rests largely on how well the host star can be characterized. Radial velocity spectroscopy (RV) has provided the bulk of mass measurements of transiting exoplanets (\citealt{win11, bat11, gill12, fres12, gau12, how13, pepe13, weis13, mar14, hay14, dres15}). However, among sub-Neptune mass planets, RV mass detections are limited to planets on short orbital periods, because the RV signal depends on the strength of planet-star interactions and declines with orbital distance. \citet{mar14} conducted an RV survey of ``bright" stars with short-period sub-Neptune-size planets in the Kepler field, and among planets below 10 $M_{\oplus}$ they found strong detections up to around 16 days for Kepler-102 e (KOI-82.01) and Kepler-96 b (KOI-261.01), as well as useful upper limits within this mass range as far out as 69 days (Kepler-409 b, KOI-1925.01). Using the known sample of  exoplanets characterized with RV, \citet{rog15} found that most planets larger than 1.6 $R_{\oplus}$ with orbital periods up to $\sim$50 days are likely volatile rich. 

Complementing this sample, transit timing variations (TTV) probe interplanetary perturbations (\citealt{assc05, hm05}) and are more readily detectable at longer orbital periods. However, the diminishing likelihood of transiting at longer orbital periods, coupled with the four-year \textit{Kepler} baseline limits the orbital period of planets that have been characterized by transit timing. TTVs have measured planetary masses below 10 $M_{\oplus}$ with orbital periods ranging from around 7.6 days (Kepler-18 c, \citealt{coch11}) to almost Venus-like distances at 191 days (Kepler-87 c, \citealt{ofir14}). TTVs have constrained the masses of non-transiting planets even further out, e.g., at Kepler-419, which has a likely massive planet beyond 660 days \citep{Dawson2012}.  

The strongest detections of dense exoplanets, mostly in RV but also including Kepler-36 b \citep{car12}, indicate a primarily rocky compositions for planets up to $\approx 1.6$ $R_{\oplus}$ in size \citep{dres15}.  However, strong mass upper limits on the weaker detections of slightly larger planets include several low-mass planets that are too low in bulk density to be rocky (e.g., Kepler-11 b and Kepler-11 f, \citealt{liss13}).

There are clear differences in the properties of planets that are characterized by TTV from those characterized by RV. Some of this may be due to target selection and/or detection biases in RV. In particular, most \textit{Kepler} targets with RV follow-up have been selected based on their size \citep{mar14}. Additionally, the majority of planets below 10 $M_{\oplus}$ that have been characterized by RV have incident fluxes above 100 times that of Earth. This sample includes low-mass planets that are unlikely to retain deep atmospheres. However, the sensitivity of TTV to low mass planets beyond the range of RV has enabled the characterization of volatile rich planets that are far enough from their hosts to have avoided significant mass loss. The precise TTV mass detections to date show a remarkably diversity in planetary density in the mass range from 2-8 $M_{\oplus}$ (\citealt{liss11a, car12, jont14, mas14, ofir14}). 

\textit{Kepler} identified over 100 multi-planet systems with transit timing variations caused by mutually perturbing, transiting exoplanets (\citealt{maz13}). Only a small fraction have been analyzed in detail using the complete seventeen quarters of \textit{Kepler} data from 2009 to 2013 (Q17) with short cadence data if available. These include the three planets of Kepler-138 \citep{jont15}. Other studies have measured precise masses using the data through Q16 (Kepler-79, \citealt{jont14}), Q14 (Kepler-11, \citealt{liss13}), or earlier datasets, as well as Q17 data in long cadence (e.g., \citealt{had15,goz15}). 

In this paper, we analyze TTVs to provide robust planetary mass measurements using the \textit{Kepler} dataset of transit times. Our method of light curve analysis and measurements of transit times (\citealt{rowe15a,rowe15b}) followed by detailed fitting of transit times with dynamical models has been demonstrated to invert the TTV signal and recover exoplanet masses that are reproducible and in many cases are insensitive to uncertainties in the data set being analyzed. In \citet{jont15}, the solutions of TTV inversion were used to generate synthetic transit times with uncertainties. The recovery of the synthesized dynamical masses and orbital parameters validated both the measurement of the transit times from the light curve and TTV inversion with dynamical fitting. Here we characterize 18 transiting planets in systems where detected TTVs are attributable to mutually perturbing transiting planets. We also test the sensitivity of our solutions to the choice of eccentricity priors and the possibility of non-Gaussian timing uncertainties (e.g., outliers). 

The structure of this paper are as follows. In Section 2, we describe our procedure for fitting transit timing data and exploring posteriors for parameter estimation, our tests for sensitivity to  outlying transit times, our tests against different priors on orbital eccentricity, tests against independently measured transit times, and our tests for long term stability. In Section 3 we present our TTV results for each planetary system. In Section 4 we explore whether there is any evidence of likely apsidal alignment among our solutions given that TTV degeneracies can favor apsidally-aligned configurations. In Section 5 we present our analysis of stellar parameters to characterize true planetary masses given our TTV solutions, and in Section 6 we compare our well-characterized planets to others on mass-radius-flux diagrams.

\section{Methods}
\subsection{Physical Model}
We simulate planetary orbits with an eighth order Dormand-Prince Runge-Kutta integrator (\citealt{fab10,liss11a, liss13, jont14, jont15}) and compare simulated transit times to the observed transit times. TTVs are expressed as the difference between the observed transit time and a calculated linear fit to the transit times (O-C). In each case, we have assumed coplanar orbits since TTV amplitudes change only to second-order in mutual inclinations (\citealt{lith12, nes14}). 

For planet pairs with large orbital period ratios, TTVs can be sensitive to mutual inclinations, but mutual inclinations do not make a major contribution to TTVs of planets near low-order mean motion resonances \citep{Payne2010}, which dominate the TTVs in our sample. Furthermore, the distribution of transit duration and orbital period ratios among \textit{Kepler}'s multi-planet systems imply that typical inclinations in \textit{Kepler}'s population of multi-transiting systems are a few degrees or less \citep{fab14}. 

We assume all the TTVs are attributable to mutual perturbations between the known transiting planet candidates. In some multi-planet models, we exclude transiting planets that have no observed and no expected TTVs given the ratio of their orbital periods with the planets that have detected TTVs. In these cases we performed integrations including all known planets to verify the accuracy of this approximation.
\subsection{Statistical Model}
For any multi-planet model, we fit five parameters per planet: the orbital period ($P$), the time of the first transit ($T_{0}$) after our chosen epoch (BJD-2,455,680), $k=e\cos\omega$, $h=e\sin\omega$, and  the ratio of the planet masses to their host, which we refer to as ``dynamical mass'' throughout. We express dynamical masses in the form $\frac{M_{p}}{M_{\oplus}} \frac{M_{\odot}}{M_{\star}}$ for easy conversion to Earth-masses for any estimate of stellar mass. 
\subsubsection{Likelihood Function}
We evaluated the goodness of fit to the transit timing data (\textbf{x}) for each simulated model $\mathcal{M}$ for a choice of parameters $\mathbf{\theta}$ using the likelihood function 
\begin{equation}
\mathcal{L}(\theta | \textbf{x}, \mathcal{M}) = \prod_{i=1}^{n}  \frac{  \exp \left[   - \frac{1}{2}  \left( \frac{ y_{i,\theta}-x_{i}}{\sigma_{i} } \right)^2 \right]} {\sqrt{ 2\pi \sigma_{i}^2}}
\label{likelihoodfn}
\end{equation}
where there are $n$ observed transit times, and \textbf{y} are simulated transit times associated with the model parameters $\mathbf{\theta}$. To minimize numerical round-off error in this product, we calculate the familiar $\chi^2 =\sum_{i=1}^{n} \left(\frac{y_{i,\theta}-x_{i}}{\sigma_{i}}\right)^{2} $ statistic for all model evaluations:
\begin{equation}
\chi^2_{eff} = -2\log(\mathcal{L}) = \sum_{i=1}^{n} \log(2\pi\sigma_{i})+ \chi^2 .
\label{chisq}
\end{equation}
In comparing any two models, the ratio of log-likelihoods is proportional to the difference in $\chi^2$, and the summation in Equation~\ref{chisq} cancels.

\subsubsection{Priors}
We adopt uniform priors in dynamical masses, assumed to be positive definite, and uniform priors in orbital periods and the time after first transit after epoch for all planets. We fixed all orbital inclinations at 90$^{\circ}$. We adopted Gaussian priors in the eccentricity vector components ($h$ and $k$), which effectively induce a Rayleigh distribution prior on scalar eccentricity.  We choose $h$ and $k$ as our eccentricity variables instead of $e$ and $\omega$, since this increases the rate of convergence for planets with small eccentricities where $\omega$ is poorly constrained (\citealt{for05,Ford2006}). Since TTVs are sensitive to orbital eccentricity, we repeated our analysis with two choices of eccentricity prior, as we explain in the following subsection.

\subsection{Reliability}

\subsubsection{Eccentricity Priors}
For our primary results in Table~\ref{tbl-results}, we have assumed a Gaussian prior on each of the eccentricity vector components ($e\sin\omega$, $e\cos\omega$) with a mean of zero and variance 0.1. This corresponds to a Rayleigh distribution with scale width 0.1 as the prior for scalar eccentricity. This is consistent with the distribution of eccentricities amongst \textit{Kepler}'s exoplanets found by \citet{moor11} using an independent analysis of measured transit durations. This wide distribution is also consistent with the known eccentricities of giant RV planets (\citealt{Kane2012,Plavchan2014}). A narrower eccentricity distribution of characteristic width$\sim0.05$ was found by \citet{Vaneylan2015} by comparing a sample of well-characterized transiting exoplanet hosts with their inferred photometric densities. 

An even narrower eccentricity distribution (of characteristic width $\sigma$=0.02) was found for a sample of compact multi-planet systems with TTVs by \citet{had14}. While this is the most relevant sample for our purposes, we adopt the wider prior to test where the TTVs can provide tight constraints on eccentricity. Where eccentricities are poorly constrained by the TTVs, we compare our dynamical masses with solutions from a narrower eccentricity prior (with scale length 0.02), to test whether our masses and orbital solutions are robust against the choice of prior for eccentricity. 

In most cases, we found that individual eccentricities are weakly constrained, with the posterior for the eccentricity vector components being strongly affected by our prior. However, we have inferred tight constraints on relative eccentricity vectors since the vector components are highly correlated. This correlation is expected for near first-order mean motion resonant TTVs, as can be seen from the analytical solutions derived by \citet{lith12}. The solution breaks the eccentricity vector up into the sum of two component vectors: a free component that is constant for timescales that are short compared to apsidal precession, and a forced component that varies over the coherence time of near-resonant perturbations. Each planet's free eccentricity vector can be expressed in complex notation ($z_{free} = k+ih=e_{free}\cos\omega+i e_{free}\sin\omega$), although the periodic TTV signal is not sensitive to these directly. Rather, the amplitude of a sinusoidal TTV signal caused by a mutually interacting pair near first-order resonance depends on the mass of the perturber, the orbital periods of both planets, and the conjugate of the complex sum of free eccentricities, $Z_{free}$,  where, given the free eccentricity of an inner planet $z_{free} = e_{free} \exp(i\omega)$, and an outer planet $z_{free}' = e_{free}' \exp(i\omega')$,
\begin{equation}
Z_{free} = f z_{free} + g z_{free}'.
\label{eqn:Zfree}
\end{equation}
Here, the coefficients $f$ and $g$ are the sums of Laplace coefficients of order unity and solved to first order by \citet{lith12} as a function of orbital period ratios. The TTVs are sensitive to the complex conjugate of the expression in Equation~\ref{eqn:Zfree}, $Z_{free}^{*}$. Since $f\approx -1$, $Z_{free}^{*}$  is the difference between eccentricity vector components scaled by the coefficients $f$ and $g$, and hence approximately proportional to the relative eccentricity. This leads to an eccentricity-eccentricity degeneracy in TTVs, where the relative eccentricities between planets may be well-constrained, but their absolute eccentricity vector components are poorly constrained. With $Z_{free}^{*}$ constant, the derivatives of the vector components of the outer planet with respect to the same component of the inner planet, depends only on Laplace coefficients. The gradients are:
\begin{equation}
 \frac{d \left( e'\cos \omega ' \right)}{d \left( e\cos \omega\right)} = \frac{d \left(e'\sin \omega ' \right)}{d\left( e\sin \omega \right)} = \frac{-f}{g}.
\label{eqn:gradient}
\end{equation}
Hence, if the orbital periods are known precisely, one can easily predict the correlation between the eccentricity vector components of neighboring planets that have detected TTVs.

\subsubsection{Non-Gaussian Uncertainties in Transit Times}
\begin{figure}[h!]
\includegraphics [height = 2.0 in]{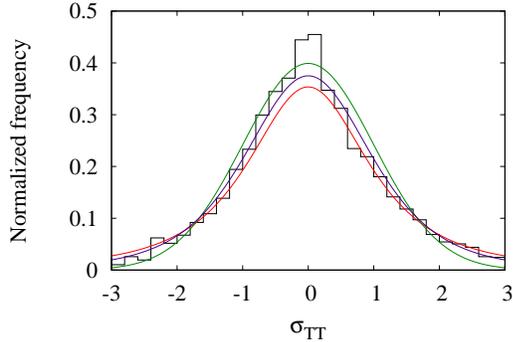}
\caption{Distribution of residuals for the combined TTV fits to all 3860 transits in this study, compared to a Gaussian (green),  Student-t$_{4}$ (blue), and Student-t$_{2}$ (red) distributions.  }
\label{fig:residuals} 
\end{figure}

Figure~\ref{fig:residuals} shows the distribution of residuals for all measured and model transit times in this study at the best fit solution for each planet, with comparisons to three normalized and symmetric statistical distributions, a Gaussian, and Student t-distributions with 2 and 4 degrees of freedom, respectively. 

The central peak slightly exceeds a Gaussian distribution, consistent with moderate over-fitting in our best fit models, but unlikely to affect our MCMC posteriors.  On the other hand, the wings of the distribution reveal far more outliers than expected from Gaussian uncertainties on measured transit times. These data have more leverage in model fitting. Hence, to check the robustness of our results against these outliers, we repeated our analysis assuming measurement uncertainties followed a Student-t distribution with 2 degrees of freedom. In this case, the likelihood function is:
\begin{equation}
\mathcal{L}(\theta | \textbf{x}, \mathcal{M}) = \prod_{i=1}^{n}  \frac{1}{2\sqrt{2}\sigma_i}\left(1+\frac{(y_{i,\theta}-x_{i})^2}{2\sigma_{i}^2} \right)^{-3/2},
\label{t2likelihood}
\end{equation}
and the log of the likelihood becomes
\begin{equation}
\log(\mathcal{L}) =  n\log \left(\frac{1}{2\sqrt{2}\sigma_i}\right ) -\sum_{i=1}^{n} \frac{3}{2} \log\left(1+\frac{(y_{i,\theta}-x_{i})^2}{2\sigma_{i}^2} \right).
\label{t2dist}
\end{equation}

\subsubsection{Independently Measured Transit Times}
For the systems studied in this paper, the majority of transit times are measured from short cadence data. All raw Kepler photometry was read from FITS files retrieved from MAST.  Q1-Q14 photometry was based on data release 21, Q15, Q16 and Q17 used data releases 20, 22 and 23 respectively.  All extracted and reported time stamps have barycentric times reported in terrestrial dynamic time (TDB). Multiple methods have been employed to detrend \textit{Kepler} light curves, fit transit models and measure transit times with their uncertainties. \citet{liss13} compared the transit times of three independent analyses of the light curve of Kepler-11 to identify outliers. For this study, we modeled the light curves using the methods described in \citet{rowe14}, using \textit{Kepler}'s complete dataset of 17 quarters including both long cadence and any available short cadence transit data. We test the sensitivity of our results by repeating our analysis using the transit times from the Q17 Holczer et al. TTV catalog, with short cadence times included where available (Holczer et al. 2015, submitted). Where dynamical masses were found to be discrepant between datasets of measured transit times, the planetary masses were deemed "not secure" and excluded from the mass-radius diagram.

\subsection{Choice of Systems}
The systems were found by identifying appropriate period ratios of all planet pairs in the online exoplanet database at the NASA Exoplanet Archive\footnote{http://exoplanetarchive.ipac.caltech.edu/}. We chose suitable systems for measuring dynamical masses by performing maximum-likelihood fits to the long cadence transit timing data of all known systems with one or more pairs of planets with near-resonant orbital periods, using the Levenberg-Marquardt algorithm, and with measured transit times from \citet{rowe15a}. The Levenberg-Marquardt algorithm efficiently converges to a local minimum near the initial values for free parameters. One can estimate uncertainties from the local curvature of the $\chi^2$ surface at the local minimum.  However, a major limitation of this method for estimating uncertainties is that it approximates the $\chi^2$ surface as parabolic and unimodal in the dimensions of all free parameters. We tested this assumption for our TTV models by searching for regions of high likelihood using repeated Levenberg-Marquardt minimization starting with initial values of $e\sin\omega$ and $e\cos\omega$ along a grid for each planet. We used a single set of values for the initial orbital periods and phases since these quantities are tightly constrained by the transit times. In all cases, we found multiple but statistically consistent solutions near the overall best fit.

Performing initial fits on this sample of systems, we sought systems with likely detections of low-mass exoplanets. We selected systems with a wide range of orbital periods, including systems with apparent synodic chopping (including Kepler-26 and Kepler-177). We rejected systems where known stellar rotation periods are nearly commensurate with a planetary orbital period, including Kepler-128 (KOI-274), whose 13.6 day rotation period has a near 5:3 commensurability with the orbital period of Kepler-128 c, since this could cause spurious TTV signals that would affect our mass measurements.

We rejected systems where the posterior did not provide useful lower bounds on planetary masses. However, we included the three-planet system Kepler-105 that included two weak detections, because the third planet was well-constrained in our Levenberg-Marquardt fits, and the planets have particularly short orbital periods compared to most planets with TTV analyses, which may make this system a viable target for RV follow-up. Kepler-26, Kepler-29, Kepler-49 and Kepler-57 were amongst the first systems to have planets confirmed via anti-correlated TTV signals (\citealt{fab12b, stef12, stef13}). We identified Kepler-177 as having a particularly strong chopping signal, deep transits and low masses, indicative of likely extremely low planetary densities. We also noted strong TTVs at Kepler-60 possibly due to librations within a three-body Laplace-like resonance chain \citep{goz15}. 

\subsection{Algorithm}
To appropriately estimate relevant parameters with their uncertainties, we explored regions of interest that had been found via our initial fits using a Differential Evolution Markov Chain Monte Carlo algorithm (DE-MCMC \citealt{ter06,nelson14a, jont15}). Using MCMC allows for the exploration a region of interest without becoming trapped in shallow local minima. Of the many variants of MCMC, the Differential Evolution MCMC algorithm is particularly efficient for exploring posteriors of correlated variables, which as we shall see, is common in the high dimensional modeling of TTV systems. It employs multiple "walkers" exploring the region of interest in parallel, with proposals calculated from the displacement vectors between other walkers chosen at random, increasing the likelihood that a proposal will be roughly parallel to the direction of correlation between variables. We employed three times as many walkers as free parameters in our model fitting; I.e, 30 walkers for the two-planet systems and 45 walkers for the three-planet systems.

\subsection{Validating Analytical Approximations}
For planet pairs near first-order mean motion resonances, the TTVs are dominated by the sinusoidal signal expressed in the approximations of \citet{lith12}, from which we identify the degeneracy in mass and eccentricity in TTV inversion. In some cases, however, the degeneracy can be broken where high frequencies TTVs are detected, like the so-called ``chopping" at synodic frequencies (\citealt{nes08, deck15, agol15, had15}). 

For two systems that we include here, Kepler-26 and Kepler-177, we test our results with the analytical solutions of \citet{agol15}. These are accurate to first-order in planet-star mass ratios, first-order in eccentricity and assume co-planarity. Accuracy to first order in eccentricities is satisfied if eccentricities $\lesssim 0.1$, which is likely given the narrow dispersion in eccentricities among Kepler's multi-planet systems (\citealt{had14, fab14}). Where there is a second-order dependence on eccentricity, e.g., near second-order mean motion resonances, the solutions of \citet{deck15b} provide an even more general analysis of TTV frequencies\footnote{The code publicly available at https://github.com/ericagol/TTVFaster}. Our analytical fits used an affine-invariant Markov Chain Monte Carlo routine in order to determine parameter estimates and uncertainties \citep{GoodmanWeare2010}. 

\subsection{Long-Term Stability}
We integrated samples of our posteriors for 1 Myr using the HNBODY symplectic integrator code \citep{rh12} to determine if the requirement of long-term stability places additional constraints on the masses and orbits of the planetary systems. Here we make a compromise between testing stability as long as numerically feasible for one solution (e.g., \citealt{liss13, jont14}), and testing for stability for a sample from our posteriors. For this study, since many of the eccentricity posteriors were poorly constrained, we chose to integrate samples of 50 sample solutions for the two-planet systems and 45 solutions for the three-planet models for 1 Myr (from the last generation of walkers in the DE-MCMC chains). We defined systems as unstable if any planet was expelled during the simulation, but none were found to be unstable.  

\subsection{Summary Statistics}
Our summary statistics from TTV analysis are the medians and the 68.3\% and 95.5\% credible intervals for dynamical masses, with high and low posterior tails of equal likelihood excluded from the credible intervals. We also report calculated radii from analysis of light curve transit profiles. We converted dynamical mass posteriors to actual mass posteriors by drawing from posteriors of the mass of each planet's host star, and the mean and variance of planetary radii are calculated by sampling posteriors of stellar radii. The resulting planetary masses, radii, and density medians and credible intervals are summarized in Table~\ref{tbl-results}. We also report credible intervals on incident flux, following sampling of stellar effective temperature and $a/R_{\star}$ posteriors measured from light curve fitting. 
\section{Results}

\subsection{Kepler-26 (KOI-250)}
\begin{figure}[h!]
\includegraphics [height = 2.1 in]{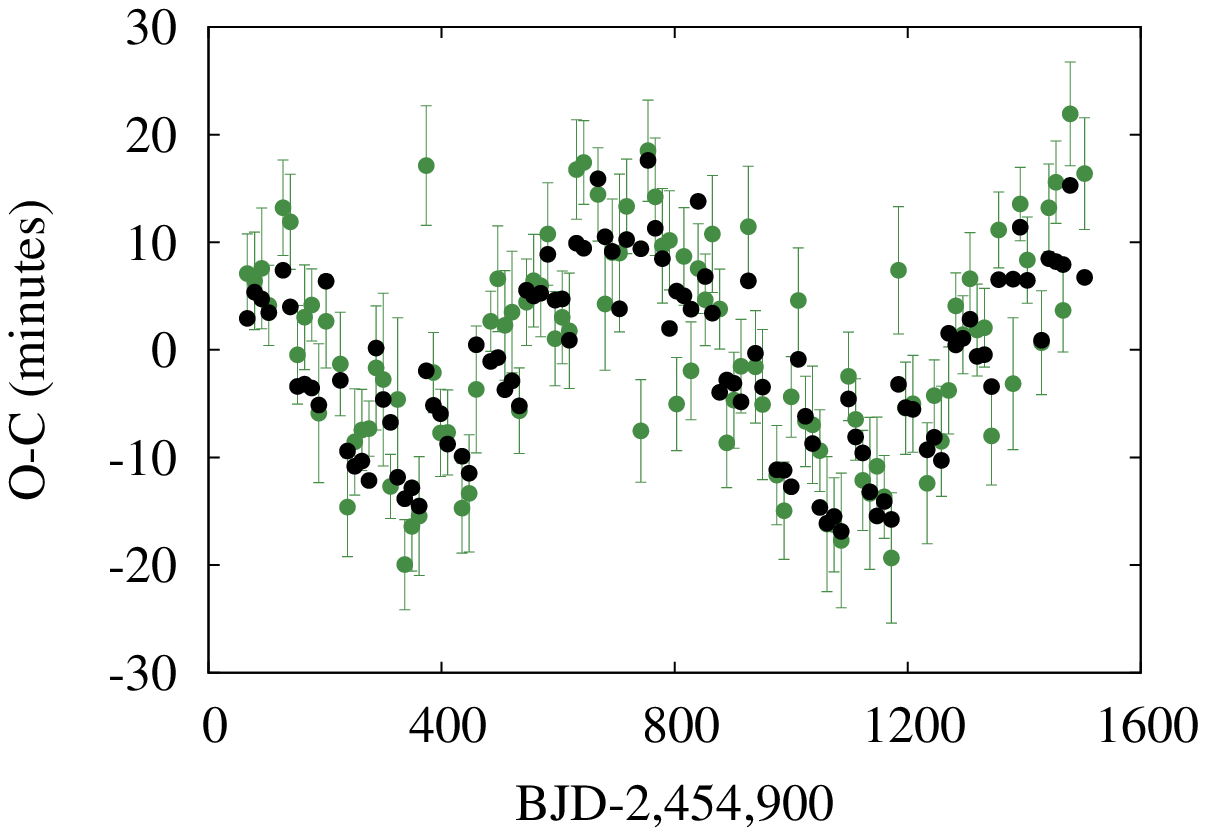}
\includegraphics [height = 2.1 in]{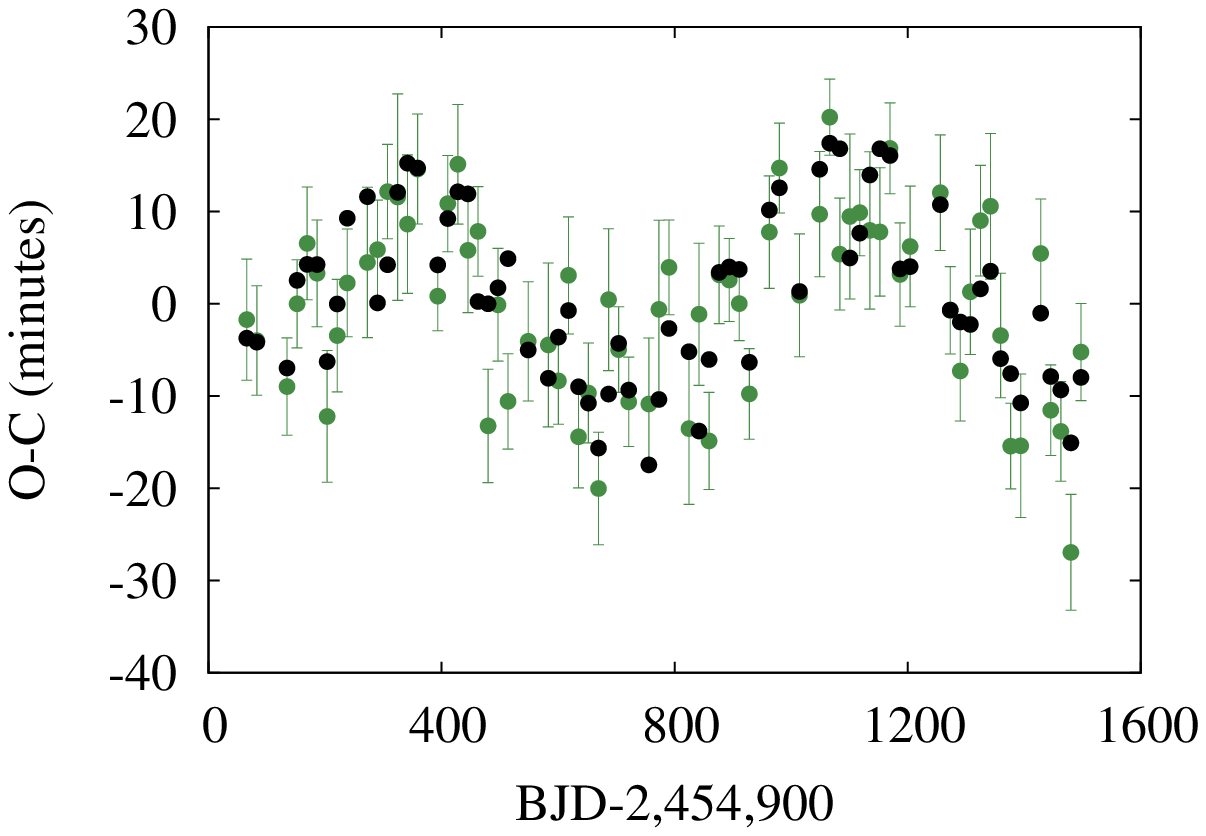}
\caption{Best fit dynamical model for Kepler-26 b (left) and c (right). In black are simulated transit times, and in green are measured transit times with their uncertainties. The detectable synodic chopping in the TTVs enables precise constraints on dynamical masses.
}
\label{fig:250-sim} 
\end{figure}

Kepler-26 has four known transiting planets (in ascending order of orbital period, they are named Kepler-26 d, b, c and e), although only the intermediate pair (Kepler-26 b and c) show significant TTVs \citep{stef12}. The period ratio between the innermost pair is 3.46 and that for the outermost pair is 2.71, which are far enough from any first or second order mean motion resonance to make any detectable TTVs unlikely. We therefore expect no TTVs in Kepler-26 d or e and we detect none. Furthermore, as these are the smallest two of the four planets, at 1.2 and 2.1 $R_{\oplus}$ respectively, they are likely to be less massive than the interacting pair, and hence we exclude them from our nominal TTV model. The intermediate pair of planets orbit near the second-order 7:5 mean motion resonance, with an expected TTV period of 658 days, easily discernible in Figure~\ref{fig:250-sim}.  

In Table~\ref{tbl-koi0250}, we include our adopted parameters from model fitting and the results of various tests on our measured dynamical masses, following posterior sampling with an alternative eccentricity prior, robust fitting with a Student-t$_{2}$ distribution on uncertainties, a four-planet model that includes the two planets that show no TTVs, analytical modeling of the chopping signal, and using the alternative measured times of the Holczer catalogue. With one exception, all tests showed close agreement with our adopted parameters. We found that the analytical formula of \citet{agol15} yielded solutions that are discrepant at the 1$\sigma$ level. However, the first-order formula for this system is likely inadequate because the planets are near the 7:5 (second-order) mean motion resonance. This requires terms at order $e^2$ to model correctly. Using an analytical model specific for the second-order resonant terms  \citep{deck15b} we found close agreement with the dynamical fits.

 \begin{table}[h!]
  \begin{center}
     \begin{tabular}{||c||}
      \hline
 \hspace{1.75 in} Adopted  parameters  with 1$\sigma$ (2$\sigma$) credible intervals     \hspace{1.75 in} \\
      \hline
    \end{tabular}  
    \begin{tabular}{||c|c|c|c|c|c||}
      \hline
 planet  & $P$(days)  &      $T_0$ (days)    &      $e\cos\omega$   &  $e\sin\omega$     &     $ \frac{ M_p  }{M_{ \oplus} } \frac{M_{\odot} } {M_{\star} }  $ ($\pm 2\sigma$)  \\   
 \hline 
  b  & \textbf{ 12.2800 }$\pm  0.0003$ &  \textbf{ 791.2497 }$\pm 0.0006$ &  \textbf{ --0.027 }$^{+ 0.053 }_{ - 0.052 }$ &  \textbf{ --0.005 }$^{+ 0.055 }_{ - 0.052 }$ &  \textbf{  9.40 }$^{+  1.09 }_{ -  1.05 }$ $\left(^{+  2.22 }_{ -  1.63 }\right)$  \\ 
 c & \textbf{ 17.2559}$\pm 0.0006$ &  \textbf{ 790.1830 } $^{+ 0.0007 }_{ - 0.0008 }$ &  \textbf{ --0.013 } $\pm 0.044$ &  \textbf{ 0.010 } $^{+ 0.046 }_{ - 0.043 }$ &  \textbf{ 11.39 } $^{+  1.10 }_{ -  1.08 }$  ($^{+  2.20 }_{ -  1.72 }$) \\ 
      \hline
    \end{tabular}
    \begin{tabular}{||c||c||c||c||c||}
      \hline
  \hspace{0.9 in}  Test 1   \hspace{0.3 in}    &   \hspace{0.3 in} Test 2  \hspace{0.25 in}  &   \hspace{0.25 in} Test 3  \hspace{0.25 in} &   \hspace{0.25 in} Test 4  \hspace{0.25 in} &   Analytic Model  \\
      \hline
    \end{tabular}   
    \begin{tabular}{||c||c||c||c||c||c||}
    \hline
 planet  \hspace{0.1 in}    &    $ \frac{ M_p  }{M_{ \oplus} } \frac{M_{\odot} } {M_{\star} }  $   \hspace{0.48 in}  &   $ \frac{ M_p  }{M_{ \oplus} } \frac{M_{\odot} } {M_{\star} }  $ \hspace{0.48 in} &     $ \frac{ M_p  }{M_{ \oplus} } \frac{M_{\odot} } {M_{\star} }  $ \hspace{0.48 in}  &     $ \frac{ M_p  }{M_{ \oplus} } \frac{M_{\odot} } {M_{\star} }  $ \hspace{0.48 in}  &     $ \frac{ M_p  }{M_{ \oplus} } \frac{M_{\odot} } {M_{\star} }  $ \hspace{0.48 in} \\
 \hline 
 b  &   9.48 $^{+  1.07 }_{ -  1.04 }$   &    9.62 $^{+  1.32 }_{ -  1.30 }$   &     9.44 $^{+  1.10 }_{ -  0.98 }$   &   8.06$^{+  0.73 }_{ -  0.70 }$ & 9.78$\pm 1.36$ \\ 
 c  &  11.52 $^{+  1.08 }_{ -  1.07}$   &    11.97 $^{+  1.35 }_{ -  1.31 }$   &    11.40 $^{+  1.11 }_{ -  0.95 }$     & 12.20 $^{+  0.85 }_{ -  0.84 }$ & 11.92$\pm 1.38$  \\ 
      \hline
    \end{tabular}    
    \caption{TTV solutions for Kepler-26 b and c (KOI-250),  the results of various tests on dynamical masses, and an analytical model: Test 1: An alternative eccentricity prior. Test 2: Robust fitting. Test 3: A four-planet model. Test 4: The Holczer catalog.  The last column in the bottom panel gives an analytical result using the approximation of \citet{deck15b} in close agreement with the dynamical fits. }\label{tbl-koi0250}
  \end{center}
\end{table}

\begin{figure}[h!]
\includegraphics [height = 1.5 in]{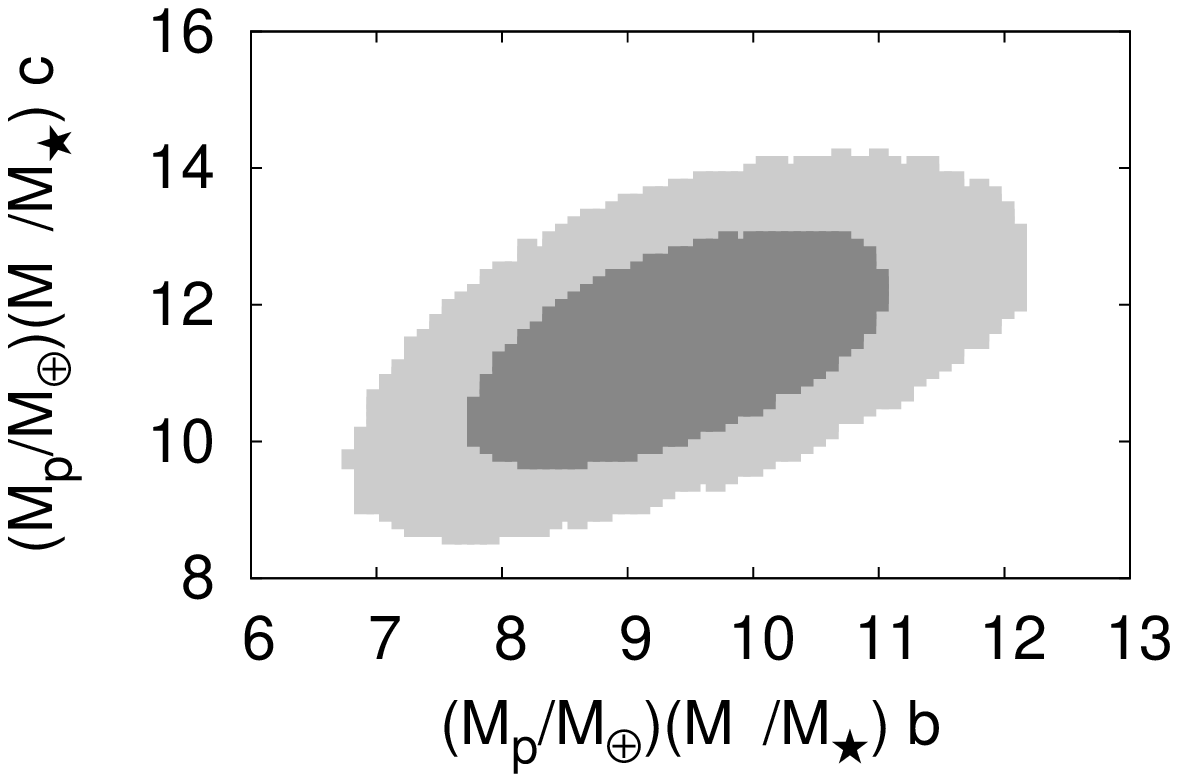}
\includegraphics [height = 1.5 in]{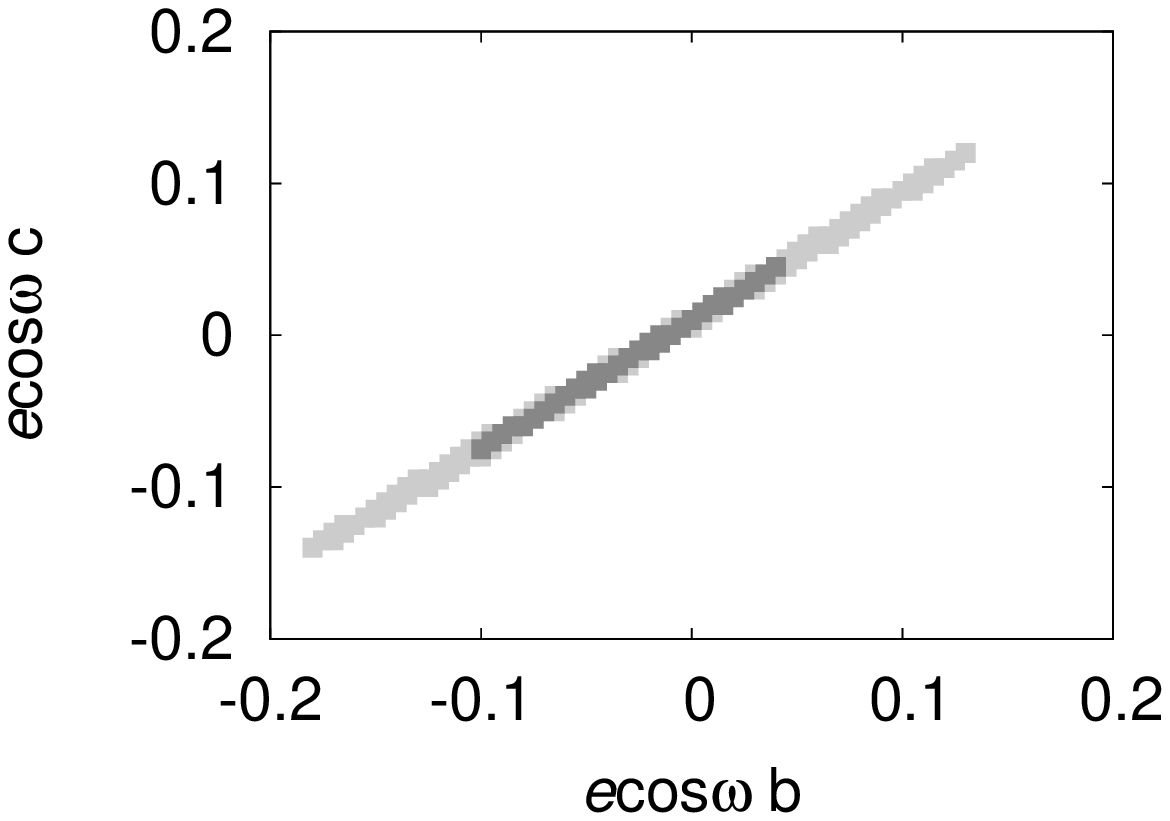}
\includegraphics [height = 1.5 in]{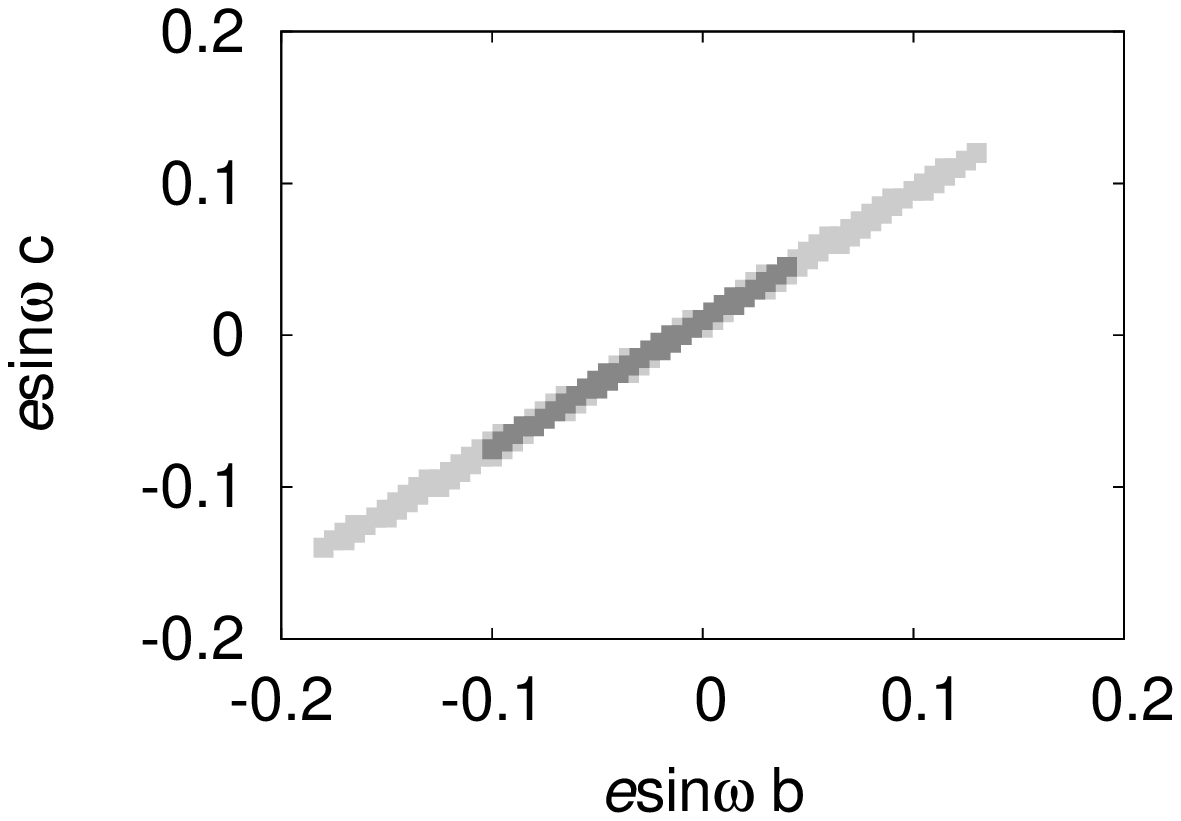}
\caption{Joint Posteriors for planetary masses relative to the host star, and eccentricity components for Kepler-26 b and c (KOI-250). 95.4\% confidence intervals are in light grey, 68.3\% confidence intervals in dark grey.}
\label{fig:250} 
\end{figure}

The joint posteriors show tight constraints on dynamical masses, but poor constraints on orbital eccentricity, as shown in Figure~\ref{fig:250}. In this case, the orbital eccentricities may be limited only by our prior, and hence are not useful. Nevertheless, the inferred masses are independent of the wide range of eccentricities that fit the data. 

To assess whether our adopted solutions for Kepler-26 b and c are long-term stable, we performed long term integrations for this system with all four planets included. For Kepler-26 d and e, we estimated masses from their measured radii using an empirical mass-radius relation for the planets of the Solar System (in Earth units, $M_{p}=R_{p}^{2.06}$; \citealt{liss11b}), and set their initial eccentricities to zero. We sampled our MCMC chains to integrate a range of masses and eccentricities from our posteriors. All simulations were stable to 1 Myr, and we adopt these as reliable dynamical masses. \citet{had15} performed an independent analysis of the TTVs of Kepler-26 using transit times measured from long cadence data. Their measured dynamical masses are closely consistent with ours.

\subsection{Kepler-29 (KOI-738)}
\begin{figure}[h!]
\includegraphics [height = 2.1 in]{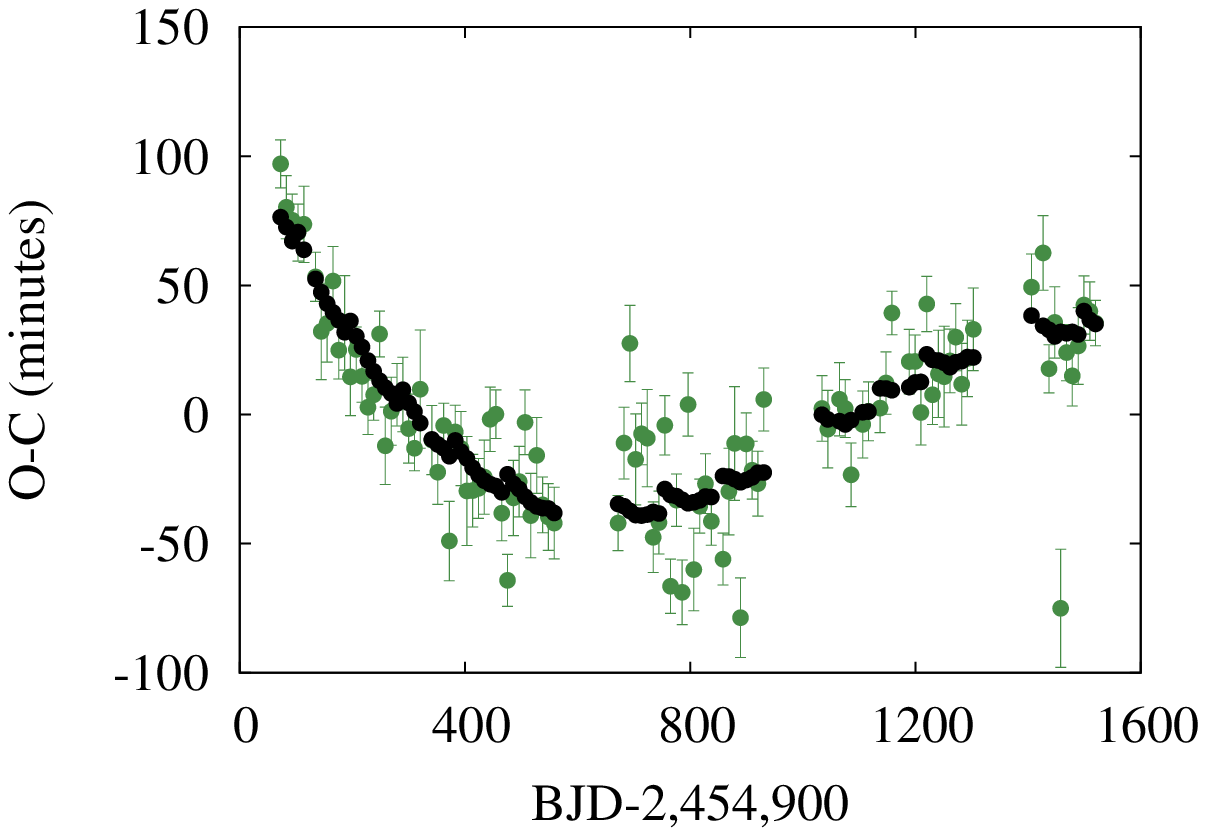}
\includegraphics [height = 2.1 in]{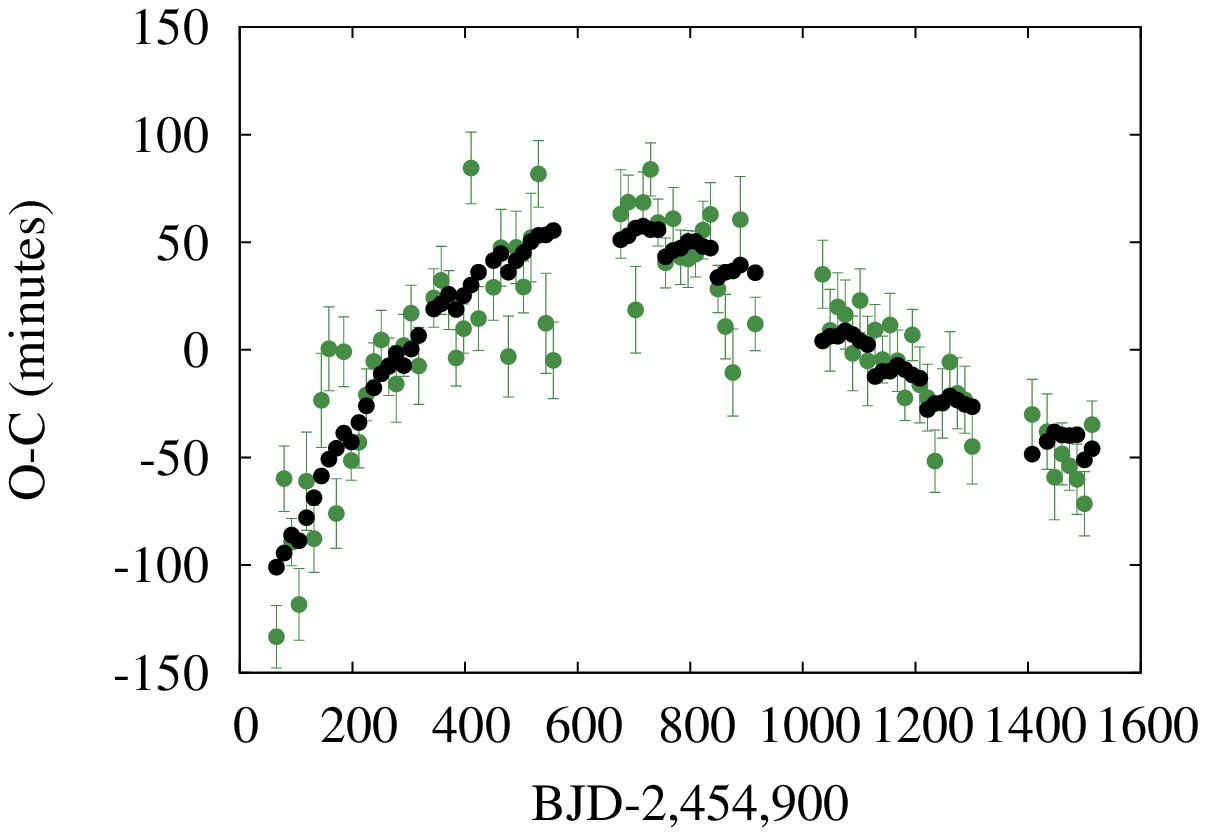}
\caption{Best fit dynamical model for Kepler-29 b (left) and c (right). In black are simulated transit times, and in green are measured transit times with their uncertainties.
}
\label{fig:738-sim} 
\end{figure}

Kepler-29 was confirmed with TTVs by \citet{fab12b} and upper limits were placed on the masses of the planets, both by the requirement of Hill Stability in the limit of low eccentricities, and with dynamical fits to the first 500 days of transit data. In that study, the TTV signal was well-fitted by a quadratic function, and the long-term TTV periodicity could not be discerned. The orbital period ratio, $\frac{P_{c}}{P_{b}} = 1.2853$, has its nearest first-order resonance at 5:4 with an expected TTV period of 94 days, commensurable with synodic chopping (93 days) and can be discerned by eye for the best fit model shown in Figure ~\ref{fig:738-sim}. The pair are much closer to and may be in the 9:7 second-order mean motion resonance, as identified by \citet{liss11b}. With the full dataset, we still only observe a fraction of the long TTV period, although the signal appears like a cubic function, and mass and orbital parameters are better constrained. 

\begin{figure}[h!]
\includegraphics [height = 1.5 in]{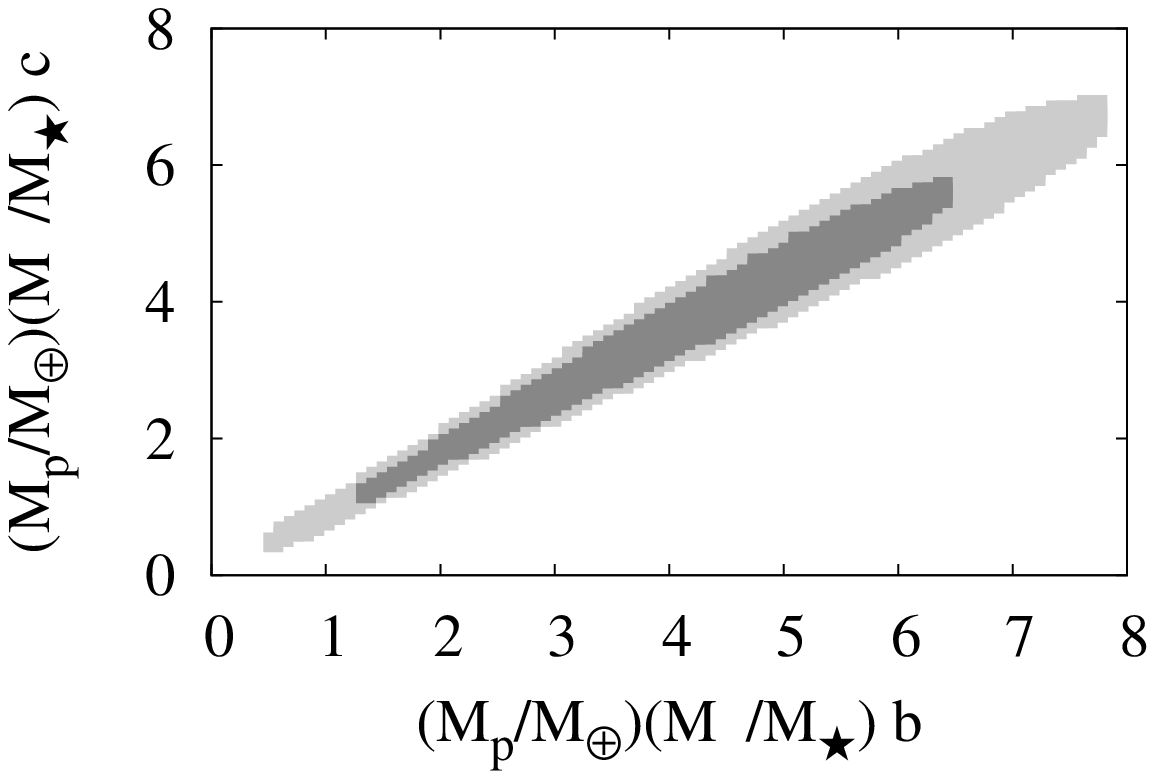}
\includegraphics [height = 1.5 in]{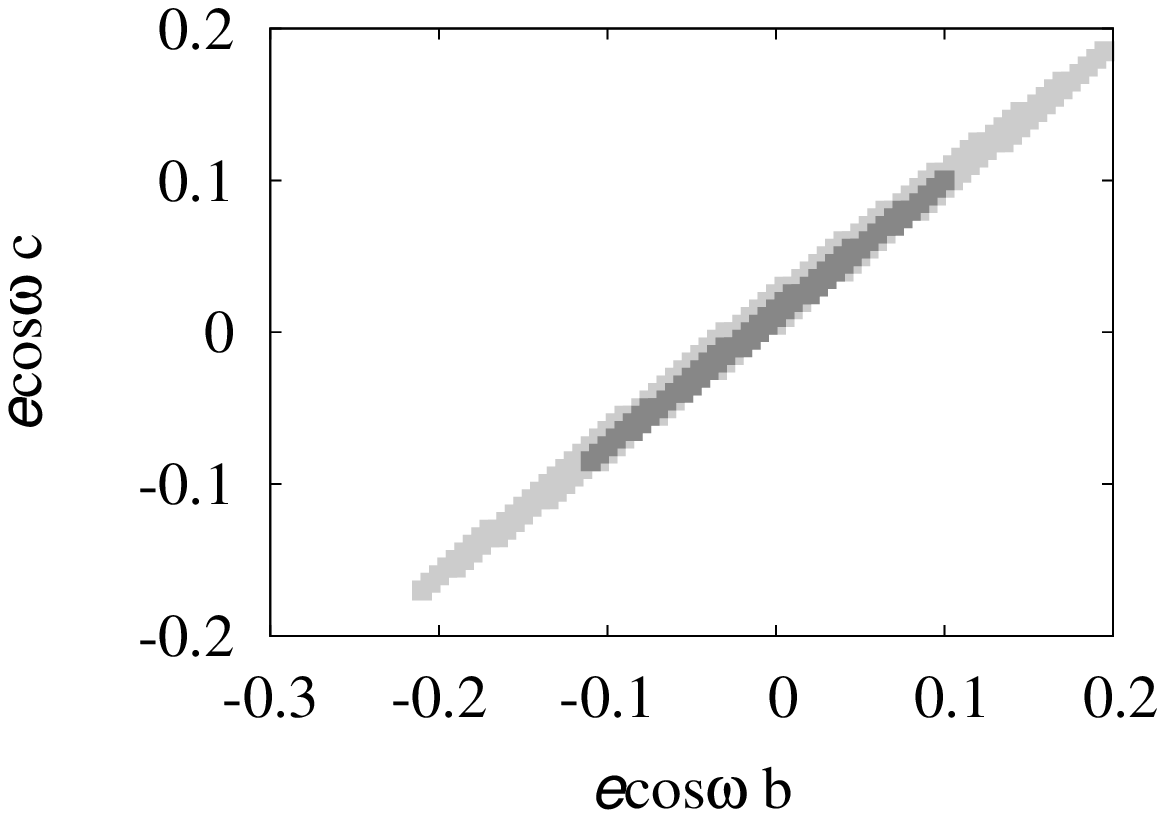}
\includegraphics [height = 1.5 in]{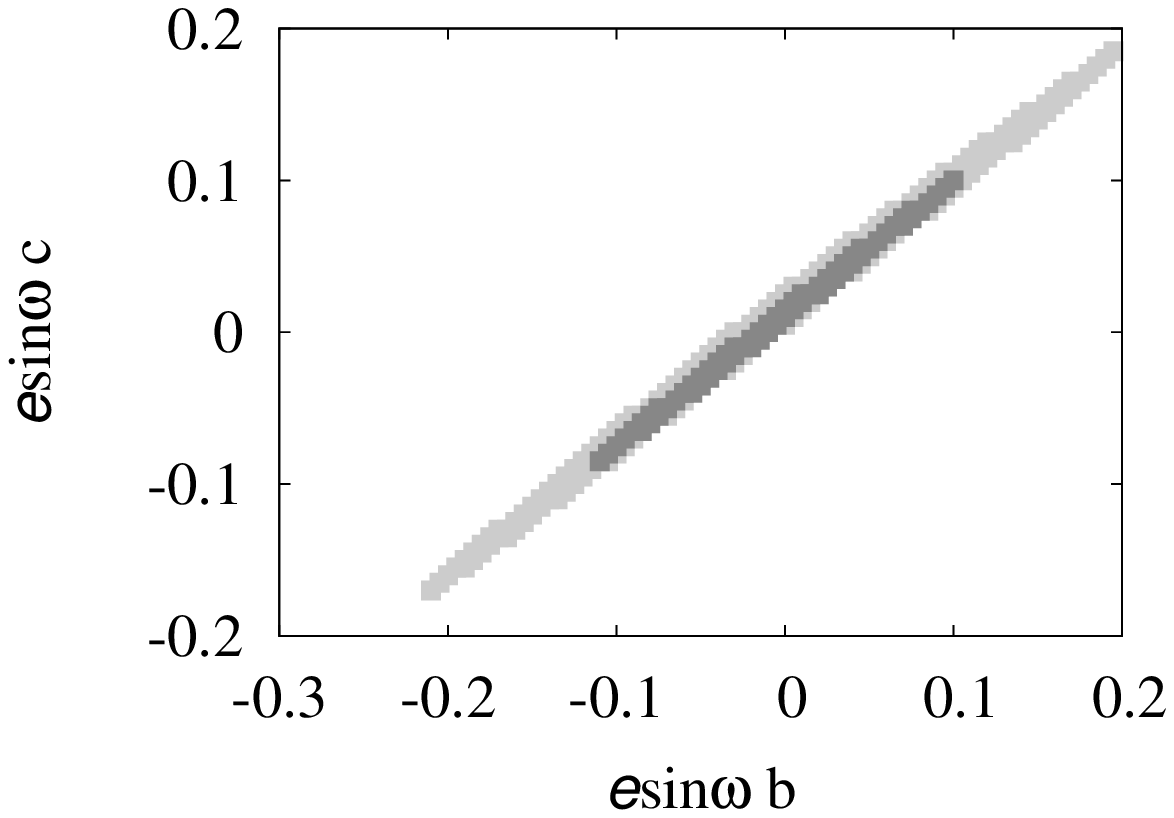}
\caption{Joint Posteriors for planetary masses relative to the host star, and eccentricity components for Kepler-29 b and c (KOI-738). 95.4\% credible intervals are in light grey, 68.3\% credible intervals in dark grey. }
\label{fig:738} 
\end{figure}

Table~\ref{tbl-koi0738} displays the results of our dynamical fits for Kepler-29. We performed three tests on our adopted results for consistency, as shown in the lower panels of Table~\ref{tbl-koi0738}; i) An alternative (narrower) prior on eccentricity, ii) Robust fitting with an assumed Student-t$_{2}$ distribution on transit timing uncertainties, iii) Fits against an alternative set of measured times from the Holczer catalog. We also tested the long term stability of a sample of from our MCMC chains, and all were stable to 1 Myr. The joint posteriors of dynamical masses and eccentricity vector components are shown in Figure~\ref{fig:738}. 

 \begin{table}[h!]
  \begin{center}
     \begin{tabular}{||c||}
      \hline
 \hspace{1.75 in} Adopted  parameters  with 1$\sigma$ (2$\sigma$) credible intervals     \hspace{1.75 in} \\
      \hline
    \end{tabular}  
    \begin{tabular}{||c|c|c|c|c|c||}
      \hline
 planet  & $P$ (days)  &      $T_0$ (days)    &      $e\cos\omega$   &  $e\sin\omega$     &     $ \frac{ M_p  }{M_{ \oplus} } \frac{M_{\odot} } {M_{\star} }  $ ($\pm 2\sigma$)  \\   
 \hline 
     b  & \textbf{ 10.3384 }$ \pm 0.0003 $ &  \textbf{ 785.7544 }$\pm 0.0014$ &  \textbf{ --0.008 }$\pm 0.072 $ &  \textbf{ --0.032 }$\pm 0.072 $ &  \textbf{  4.59 }$^{+  1.43 }_{ -  1.47 }$  $\left(^{+  2.88 }_{ -  2.28 }\right)$   \\ 
  c & \textbf{ 13.2884 }$\pm 0.0005$ &  \textbf{ 782.7818 }$ \pm 0.0019$ &  \textbf{ 0.006 }$^{+ 0.063 }_{ - 0.062 }$ &  \textbf{ --0.023 }$^{+ 0.063 }_{ - 0.062 }$ &  \textbf{  4.06 }$^{+  1.25 }_{ -  1.29 }$ $\left(^{+  2.51 }_{ -  2.03 }\right)$   \\ 
\hline
      \hline
    \end{tabular}
    \begin{tabular}{||c||c||c||}
      \hline
  \hspace{0.48 in}  Test 1   \hspace{0.48 in}    &   \hspace{0.48 in} Test 2  \hspace{0.48 in}   &   \hspace{0.48 in} Test 3  \hspace{0.48 in}    \\
      \hline
    \end{tabular}   
    \begin{tabular}{||c||c||c|c||c|c||}
 planet  \hspace{0.08 in}    &    $ \frac{ M_p  }{M_{ \oplus} } \frac{M_{\odot} } {M_{\star} }  $   \hspace{0.08 in}  &  planet  \hspace{0.08 in}  &    $ \frac{ M_p  }{M_{ \oplus} } \frac{M_{\odot} } {M_{\star} }  $ \hspace{0.08 in} &  planet  \hspace{0.08 in}  &    $ \frac{ M_p  }{M_{ \oplus} } \frac{M_{\odot} } {M_{\star} }  $ \hspace{0.08 in}  \\
 \hline 
  b  &   4.69 $\pm 1.44$   &  b  &   5.89 $^{+  1.86 }_{ -  1.92 }$  &   b  &   3.35 $^{+  1.79 }_{ -  1.71 }$    \\ 
  c  &  4.16 $^{+  1.26 }_{ -  1.27 }$   &  c  &   5.11 $^{+  1.54 }_{ -  1.61 }$ & c  &   2.58 $^{+  1.39 }_{ -  1.31 }$      \\ 
      \hline
    \end{tabular}    
    \caption{TTV solutions for Kepler-29 b and c (KOI-738), and  the results of various tests on dynamical masses. Test 1: An alternative eccentricity prior. Test 2: Robust fitting. Test 3. Holczer catalog. }\label{tbl-koi0738}
  \end{center}
\end{table}

Since these results pass all tests, we adopt our nominal dynamical masses as being secure for Kepler-29 b and c.

\subsection{Kepler-49 (KOI-248)}
Kepler-49 has four transiting planets. While the intermediate pair were confirmed by \citet{stef13} with anti-correlated TTVs at the expected periodicity, the innermost and outermost planets awaited validation by \citet{rowe14} and have been named Kepler-49 d and Kepler-49 e respectively. The period ratio of the innermost pair is 2.80, far from any first or second-order resonance. The outer pair have a period ratio of 1.7 close to the 5:3 resonance. However, being second order, this resonance is fairly weak in the regime of low orbital eccentricities and is unlikely to induce strong TTVs. To test this, we compared 4-planet models to models of just the middle pair, Kepler-49 b and c. The expected TTV periodicity of Kepler-49 b and c is around 370 days, very close to the period detected by \citet{stef13}. With dynamical fitting using a four-planet model, using Levenberg-Marquardt minimization over a grid of input parameters, our best fit models do not usefully constrain the masses for the innermost and outermost planets, but we do find strong mass detections for the intermediate planets. We include the credible intervals of the masses of the middle pair from our four-planet model in Table~\ref{tbl-koi-248} and adopt the two-planet model as our nominal result.

\begin{figure}[h!]
\includegraphics [height = 2.1 in]{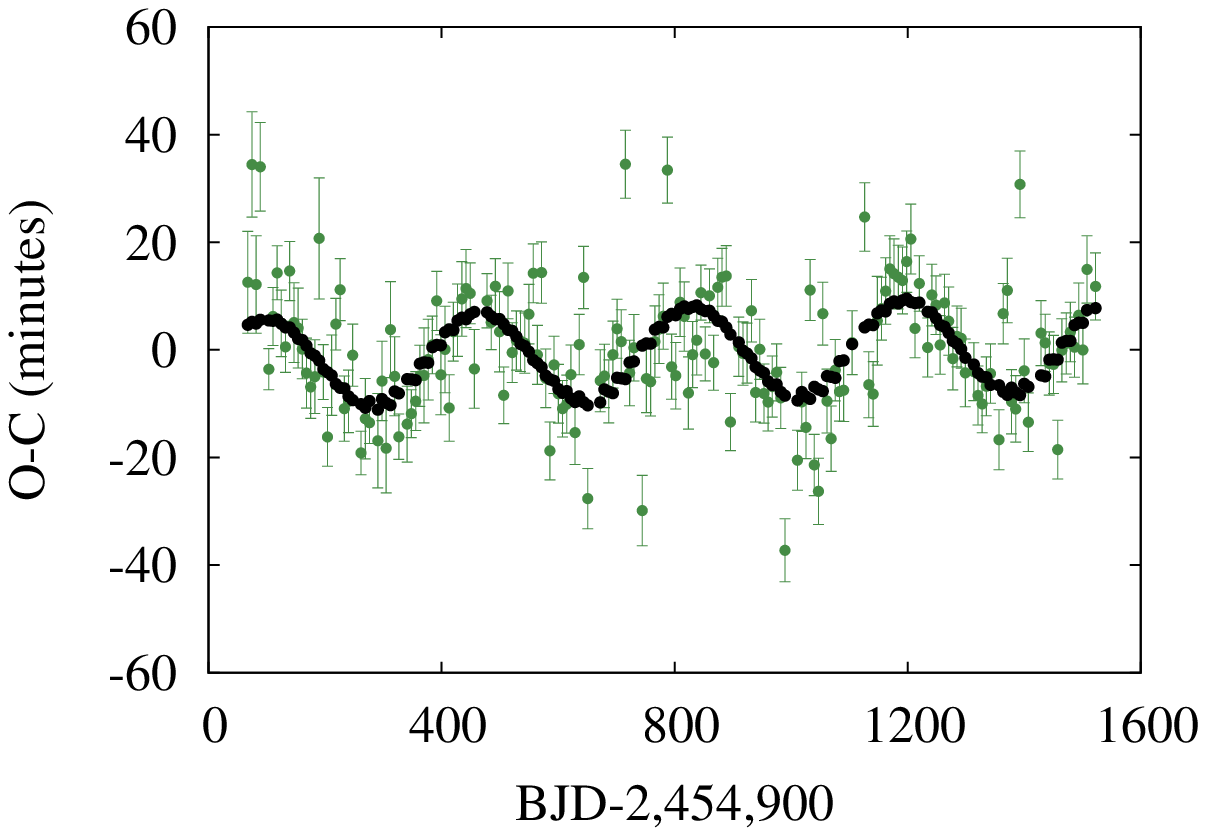}
\includegraphics [height = 2.1 in]{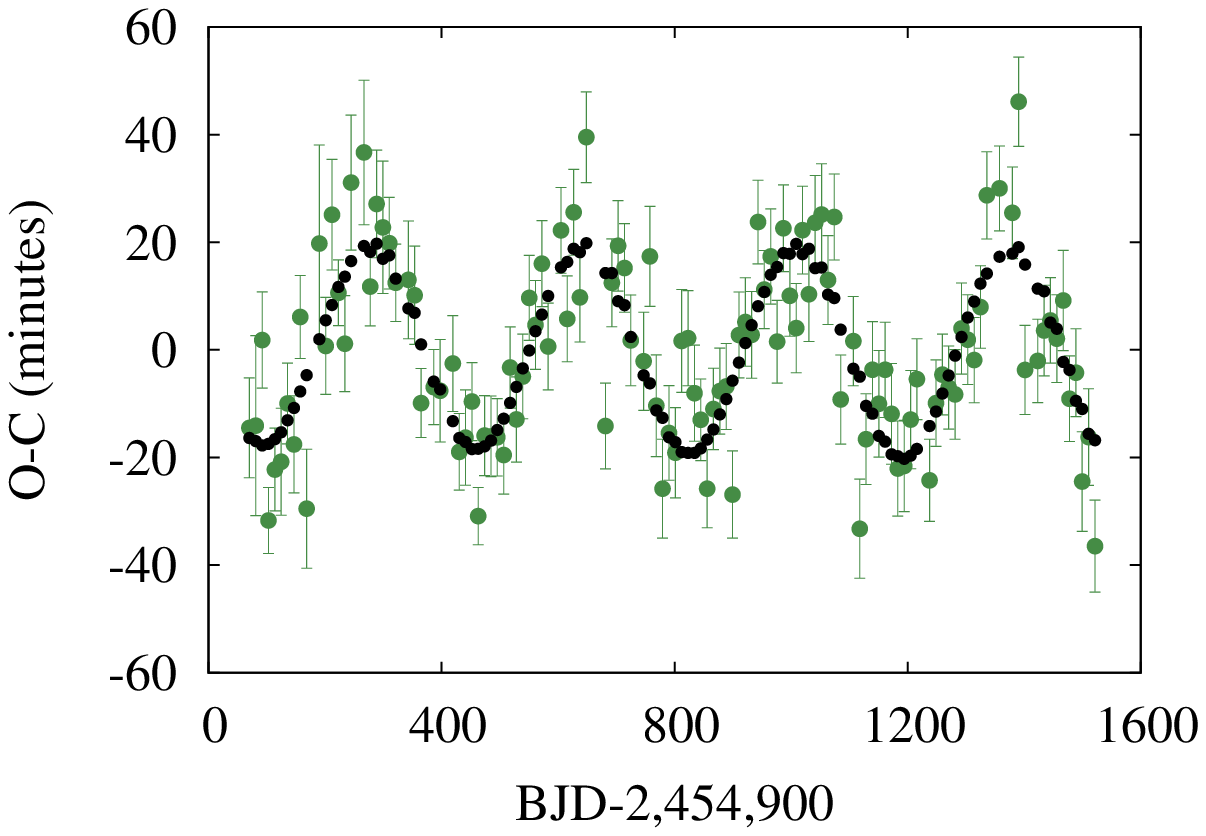}
\caption{Best fit dynamical model for Kepler-49 b (left) and c (right). In black are simulated transit times, and in green are measured transit times with their uncertainties.
}
\label{fig:248-sim} 
\end{figure}

\begin{figure}[h!]
\includegraphics [height = 1.5 in]{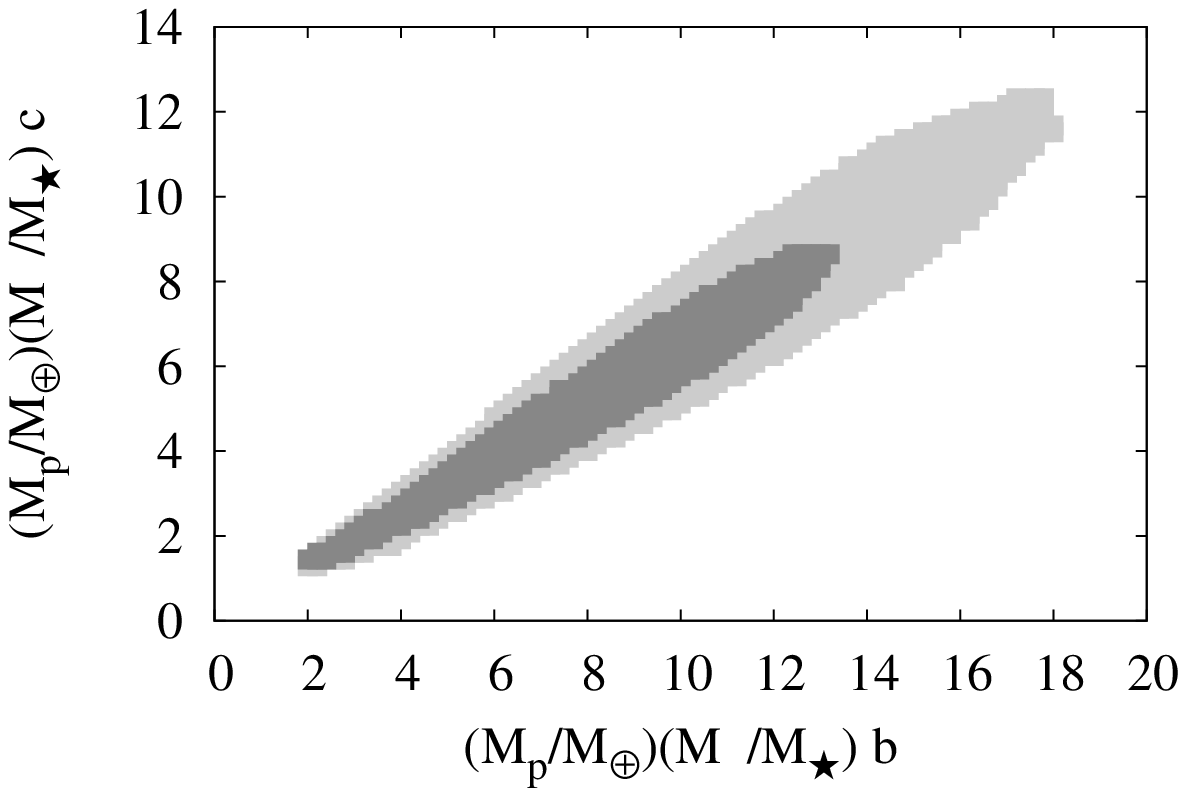}
\includegraphics [height = 1.5 in]{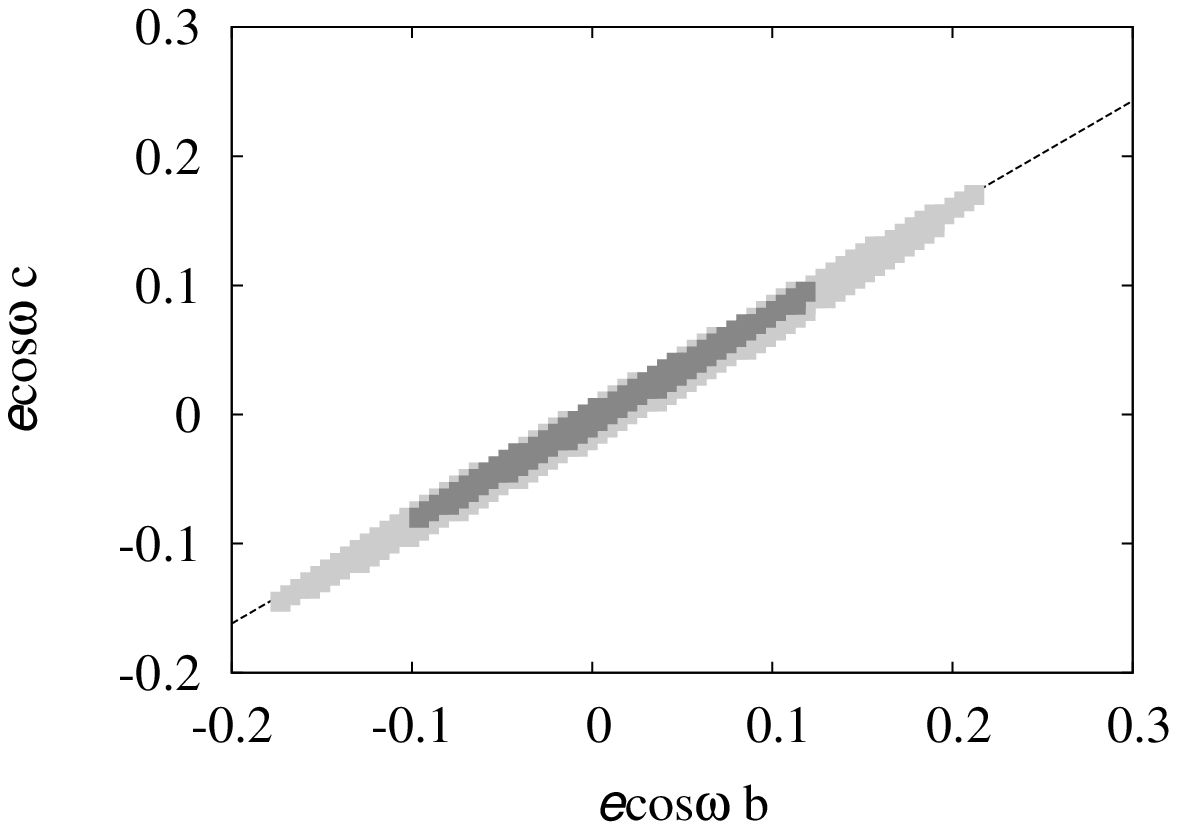}
\includegraphics [height = 1.5 in]{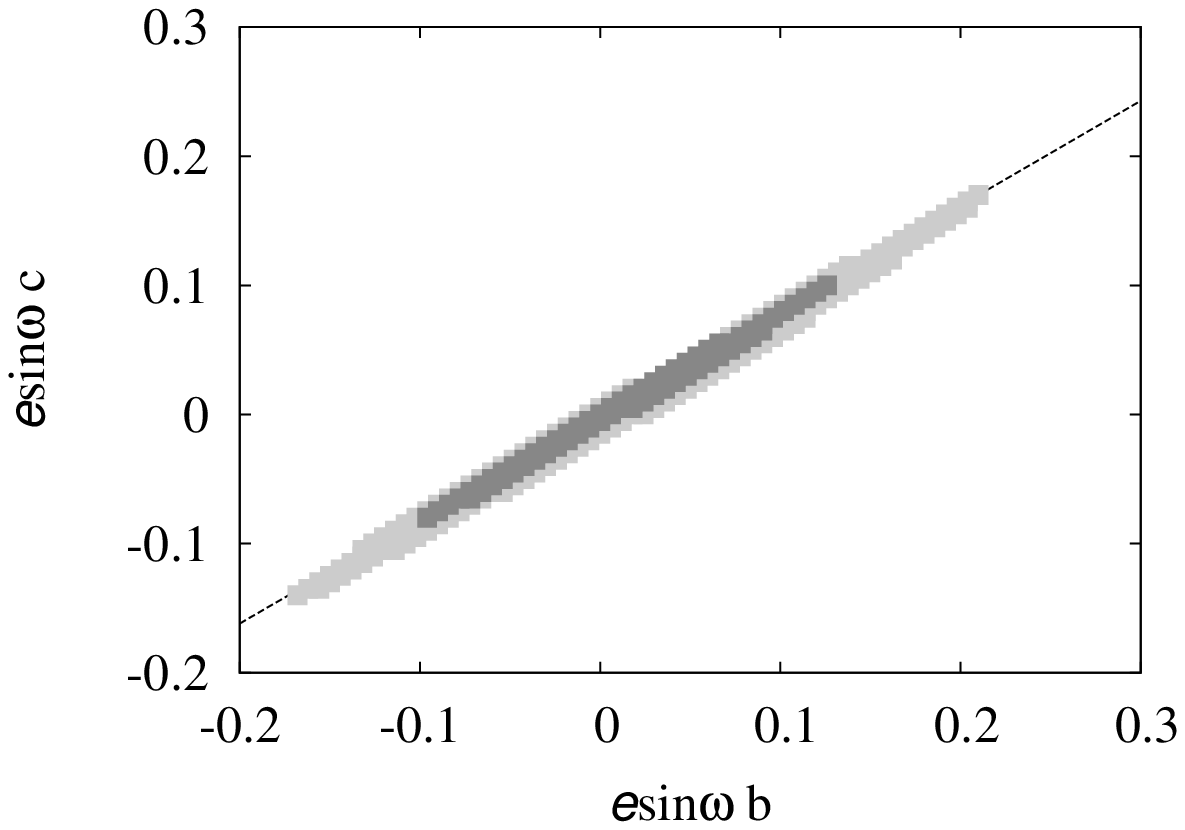}
\caption{Joint Posteriors for planetary masses relative to the host star, and eccentricity components for Kepler-49 b and c (KOI-248). 95.4\% confidence intervals are in light grey, 68.3\% confidence intervals in dark grey. The dashed lines mark the expected correlation between eccentricity vector components from Equation~\ref{eqn:gradient}.}
\label{fig:248} 
\end{figure}

Our best-fit 2-planet TTV model is displayed in Figure~\ref{fig:248-sim}, and the results of our MCMC analysis is in Table~\ref{tbl-koi-248}. Once again, the eccentricities are poorly constrained, although we infer strong constraints on dynamical masses. Figure~\ref{fig:248} shows that the masses of the two planets are correlated, and that the eccentricity vector components are likely limited only by their prior, although the relative eccentricities are tightly constrained.  

%Stellar parameters updated in \citet{swift15}.
 \begin{table}[h!]
  \begin{center}
       \begin{tabular}{||c||}
      \hline
 \hspace{1.75 in} Adopted  parameters  with 1$\sigma$ (2$\sigma$) credible intervals     \hspace{1.75 in} \\
      \hline
    \end{tabular} 
    \begin{tabular}{||c|c|c|c|c|c||}
      \hline
 planet  & $P$ (days)  &      $T_0$ (days)    &      $e\cos\omega$   &  $e\sin\omega$     &     $ \frac{ M_p  }{M_{ \oplus} } \frac{M_{\odot} } {M_{\star} }  $($\pm 2\sigma$)  \\   
 \hline 
    b  & \textbf{ 7.2040 }$\pm 0.0002$ &  \textbf{ 780.4529 }$\pm 0.0006 $ &  \textbf{ 0.011 }$^{+ 0.074 }_{ - 0.067 }$ &  \textbf{ 0.037 }$^{+ 0.062 }_{ - 0.068 }$ &  \textbf{  9.16 }$^{+  3.75 }_{ -  3.46 }$ $\left(^{+  8.03 }_{ -  5.31 }\right)$   \\ 
  c & \textbf{ 10.9123 }$\pm 0.0006$ &  \textbf{ 790.3470 }$\pm 0.0011 $ &  \textbf{ 0.006 }$^{+ 0.059}_{ - 0.054 }$ &  \textbf{ 0.027 }$^{+ 0.051 }_{ - 0.057 }$ &  \textbf{  5.91 }$^{+  2.65 }_{ -  2.33 }$$\left(^{+  5.88 }_{ -  3.52 }\right)$   \\
    \hline
    \end{tabular}
        \begin{tabular}{||c||c||c||c||}
      \hline
  \hspace{0.5 in}  Test 1   \hspace{0.5 in}    &   \hspace{0.5 in} Test 2  \hspace{0.5 in} &   \hspace{0.5 in} Test 3  \hspace{0.5 in}  &   \hspace{0.5 in} Test 4  \hspace{0.5 in}  \\
      \hline
    \end{tabular}   
    \begin{tabular}{||c|c||c|c||c|c||c|c||}
 planet  \hspace{0.2 in}    &    $ \frac{ M_p  }{M_{ \oplus} } \frac{M_{\odot} } {M_{\star} }  $   \hspace{0.2 in}  &  planet  \hspace{0.25 in}  &    $ \frac{ M_p  }{M_{ \oplus} } \frac{M_{\odot} } {M_{\star} }$  & planet  \hspace{0.25 in}  &     $ \frac{ M_p  }{M_{ \oplus} } \frac{M_{\odot} } {M_{\star} }  $ \hspace{0.2 in} & planet  \hspace{0.2 in}  &     $ \frac{ M_p  }{M_{ \oplus} } \frac{M_{\odot} } {M_{\star} }  $ \hspace{0.2 in}  \\
 \hline 
  b  &   9.65 $^{+  3.29 }_{ -  3.19 }$   &  b  &   9.26 $^{+  4.11 }_{ -  3.50 }$ &  b  &   15.35 $^{+  5.03 }_{ -  3.96 }$  &  b  &   11.12 $^{+  5.93 }_{ -  2.78 }$       \\ 
  c  &  6.28 $^{+  2.36 }_{ -  2.18 }$   &  c  &   7.08 $^{+  3.55 }_{ -  2.80 }$  & c  &   14.39 $^{+  5.71 }_{ -  4.24 }$ & c  &   7.32 $^{+  4.49 }_{ -  2.21 }$      \\ 
      \hline
    \end{tabular}  
    \caption{TTV solutions for Kepler-49 b and c (KOI-248) and  the results of various tests on dynamical masses. Test 1: An alternative eccentricity prior. Test 2: Robust fitting  Test 3. The Holczer catalog.  Test 4: a four-planet model.}\label{tbl-koi-248}
  \end{center}
\end{table}

Our solutions were insensitive to the choice of prior and the outliers, although we find higher masses with the Holczer catalog of transit times. We also found that a four-planet model allowed agreement at the 1$\sigma$ level for the dynamical masses of Kepler-49 b and c in a two-planet model. Both transit time catalogs were plagued with outliers for Kepler-49. Furthermore, our light-curve analysis of Kepler-49 c yields an inconsistent measure of the stellar bulk density compared to the three other transiting planets, as discussed in more detail in Section 5. Since the TTVs of this discrepant transit model were used in this TTV analysis, we flag this system as one requiring more analysis to obtain reliable results. Hence, we exclude Kepler-49 b and c from our list of reliably measured planetary masses.

\newpage
\subsection{Kepler-57 (KOI-1270)}
\begin{figure}[h!]
\includegraphics [height = 2.1 in]{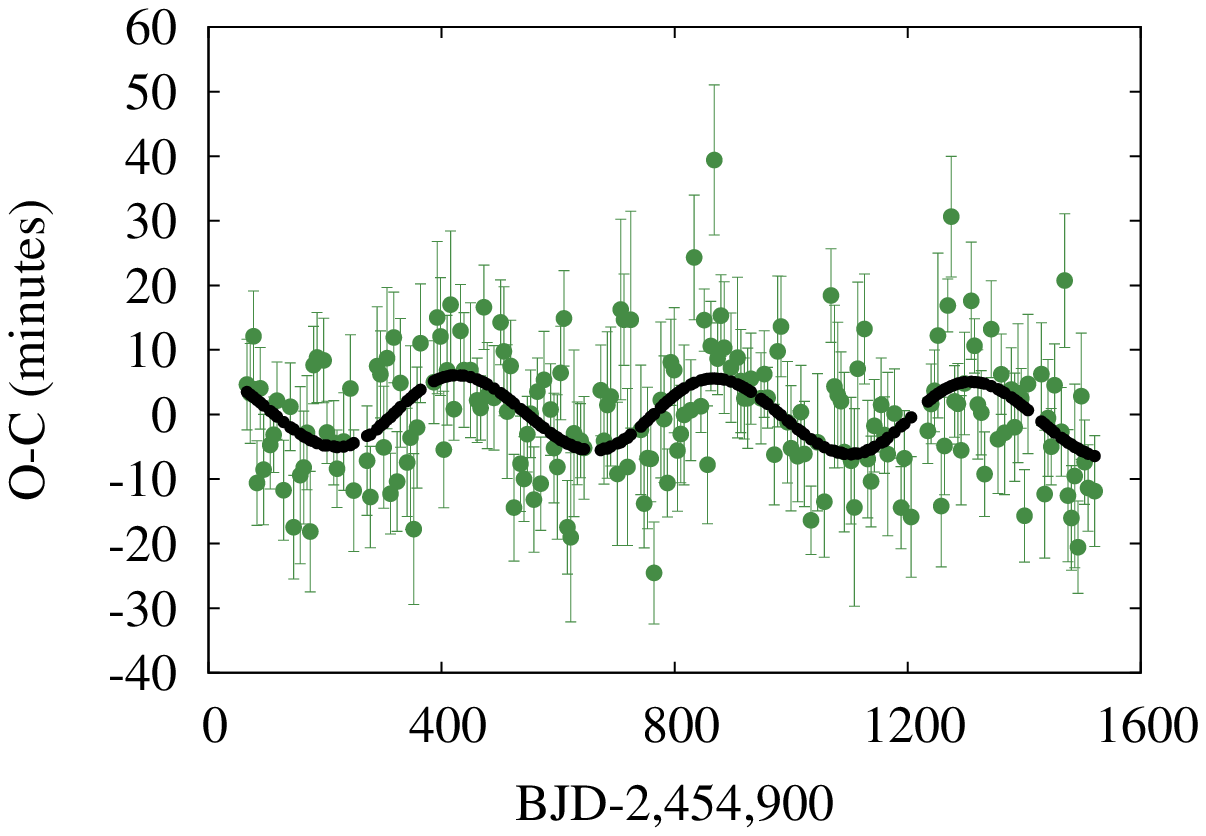}
\includegraphics [height = 2.1 in]{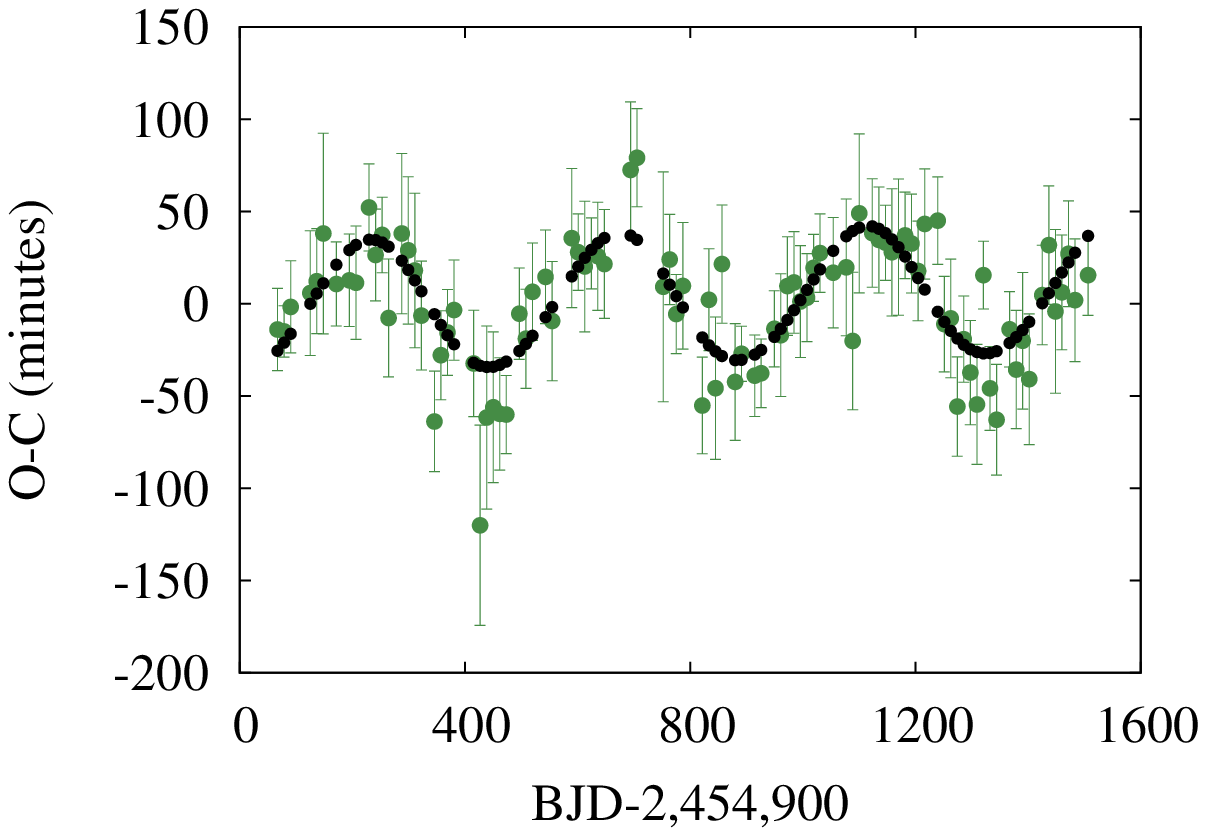}
\caption{Best fit dynamical model for Kepler-57 b (left) and c (right). In black are simulated transit times, and in green are measured transit times with their uncertainties.
}
\label{fig:1270-sim} 
\end{figure}

Two transiting planets with anti-correlated TTVs are known to orbit Kepler-57 (\citealt{stef12,had14}). Given their proximity to the first order 2:1 mean motion resonance, their expected TTV period of 456 days is easily seen in Figure~\ref{fig:1270-sim}. 

 \begin{table}[h!]
  \begin{center}
       \begin{tabular}{||c||}
      \hline
 \hspace{1.75 in} Adopted  parameters  with 1$\sigma$ (2$\sigma$) credible intervals     \hspace{1.75 in} \\
      \hline
    \end{tabular} 
    \begin{tabular}{||c|c|c|c|c|c||}
      \hline
 planet  & $P$ (days)  &      $T_0$ (days)    &      $e\cos\omega$   &  $e\sin\omega$     &     $ \frac{ M_p  }{M_{ \oplus} } \frac{M_{\odot} } {M_{\star} }  $($\pm 2\sigma$)  \\   
 \hline 
   b  & \textbf{ 5.7295 }$\pm 0.0006 $ &  \textbf{ 781.9966 }$\pm 0.0006 $ &  \textbf{ 0.018 }$^{+ 0.009 }_{ - 0.012 }$ &  \textbf{ --0.019 }$^{+ 0.026 }_{ - 0.010 }$ &  \textbf{ 27.81 }$^{+ 11.40 }_{ - 10.61 }$ $\left(^{+  22.73 }_{ -  15.85 }\right)$   \\ 
  c & \textbf{ 11.6065 }$\pm  0.0006$ &  \textbf{ 786.7562 }$^{+ 0.0020 }_{ - 0.0021 }$ &  \textbf{ 0.030 }$^{+ 0.014 }_{ - 0.055 }$ &  \textbf{ --0.036 }$^{+ 0.076 }_{ - 0.020 }$ &  \textbf{  6.62 }$^{+  3.10 }_{ -  2.80 }$   $\left(^{+  5.97 }_{ -  4.14 }\right)$ \\
    \hline
    \end{tabular}
        \begin{tabular}{||c||c||c||}
      \hline
  \hspace{0.78 in}  Test 1   \hspace{0.78 in}    &   \hspace{0.78 in} Test 2  \hspace{0.78 in} &   \hspace{0.78 in} Test 3  \hspace{0.78 in}   \\
      \hline
    \end{tabular}   
    \begin{tabular}{||c|c||c|c||c|c||}
 planet  \hspace{0.48 in}    &    $ \frac{ M_p  }{M_{ \oplus} } \frac{M_{\odot} } {M_{\star} }  $   \hspace{0.48 in}  &  planet  \hspace{0.48 in}    &    $ \frac{ M_p  }{M_{ \oplus} } \frac{M_{\odot} } {M_{\star} }  $   \hspace{0.48 in}  &  planet  \hspace{0.48 in}    &    $ \frac{ M_p  }{M_{ \oplus} } \frac{M_{\odot} } {M_{\star} }  $   \hspace{0.48 in}   \\
 \hline 
  b  &   33.79 $^{+  10.25 }_{ -  10.83 }$   &  b  &   16.97 $^{+  12.67 }_{ -  9.54 }$    &  b  &   17.24 $^{+  9.93 }_{ -  7.33 }$       \\ 
  c  &  7.80 $^{+  2.85 }_{ -  2.83 }$   &  c  &   3.98 $^{+  3.66 }_{ -  2.39 }$  &  c  &   3.22 $^{+  2.01 }_{ -  1.43 }$     \\ 
      \hline
    \end{tabular}  
    \caption{TTV solutions for Kepler-57 b and c (KOI-1270) and  the results of various tests on dynamical masses. Test 1: An alternative eccentricity prior. Test 2: Robust fitting.  Test 3. Holczer catalog }\label{tbl-koi-1270}
  \end{center}
\end{table}

\begin{figure}[h!]
\includegraphics [height = 1.5 in]{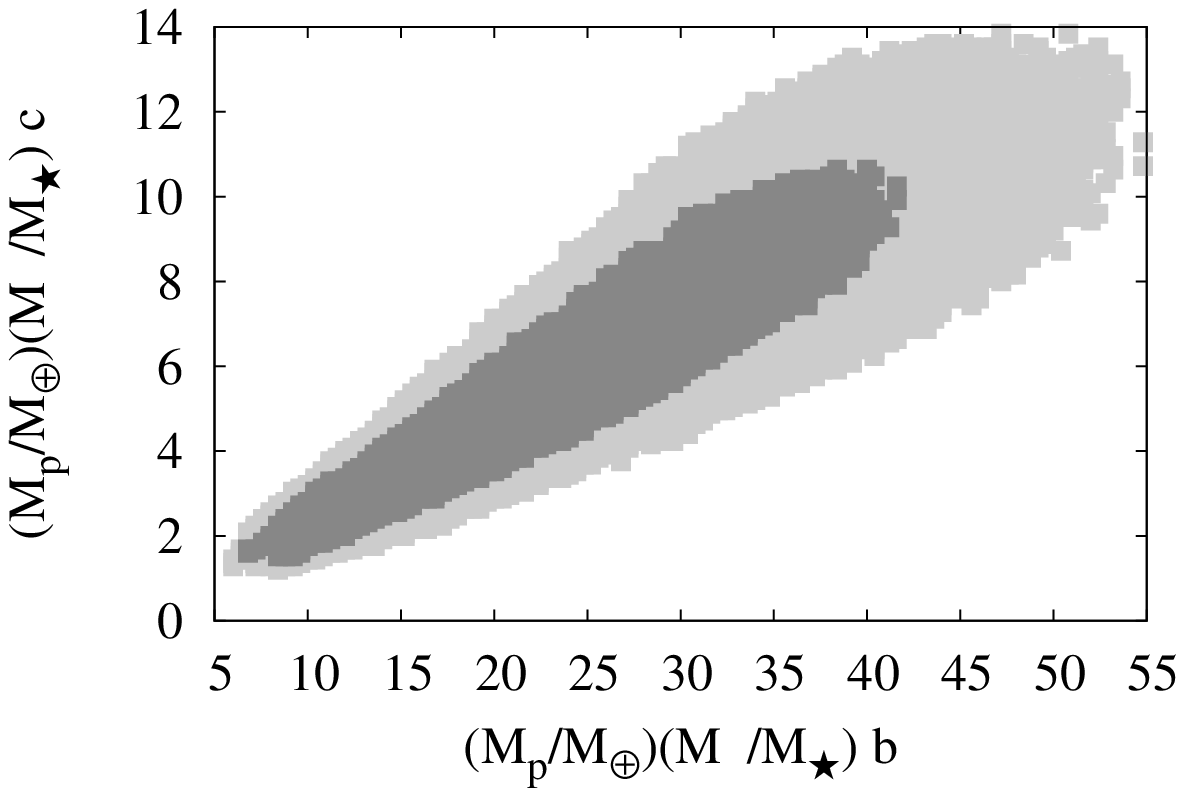}
\includegraphics [height = 1.5 in]{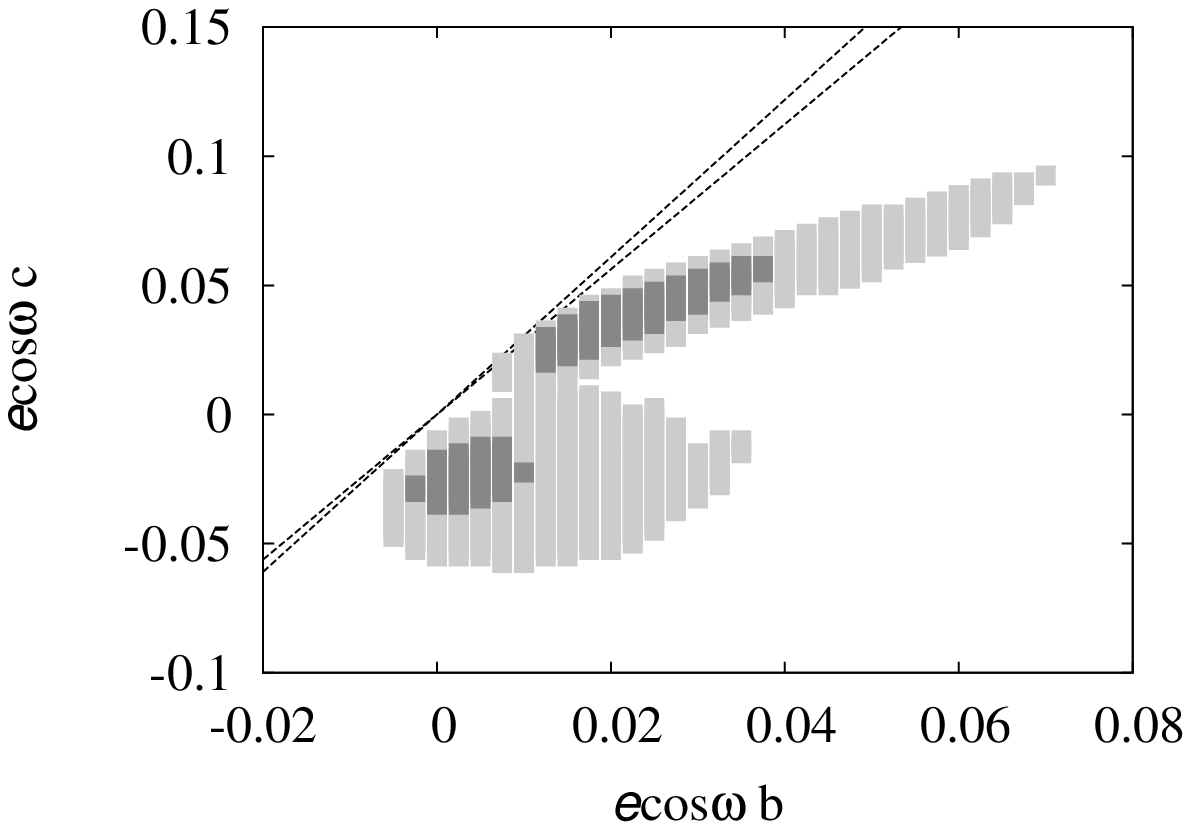}
\includegraphics [height = 1.5 in]{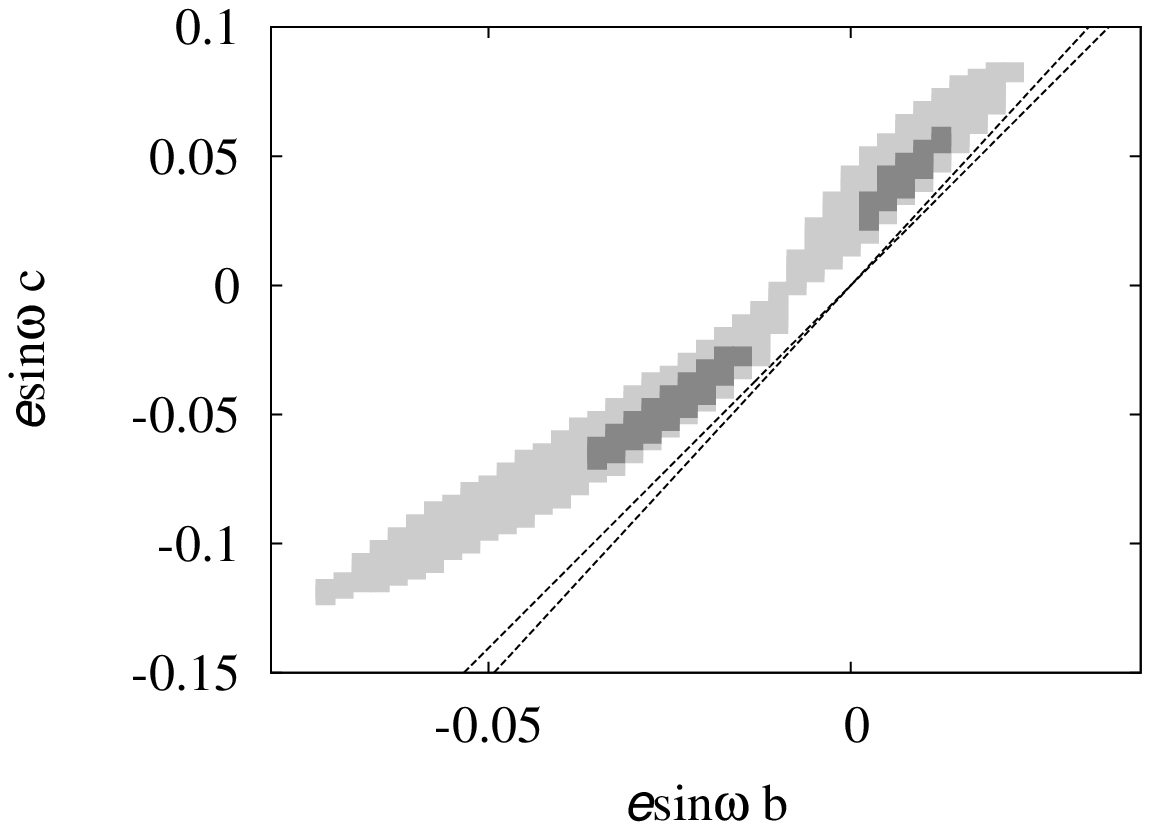}
\caption{Joint Posteriors for planetary masses relative to the host star, and eccentricity components for Kepler-57 b and c (KOI-1270). 95.4\% credible intervals are in light grey, 68.3\% credible intervals in dark grey.}
\label{fig:1270} 
\end{figure}

Our results in Table~\ref{tbl-koi-1270} and the joint posteriors shown in Figure~\ref{fig:1270} show that the dynamical mass of Kepler-57 b is less well constrained than that of its neighbor `c'. We also note that the bimodal posterior for the eccentricity vector components makes the TTV modeling for this system rather slow to converge. Our nominal solutions here result in bulk densities for both planets that are consistent with rock. However, our results are not robust against outliers, and there is moderate disagreement in inferred dynamical masses from our independent datasets of measured transit times. In addition, as noted in Section 5, transit models for Kepler-57 b and c give an estimate for the stellar density that is inconsistent spectral observations, perhaps due to a low-mass stellar companion. We therefore omit these planets from the mass-radius diagram.

%For Kepler-57 (KOI-1270), all sample solutions were stable to 1Myr.

\subsection{Kepler-60 (KOI-2086)}
\begin{figure}[h!]
\includegraphics [height = 1.5 in]{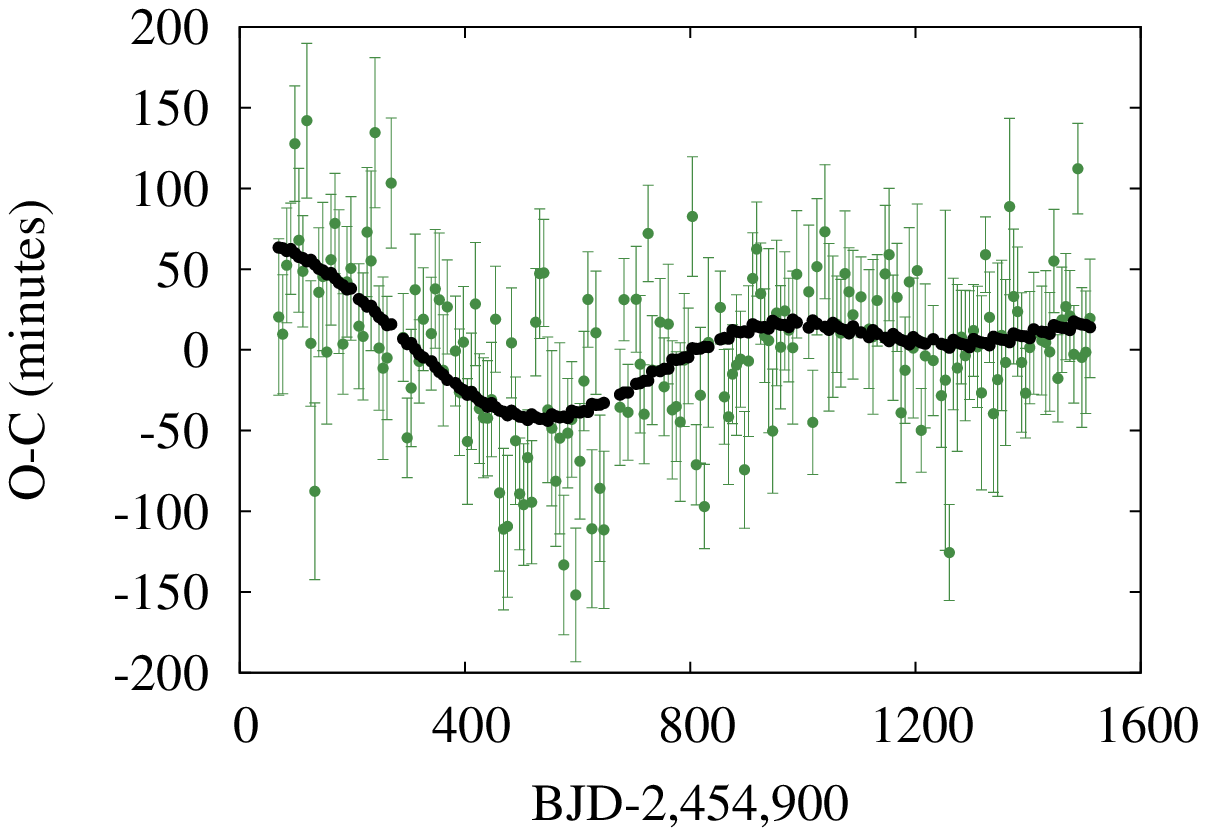}
\includegraphics [height = 1.5 in]{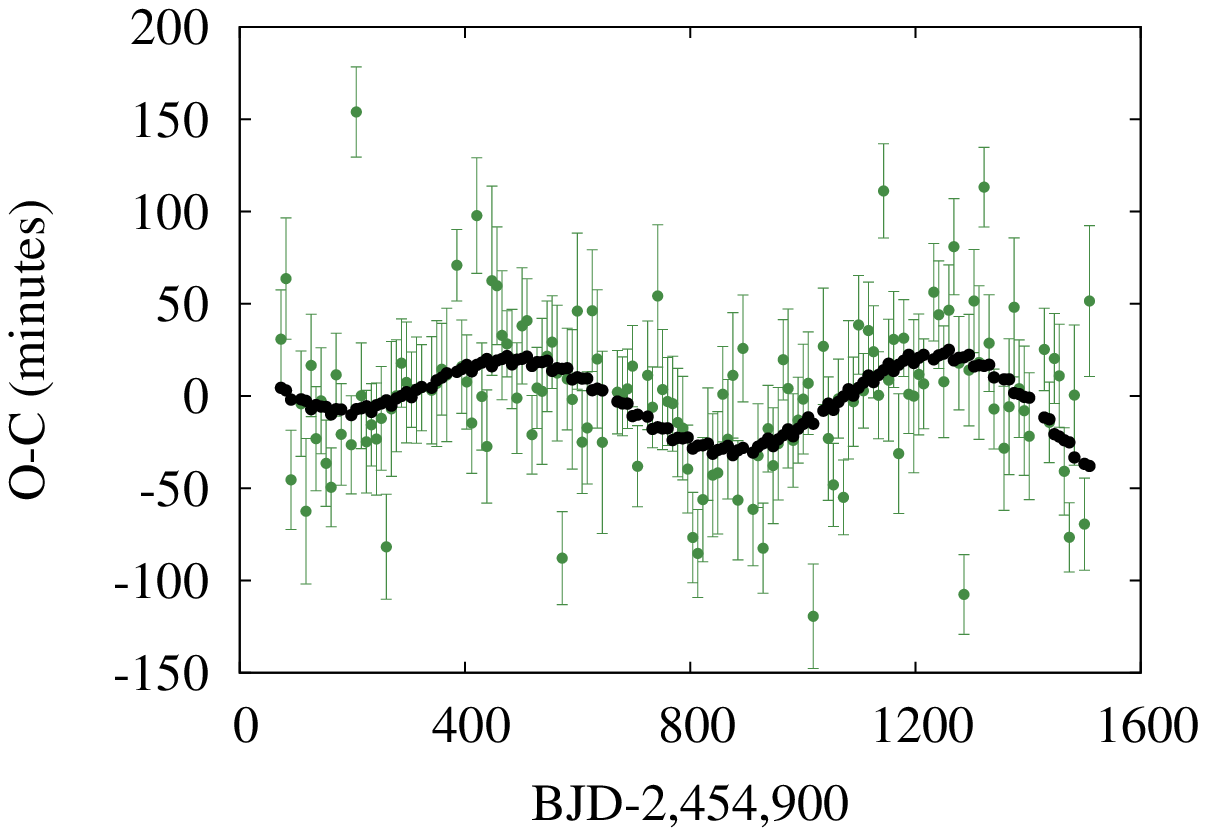}
\includegraphics [height = 1.5 in]{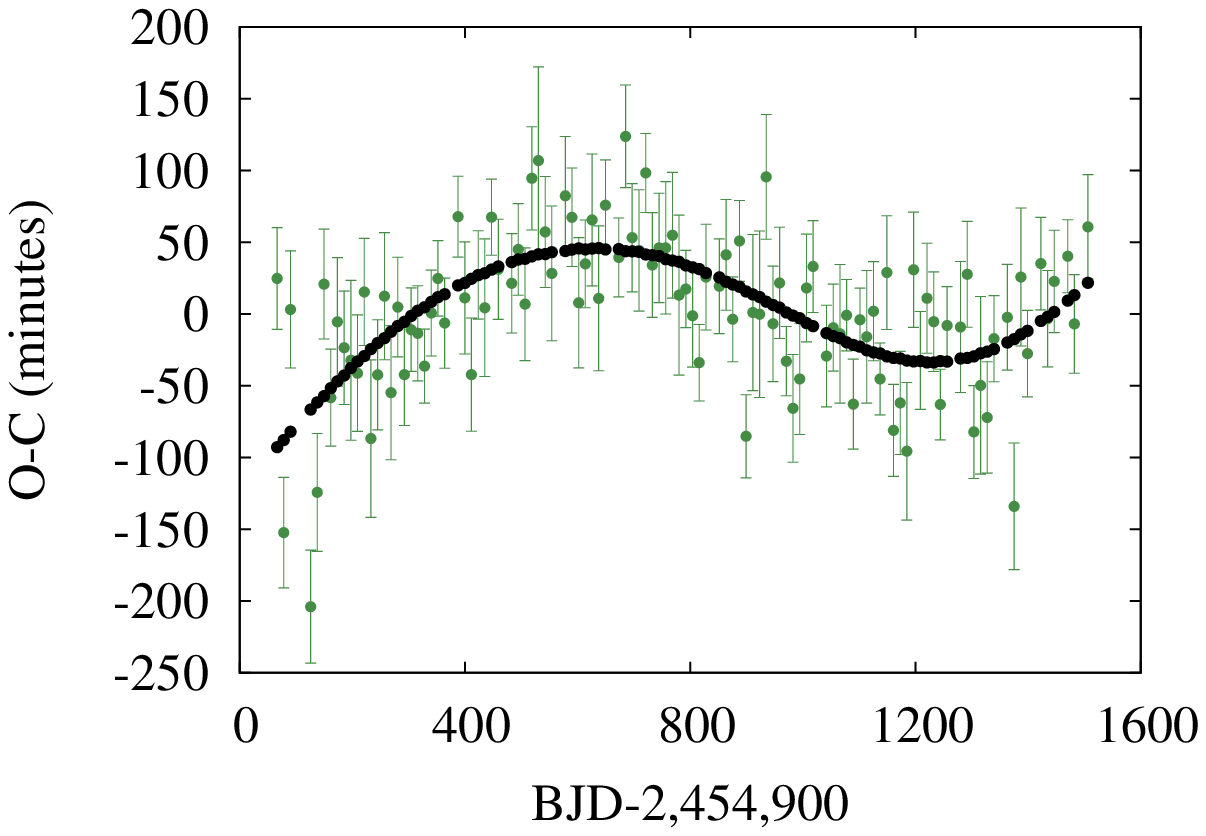}
\caption{Best fit dynamical model for Kepler-60 b (left), c (middle) and d (right). In black are simulated transit times, and in green are measured transit times with their uncertainties.
}
\label{fig:2086-sim} 
\end{figure}

Kepler-60 has three confirmed planets in a very compact configuration \citep{ste12}. The inner pair orbit near the 5:4 mean motion resonance, and the outer pair are near the 4:3 resonance, leading to a near 5:4:3 chain of commensurability in mean motions, where the resonant quantity $|n_b-2n_c+n_d| \lesssim$ 0.005$^{\circ}$/day. The TTVs for all three planets are plotted Figure~\ref{fig:2086-sim}.

 \begin{table}[h!]
  \begin{center}
       \begin{tabular}{||c||}
      \hline
 \hspace{1.75 in} Adopted  parameters  with 1$\sigma$ (2$\sigma$) credible intervals  \hspace{1.75 in} \\
      \hline
    \end{tabular} 
    \begin{tabular}{||c|c|c|c|c|c||}
      \hline
 planet  & $P$ (days)  &      $T_0$ (days)    &      $e\cos\omega$   &  $e\sin\omega$     &     $ \frac{ M_p  }{M_{ \oplus} } \frac{M_{\odot} } {M_{\star} }  $($\pm 2\sigma$)  \\   
 \hline 
    b  & \textbf{ 7.1334 }$\pm 0.0001$ &  \textbf{ 782.2796 }$^{+ 0.0034 }_{ - 0.0033 }$ &  \textbf{ 0.023 }$^{+ 0.067 }_{ - 0.069 }$ &  \textbf{ 0.008 }$^{+ 0.060 }_{ - 0.059 }$ &  \textbf{  4.02 }$^{+  0.43 }_{ -  0.42 }$   $\left(^{+  0.88 }_{ -  0.66 }\right)$   \\ 
  c & \textbf{ 8.9187 }$\pm 0.0002 $ &  \textbf{ 786.5827 }$\pm 0.0032 $ &  \textbf{ --0.003 }$^{+ 0.062 }_{ - 0.063 }$ &  \textbf{ 0.034 }$^{+ 0.054 }_{ - 0.053 }$ &  \textbf{  3.70 }$^{+  0.72 }_{ -  0.73 }$  $\left(^{+  1.44 }_{ -  1.16 }\right)$    \\ 
  d  & \textbf{ 11.8981 }$\pm 0.0002$ &  \textbf{ 780.2779 }$^{+ 0.0038 }_{ - 0.0037 }$ &  \textbf{ 0.021 }$^{+ 0.052 }_{ - 0.053 }$ &  \textbf{ 0.002 }$^{+ 0.047 }_{ - 0.046 }$ &  \textbf{  4.00 }$^{+  0.73 }_{ -  0.66 }$  $\left(^{+  1.58 }_{ -  1.01 }\right)$    \\
    \hline
    \end{tabular}
        \begin{tabular}{||c||c||}
      \hline
  \hspace{0.78 in}  Test 1   \hspace{0.78 in}    &   \hspace{0.78 in} Test 2  \hspace{0.78 in}   \\
      \hline
    \end{tabular}   
    \begin{tabular}{||c|c||c|c||}
 planet  \hspace{0.48 in}    &    $ \frac{ M_p  }{M_{ \oplus} } \frac{M_{\odot} } {M_{\star} }  $   \hspace{0.48 in}  &  planet  \hspace{0.48 in}  &    $ \frac{ M_p  }{M_{ \oplus} } \frac{M_{\odot} } {M_{\star} }  $ \hspace{0.48 in}  \\
 \hline 
  b  &   4.02 $^{+  0.44 }_{ -  0.42 }$   &  b  &   4.07 $^{+  0.61 }_{ -  0.59 }$         \\ 
c  &  3.57 $^{+  0.72 }_{ -  0.74 }$   &  c  &   3.93 $^{+  0.88 }_{ -  0.92 }$      \\ 
d  &  3.95 $^{+  0.73 }_{ -  0.66 }$   &  d  &   4.14 $^{+  1.11 }_{ -  0.92 }$      \\ 
      \hline
    \end{tabular}  
    \caption{TTV solutions for Kepler-60 b, c and d (KOI-2086) and  the results of various tests on dynamical masses. Test 1: An alternative eccentricity prior. Test 2: Robust fitting.  }\label{tbl-koi2086}
  \end{center}
\end{table}

The results of our MCMC analysis for the TTVs of Kepler-60 are in Table~\ref{tbl-koi2086}. Our results were robust against the choice of eccentricity prior and outlying transit times. The Holczer catalog does not include measured transit times for all three planets of Kepler-60, although a three-planet model is required to study this system. Hence we have only the tests shown in Table~\ref{tbl-koi2086} for this system. \citet{goz15} has performed an independent TTV analysis of this system, using long cadence data. With short cadence data, we find more precise bounds on the dynamical masses of the planets. Our results are closely consistent, with differences at the 1$\sigma$ level only for the middle planet, Kepler-60 c. We have also significantly improved upon the precision of the stellar parameters for this system and place these planets on the mass-radius diagram in Section 6. 

\begin{figure}[!htb]
\includegraphics [height = 1.5 in]{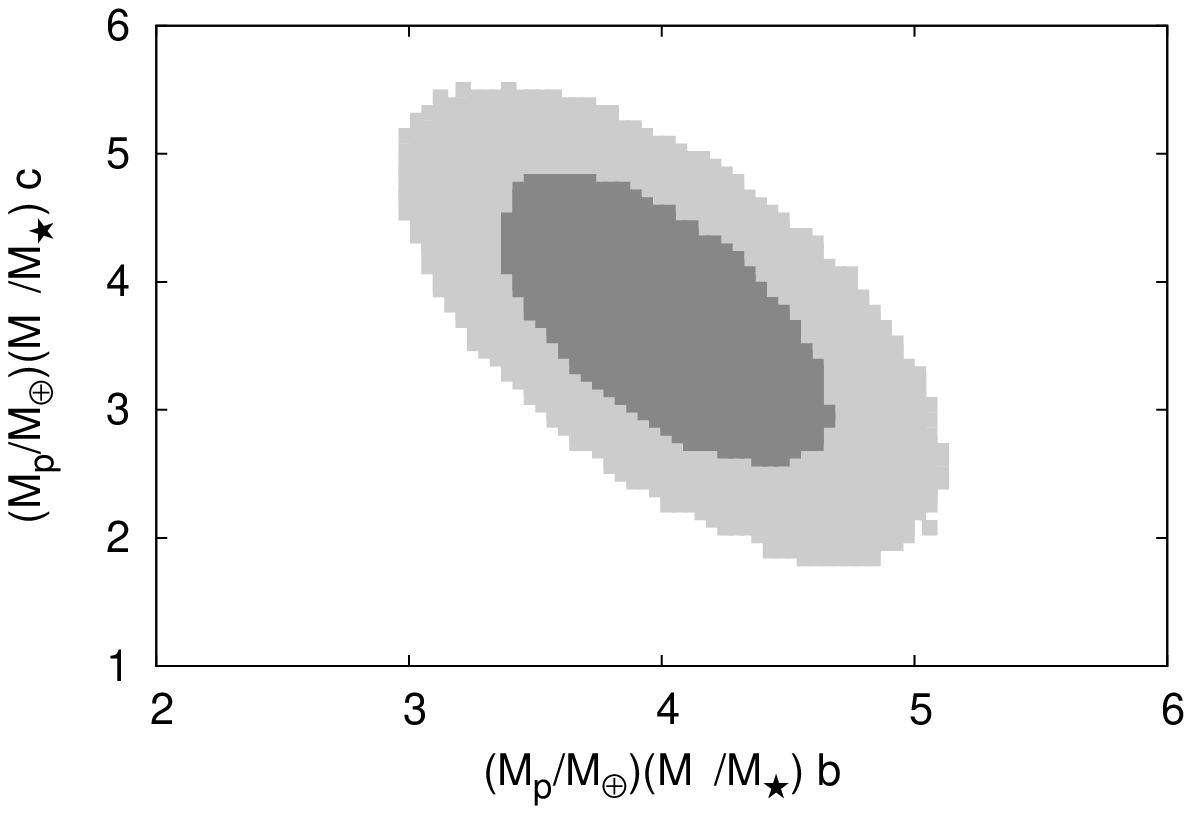}
\includegraphics [height = 1.5 in]{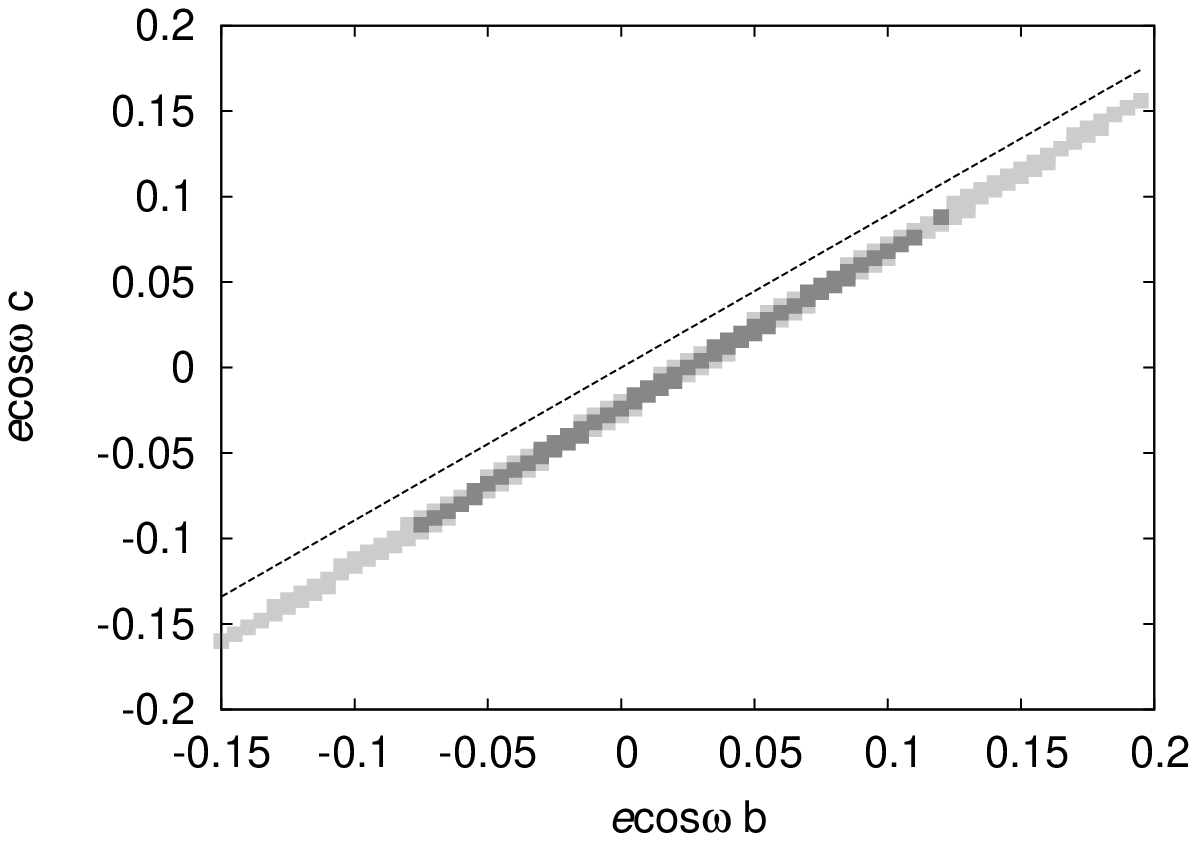}
\includegraphics [height = 1.5 in]{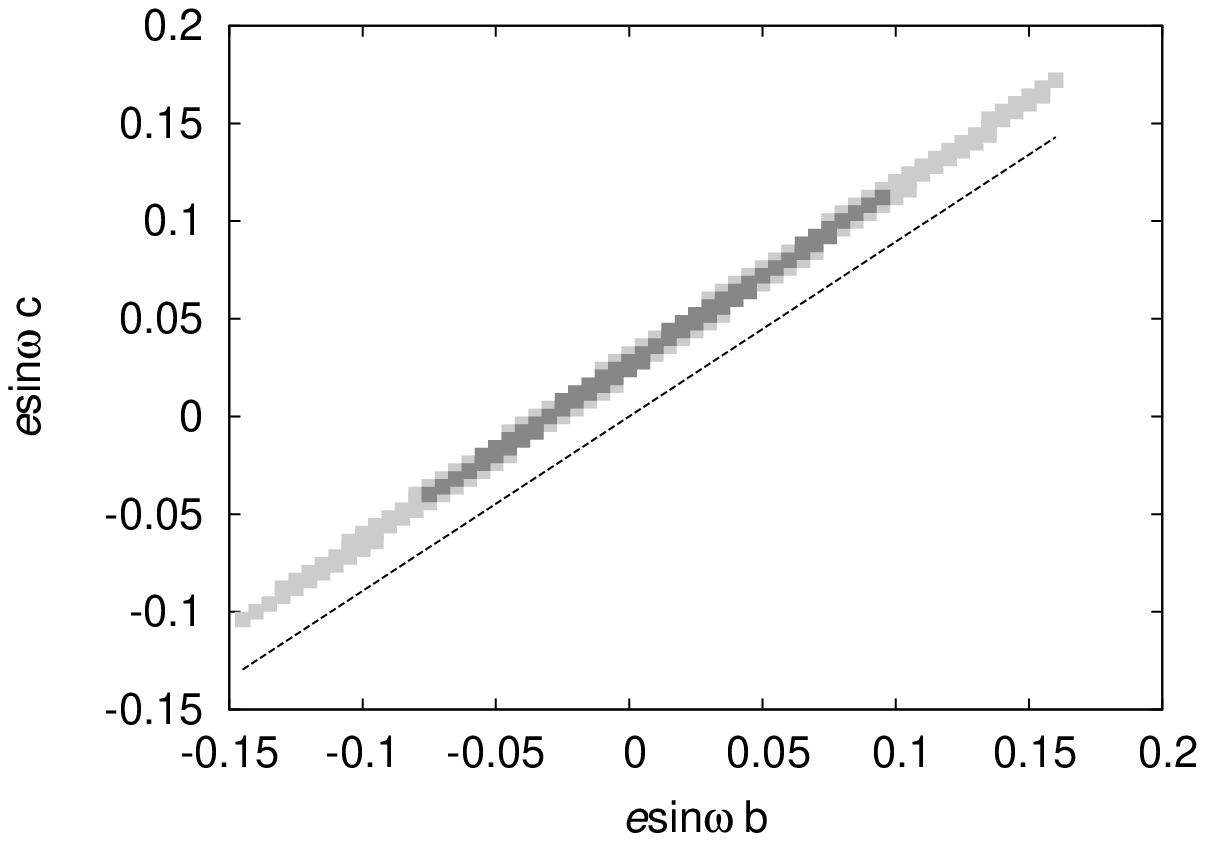}
\includegraphics [height = 1.5 in]{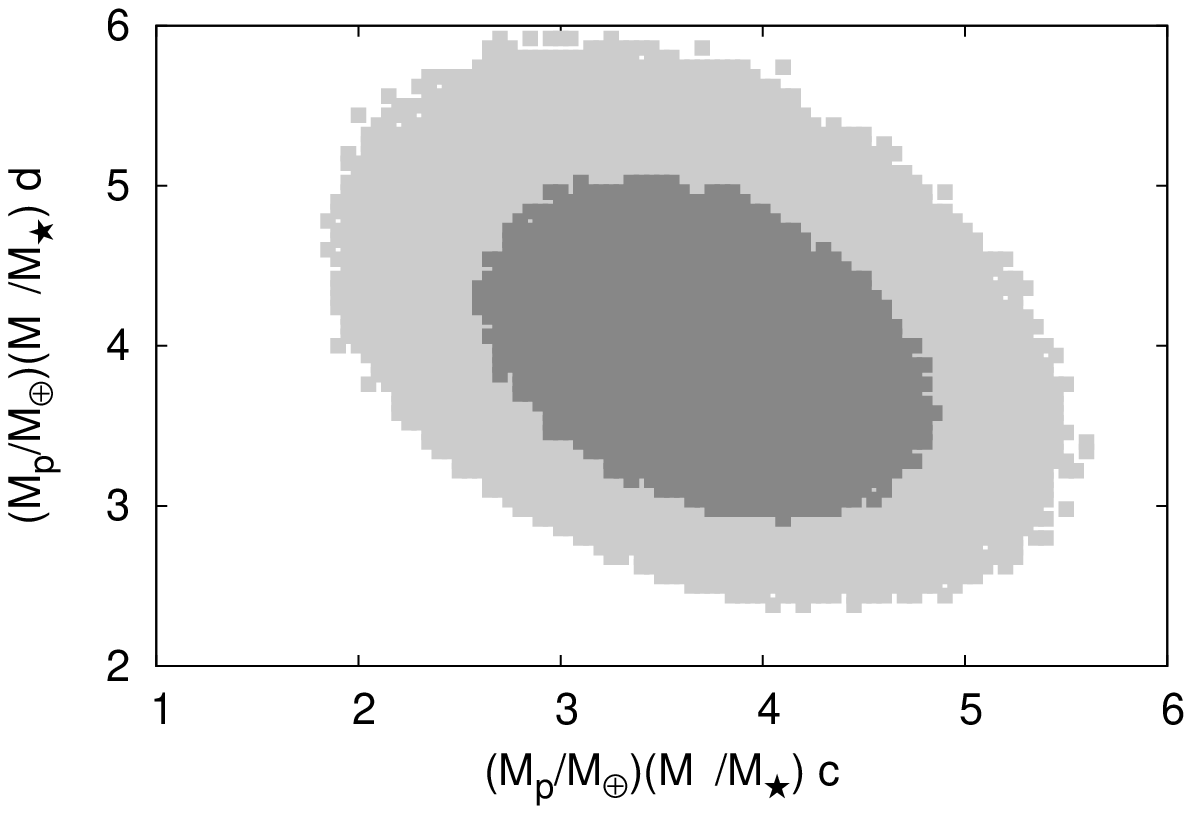}
\includegraphics [height = 1.5 in]{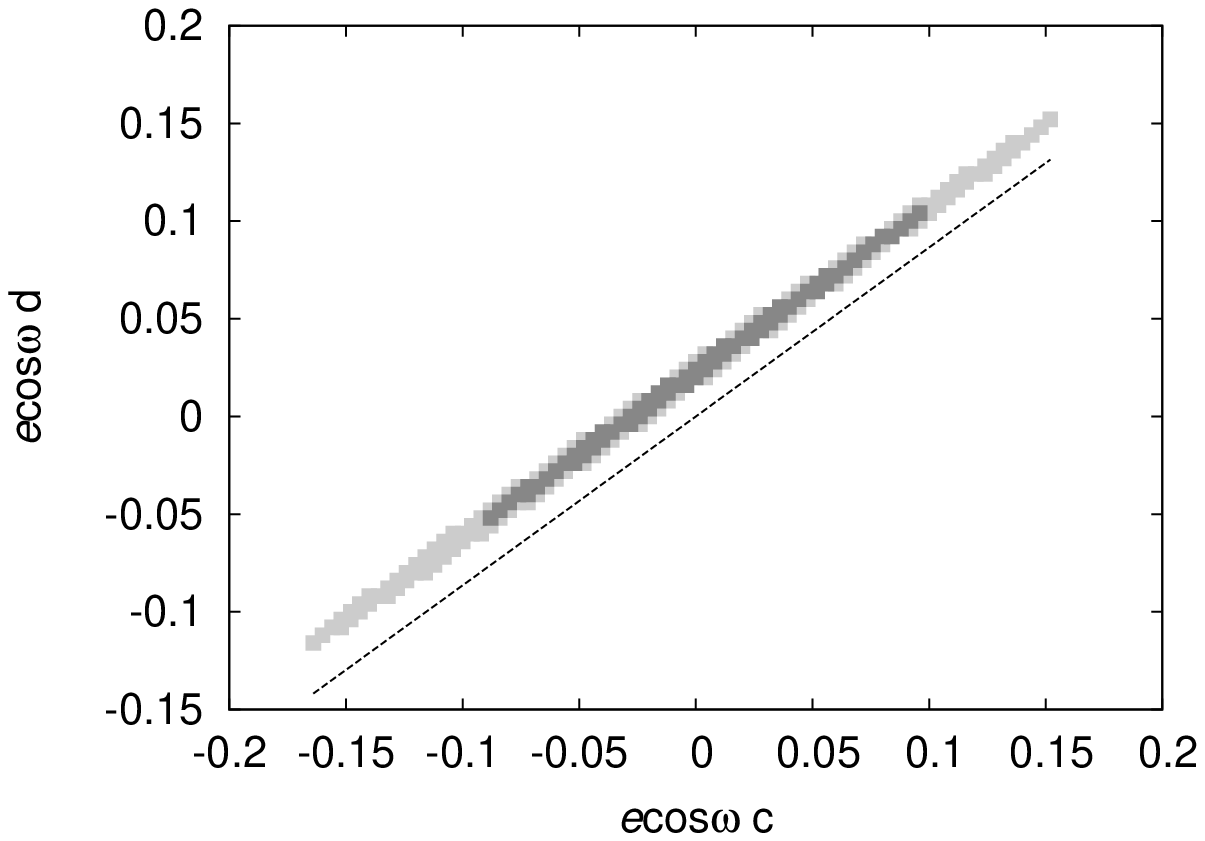}
\includegraphics [height = 1.5 in]{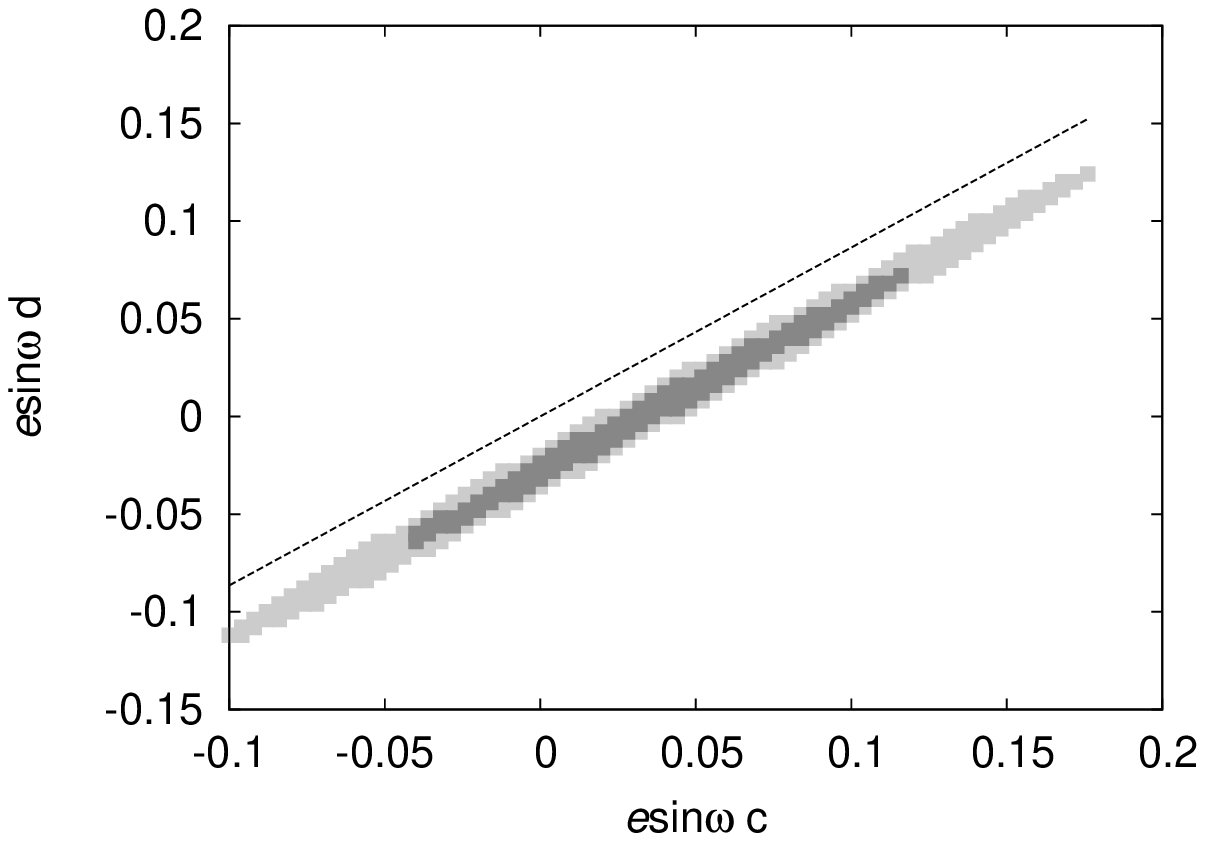}
\includegraphics [height = 1.5 in]{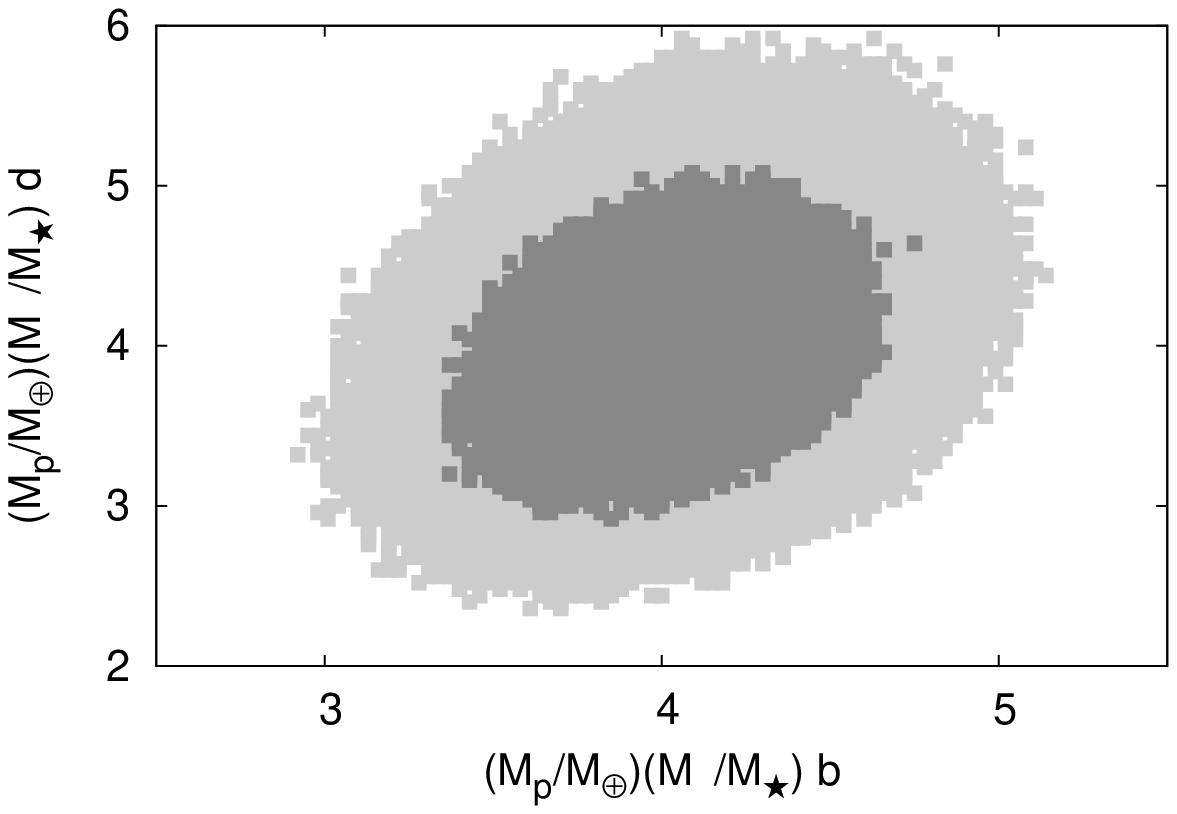}
\includegraphics [height = 1.5 in]{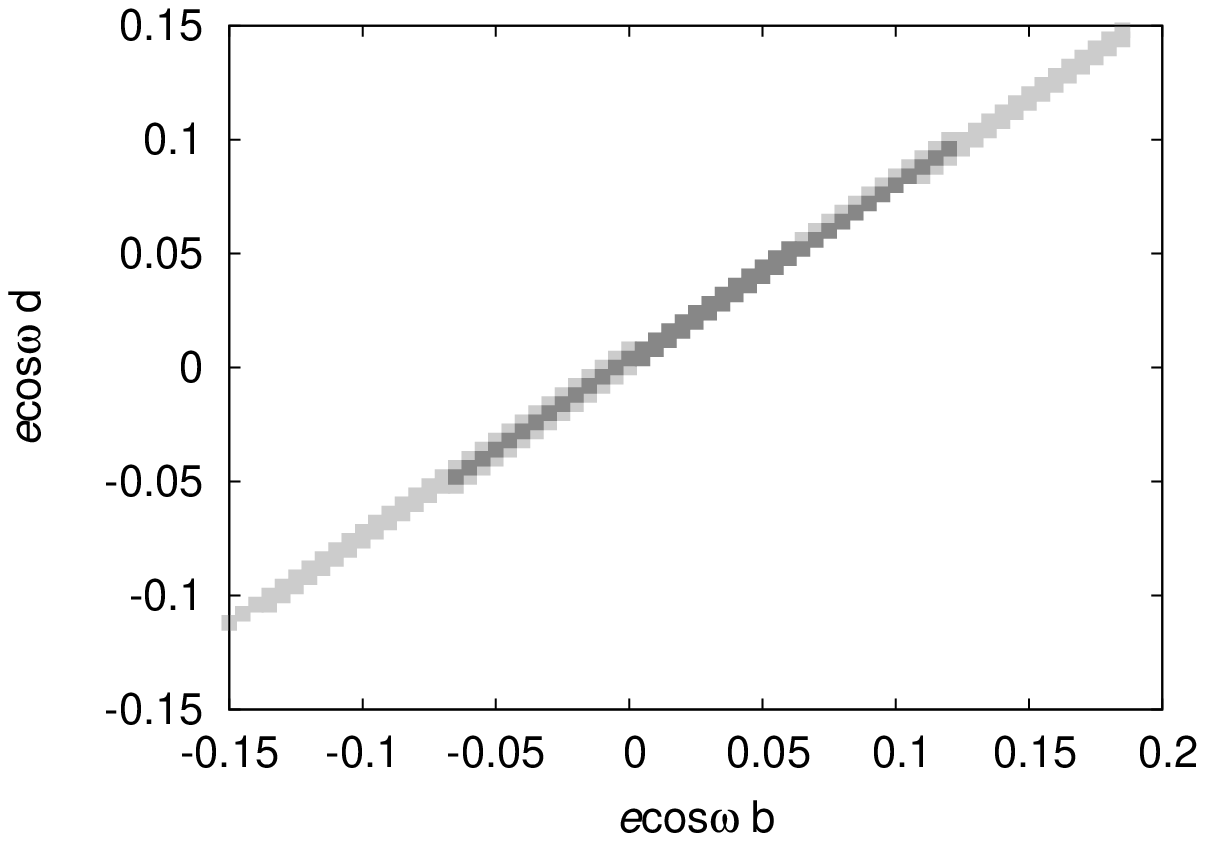}
\includegraphics [height = 1.5 in]{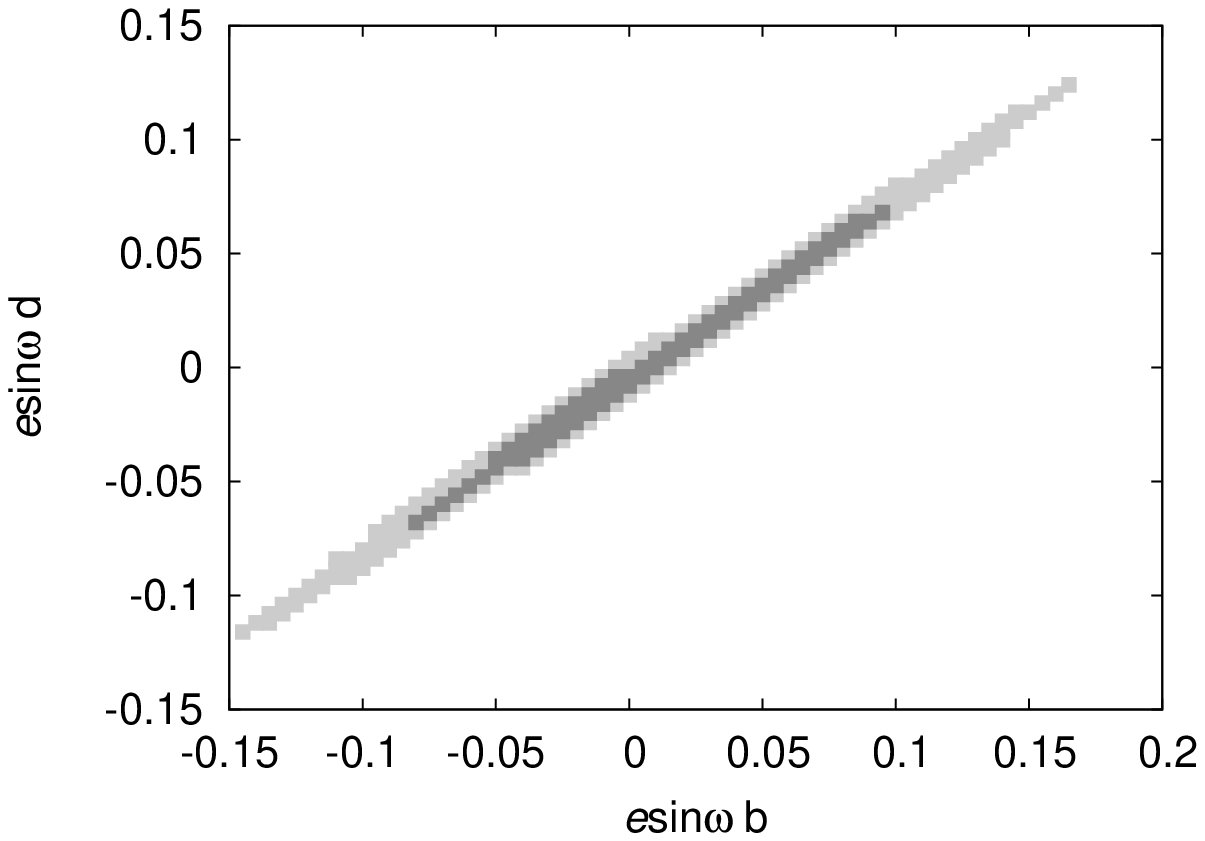}
\caption{Joint Posteriors for planetary masses relative to the host star, and eccentricity components for Kepler-60 b, c and d (KOI-2086). 95.4\% confidence intervals are in light grey, 68.3\% confidence intervals in dark grey. The dashed lines mark the expected correlation between eccentricity vector components from Equation~\ref{eqn:gradient}.}
\label{fig:2086} 
\end{figure}

Joint posteriors for dynamical masses and eccentricity vector components are shown in Figure~\ref{fig:2086}. Unlike in all two-planet interactions analyzed with TTVs, the dynamical masses of neighboring planets in Kepler-60 are moderately anti-correlated. %This may be due to the multi-planet interactions that cause the TTVs. For example, if a solution with a higher mass for 'b' is proposed, a lower mass for 'c' can provide an acceptable solution if a higher mass for 'd' can compensate. 

While the coefficients of Equation~\ref{eqn:gradient} match the gradient seen in the joint posteriors of eccentricities vectors, we see that there is an offset in the correlation, excluding the possibility of zero relative eccentricity. The offset in the joint posteriors of eccentricity vector components relative to the lines passing through the origin shown in Figure~\ref{fig:2086}, gives us a minimum relative eccentricity for each adjacent planet pair at Kepler-60 of 0.02; I.e, $|\vec{e_c}-\vec{e_b} | > 0.02$ and $|\vec{e_d}-\vec{e_c} | > 0.02$. 

\begin{figure}[h!]
\includegraphics [height = 2.5 in]{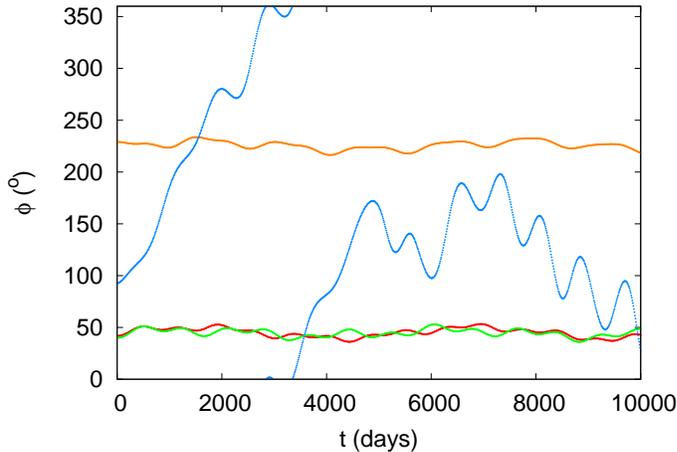}
\caption{Libration in the 3:4:5 resonant chain of Kepler-60. The three-body argument, $\phi = \lambda_{b}-2\lambda_{c}+\lambda_{d}$, is shown over time for a small sample of solutions from our low-eccentricity posteriors. In red, green and orange are examples of stable librations, while the other example (blue) shows drift in the resonant argument.}
\label{fig:resarg} 
\end{figure}

We tested what fraction of our posterior samples show stable librations in the 3:4:5 resonance. We tested whether the solutions were in stable librations in the Laplace-like 3:4:5 resonant chain, where the resonant argument is defined as $\phi \equiv \lambda_{b}-2\lambda_{c}+\lambda_{d}$. We integrated 50 solutions from our low eccentricity posterior samples for 1 Myr and found that both librating and non-librating solutions were all stable. Hence, whether or not the system is trapped in libration is still uncertain. We show examples of libration and non-libration over 10,000 days in Figure~\ref{fig:resarg}. Of our sample, we found 40 of 50 (80$\%$) are in libration in the Laplace-like resonance. All but one of these librated around the resonant argument $\phi = 45^{\circ}$, as found by \citet{goz15}. The exception librated about $\phi =$225$^{\circ}$. The mean and standard deviation of the dynamical masses of the librating  sample were: $m_b = 4.08 \pm 0.11$ $M_{\oplus} \frac{M_{\odot}}{M_{\star}}$, $m_c = 3.69 \pm 0.34$ $M_{\oplus} \frac{M_{\odot}}{M_{\star}}$, and $m_d = 3.96 \pm 0.35$ $M_{\oplus} \frac{M_{\odot}}{M_{\star}}$. Of the sample that were not in libration, the dynamical masses were: $m_b = 3.81 \pm 0.37$ $M_{\oplus} \frac{M_{\odot}}{M_{\star}}$, $m_c = 4.03 \pm 0.43$ $M_{\oplus} \frac{M_{\odot}}{M_{\star}}$, and $m_d = 3.99 \pm 0.65$ $M_{\oplus} \frac{M_{\odot}}{M_{\star}}$. These are in close agreement with the sample in libration, and hence we conclude that the condition of being in the Laplace-like resonance imposes no significant additional constraints on the masses.

\subsection{Kepler-105 (KOI-115)}
There are three planet candidates in a compact configuration at Kepler-105, orbiting every 3.44 days, 5.41 days and 7.12 days. Since TTVs scale with orbital period, this makes the detection of TTVs among short period planets quite rare. Nevertheless, our preliminary fits indicate a detection of perturbations due to the outer planet in the transit times of the middle planet, Kepler-105 b. In orbital period, Kepler-105 b orbits closer than other low-mass planets with published TTV analyses. Kepler-18 c with an orbital period of 7.6 days has strongly detected TTVs \citep{coch11} which constrained the mass of its neighbor orbiting every 14.9 days. WASP-47b, a jovian mass planet with an orbital period of 4.2 days has both strongly detected TTVs and a measurable mass from the TTVs it induces on its outer neighbor \citep{beck15}. Here we explore whether a TTV model can usefully constrain the mass of a planet just 1.3$R_{\oplus}$ in size with an orbital period of 7.12 days.  

The inner-most candidate of Kepler-105 (KOI-115.03) has a transit S/N of just 9, and hence remains unconfirmed. It is likely very small (0.7 $R_{\oplus}$) and has a transit depth of just 23 ppm. Thus KOI-115.03 has poorly constrained individual transit times. The inner pair have period ratio of 1.57, far from any first-order mean motion resonance, whilst the outer pair have a period ratio of 1.32, close enough to the first order 4:3 mean motion resonance for an expected TTV period of 145 days. The inner candidate is close to the 2:1 resonance with the outer planet.

The expected periodicity of this near-resonance is 97 days, and it is expected to have a low amplitude. We estimated the expected minimum amplitude of the TTVs induced between planet pairs of Kepler-105, using the equations of \citet{lith12} and assuming circular orbits. For masses, we assumed the mass radius relation of \citet{liss11b}, with $M_p/M_{\oplus} = (R_p/R_{\oplus})^{2.06}$ for planets larger than Earth (Kepler-105 b and c), and $M_p/M_{\oplus} =( R_p/R_{\oplus})^{3}$ for planets smaller than Earth (KOI-115.03). 

We compared these expected amplitudes to the likely uncertainty of transit times in phase-folded TTV curves with ten bins. We take this effective uncertainty as $\sigma_{eff} = \frac{\sigma_{med}}{\sqrt{N/10}}$ where $\sigma_{med}$ is the median timing uncertainty for the planet and $N$ is the number of measured transit times. 

The estimates for all six possible TTV inducing interactions between the planets at Kepler-105 are tabulated below. These estimates indicate that the interactions between the outer pair should be readily detectable in the measured transit times. While detectable interactions with the inner planet appear unlikely from our estimates in Table~\ref{tbl-koi115signals}, we nevertheless include all three planets in our dynamical fits, since the inner planet is near 2:1 with the outer planet and there are many uncertainties in the approximations used to estimate the signal strength. 
 \begin{table}[h!]
  \begin{center}
       \begin{tabular}{||c|c|c|c||}
       \hline
       Interaction  & $\sigma_{med}$ (mins) & $\sigma_{eff}$ (mins) & Expected Min. TTV ampl. (mins)   \\
              \hline
              .03 $\rightarrow$ b &  2.6 & 0.5 & 0.04  \\
              b $\rightarrow$  .03   &  43  &  6.9 & 0.3  \\
              b $\rightarrow$  c   & 7.4  &  1.7   & 3.7    \\
              c $\rightarrow$  b     & 2.6  & 0.5   &  0.9  \\
              .03 $\rightarrow$  c   & 7.4  & 1.7  & 0.02 \\
              c $\rightarrow$  .03   & 43   & 6.9 & 0.16  \\
              \hline
      \end{tabular} 
    \caption{Expected information content in phase-folded TTV signals of planet pairs at Kepler-105. The first column lists the perturber and perturbee in each interaction. The remaining columns give the median measurement uncertainty in the perturbee, $\sigma_{med}$; the effective uncertainty in the times of the perturbee given the number of transits, $\sigma_{eff}$; and the minimum expected TTV amplitude, all measured in minutes. Where the expected signals in the last column exceed the effective uncertainties in the third column, strong upper and lower limits on perturbing masses are are likely obtainable.}\label{tbl-koi115signals}
  \end{center}
\end{table}

\begin{figure}[!htb]
\includegraphics [height = 1.5 in]{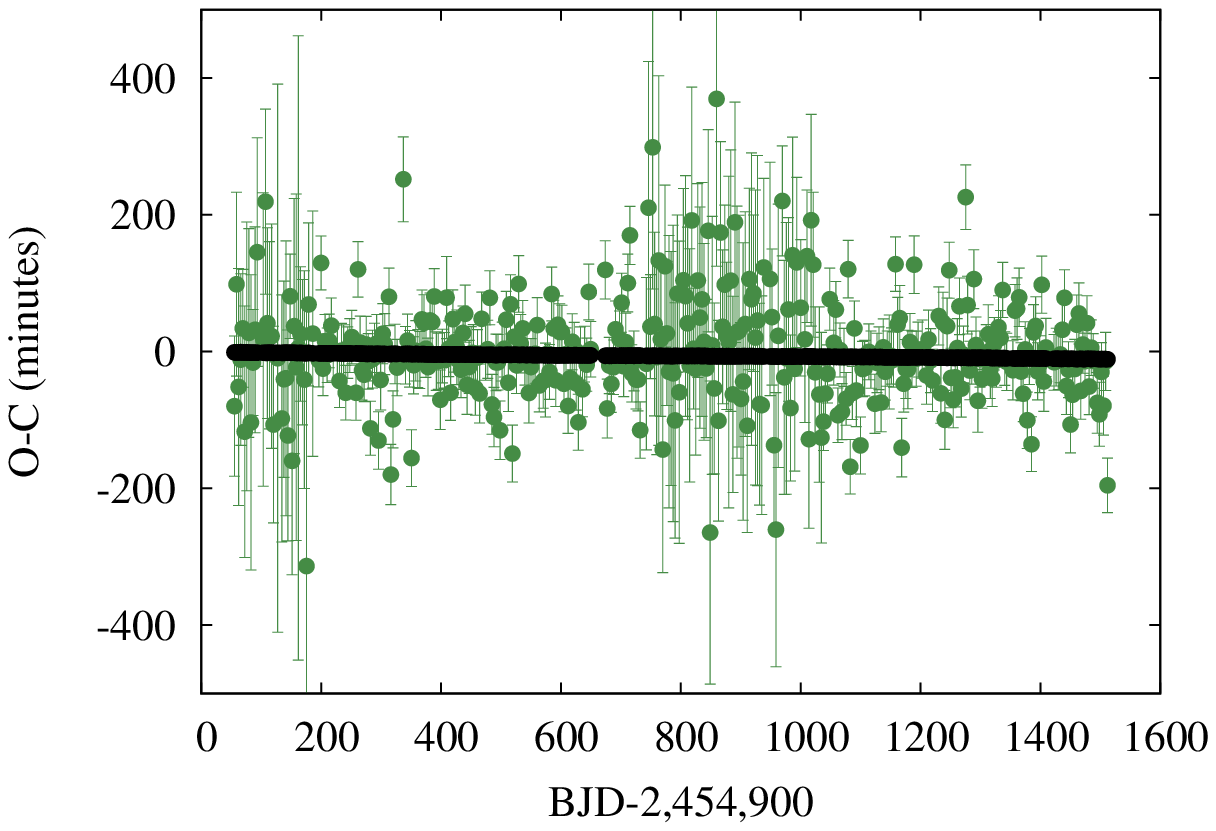}
\includegraphics [height = 1.5 in]{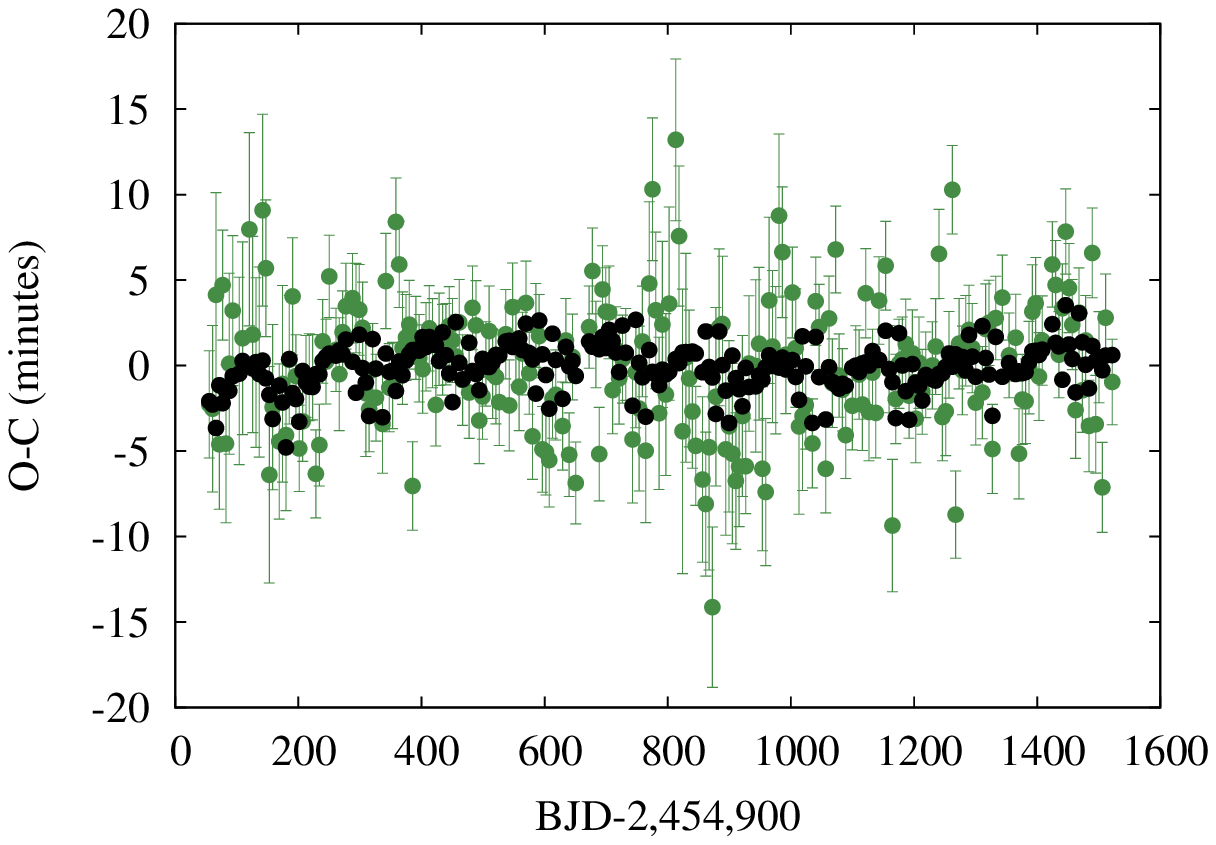}
\includegraphics [height = 1.5 in]{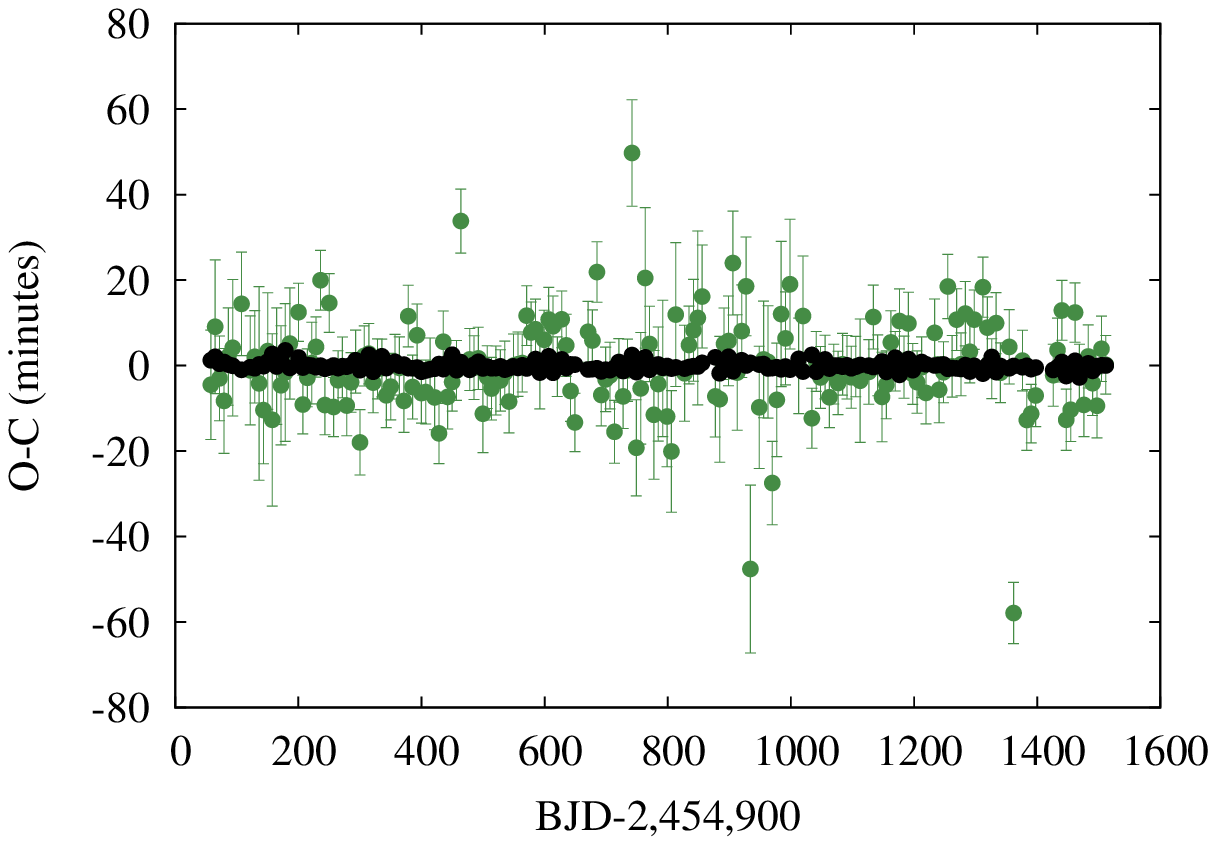}
\caption{Best fit TTV models to Kepler-105 b, c (middle and right panels) and the unconfirmed inner candidate (KOI-115.03, left panel). In black are simulated transit times, and in green are measured transit times with their uncertainties. Although the scatter in TTVs looks high, we detect signals in planets `b' and `c' at the expected TTV period of 145 days. The non-detection of TTVs in KOI-115.03 is consistent with the large measurement uncertainties in its transit times.}
\label{fig:115-sim} 
\end{figure}

We plot the observed TTVs and our best fitting model in Figure~\ref{fig:115-sim}. The model shows a reasonable detection at Kepler-105 b and what may be a weak signal in Kepler-105 c, although there are several outlying transit times in Kepler-105 c which may cause a spurious detection notwithstanding the promising expected TTV amplitude for this planet shown in Table~\ref{tbl-koi115signals}. In this case, the underestimated uncertainties in Kepler-105 c (as indicated by the excess of outliers) may cause our estimate of the effective timing uncertainty for binned data to be too low also.

%% Estimate amplitudes of TTV amplitude of .03c, .03b, bc, 

\begin{figure}[!htb]
\includegraphics [height = 2.5 in]{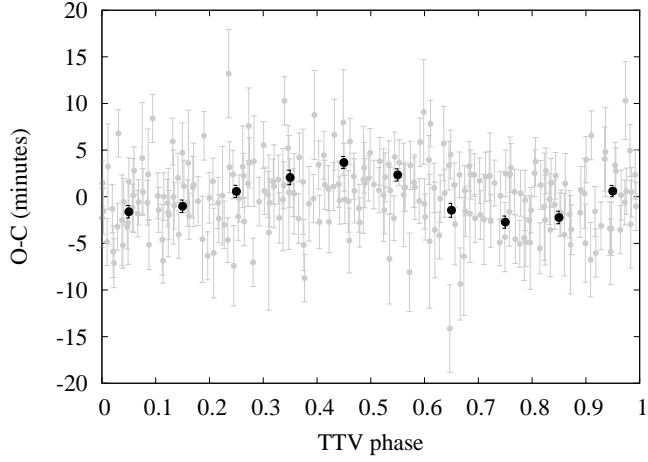}
\caption{Phase-folded TTV curve for Kepler-105 b (the middle planet candidate). The TTV data here (grey points) are phase-folded, and averaged in ten bins (black points). Uncertainties are the sum of measurement uncertainties in each bin added in quadrature divided by the number of transit times in each bin. The TTV phase curve here validates the signal detected in our dynamical fits, despite the relatively short orbital period and small TTV amplitude.}
\label{fig:115phase} 
\end{figure}

\begin{figure}[!h]
\includegraphics [height = 1.5 in]{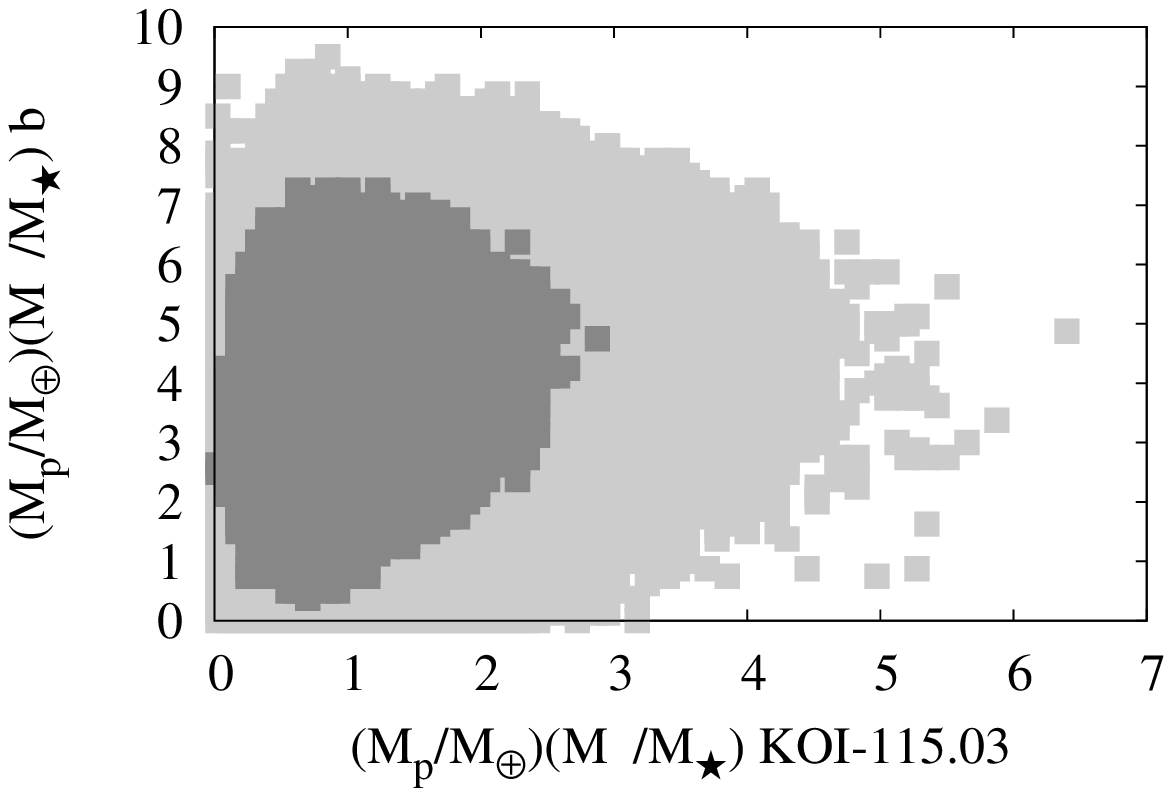}
\includegraphics [height = 1.5 in]{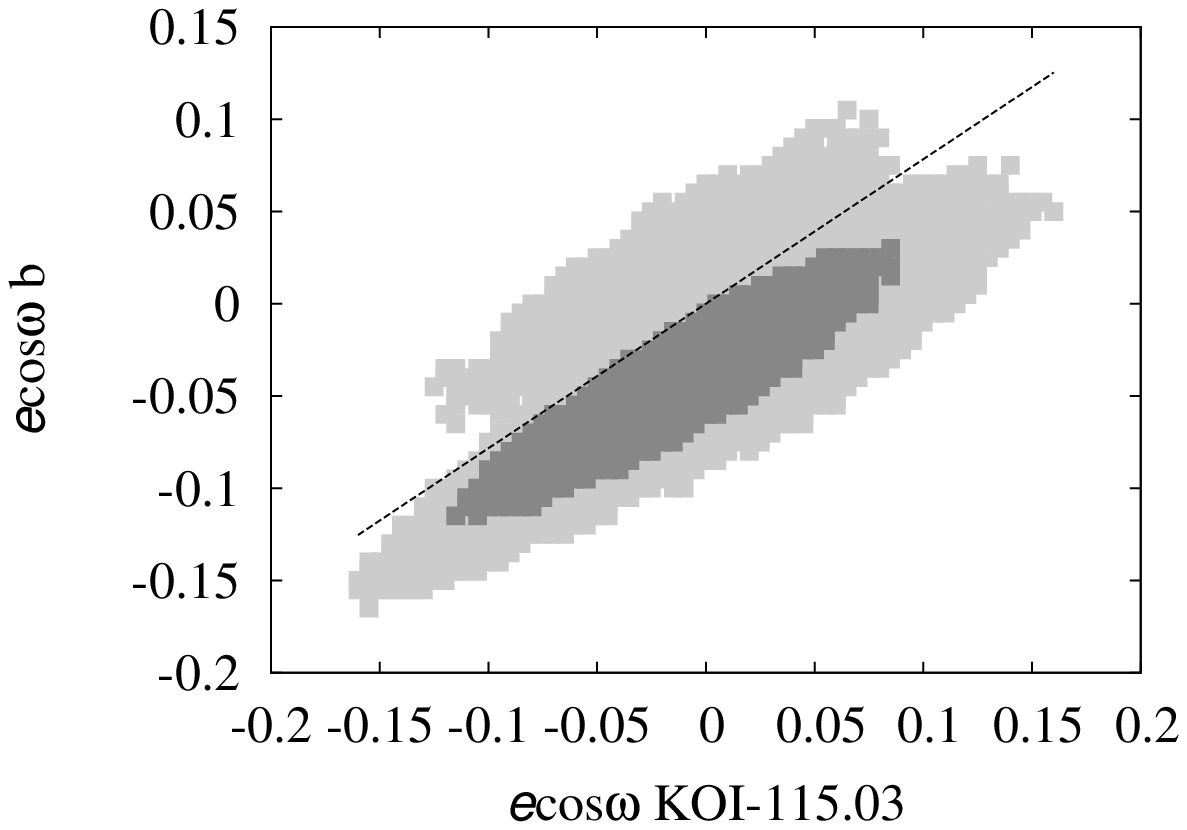}
\includegraphics [height = 1.5 in]{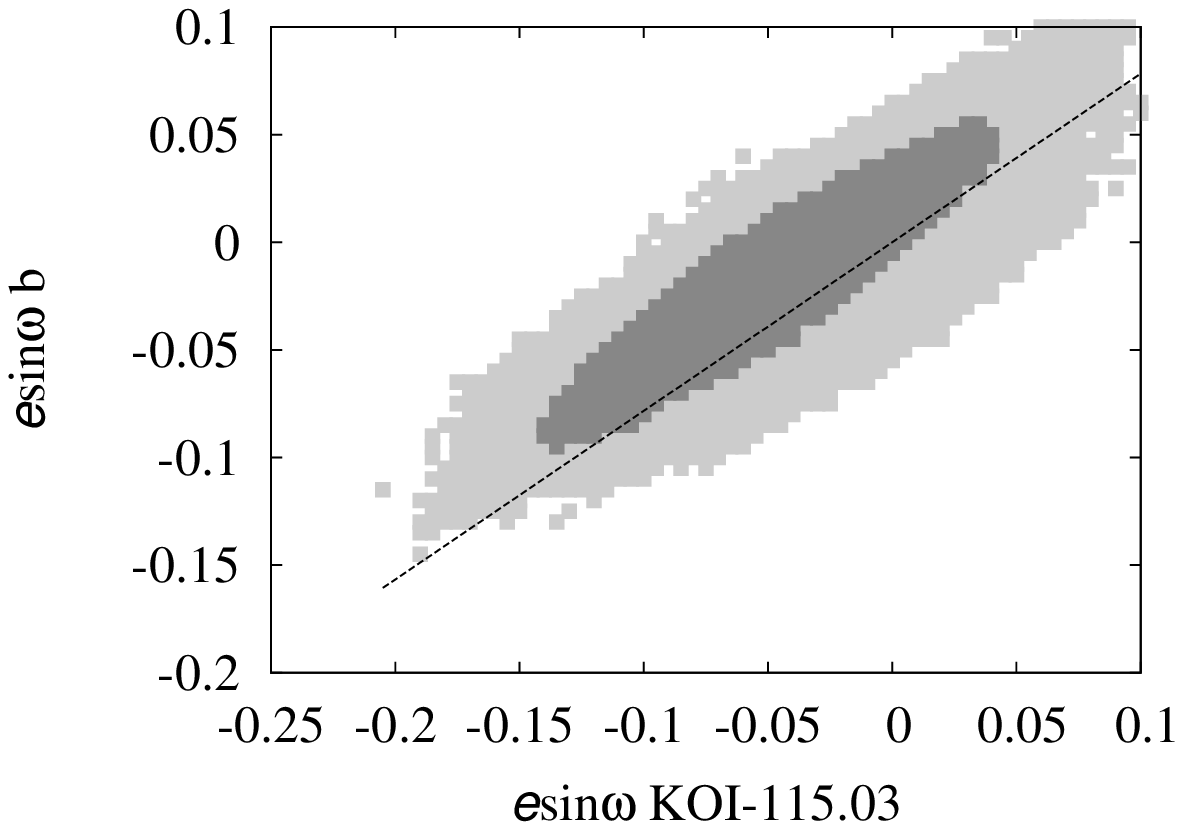}
\includegraphics [height = 1.5 in]{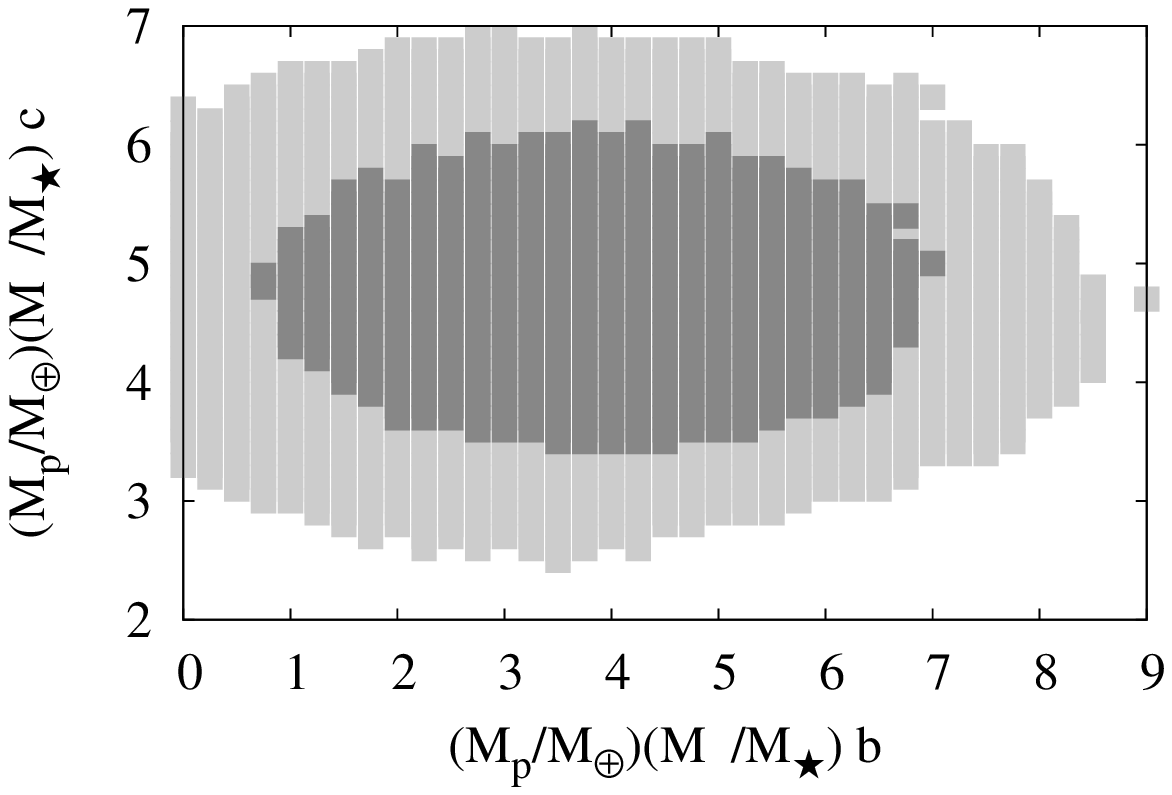}
\includegraphics [height = 1.5 in]{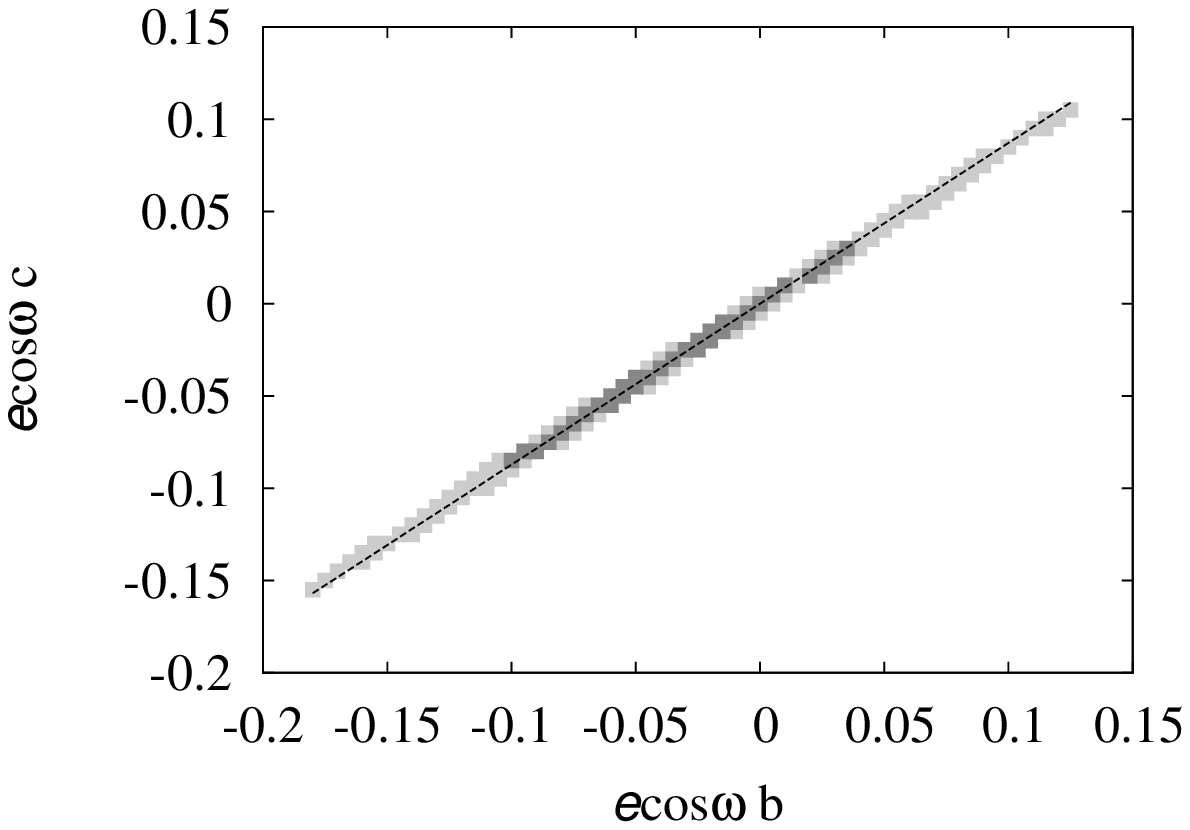}
\includegraphics [height = 1.5 in]{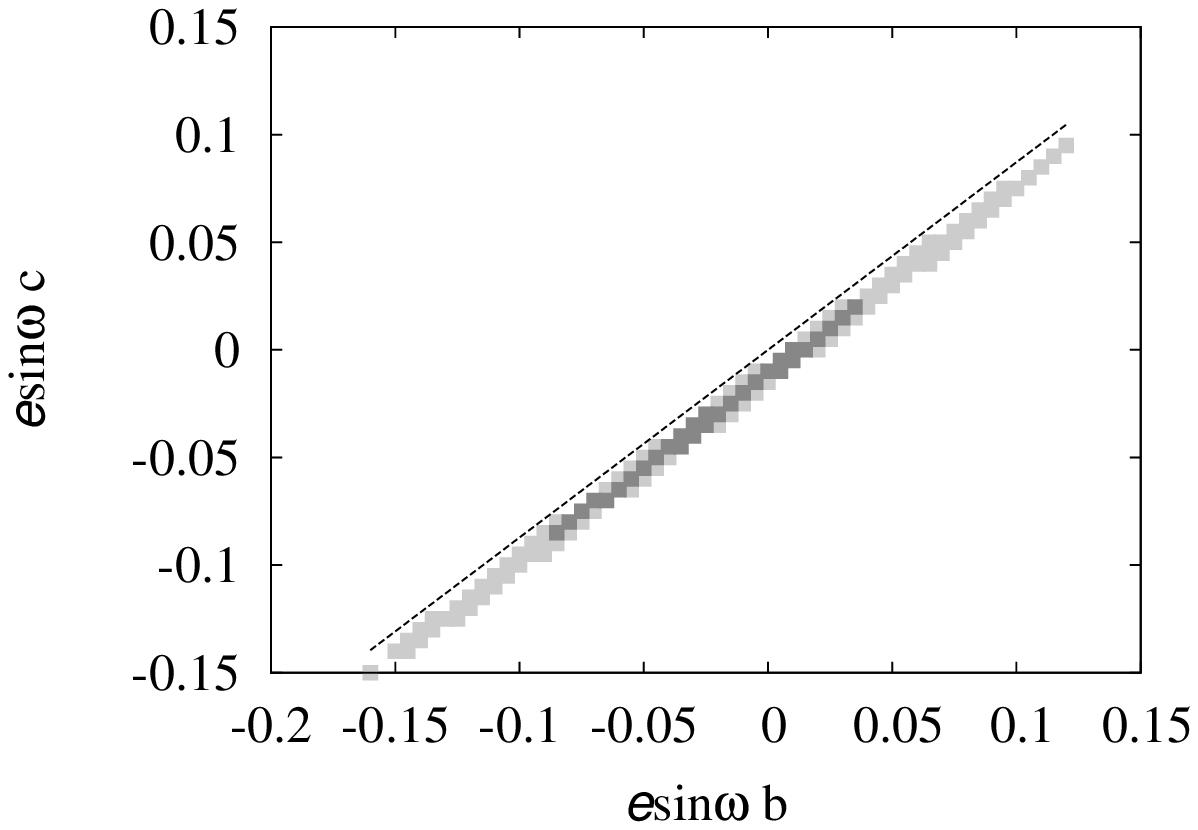}
\caption{Joint posteriors for planetary masses relative to the host star, and eccentricity components for Kepler-105 b, c and KOI-115.03. 95.4\% confidence intervals are in light grey, 68.3\% confidence intervals in dark grey. The dashed lines mark the expected correlation between eccentricity vector components from Equation~\ref{eqn:gradient}}
\label{fig:115} 
\end{figure}

To test whether the TTV period in Kepler-105 b could be the result of fitting noise (middle panel of Figure~\ref{fig:115-sim}), we phase-folded the TTV data at the expected TTV period from the date of the first transit and binned at ten phases. The resulting phase TTVs and their uncertainties are plotted in Figure~\ref{fig:115phase}. In this case, the shape of the phase curve matches the simulated TTVs shown in Figure~\ref{fig:115-sim}b. This agreement is consistent with the mass detection of Kepler-105 c shown in the posteriors in Figure~\ref{fig:115}.

We find strong upper limits for all three planets, but the upper limit on the dynamical mass of KOI-115.03 is not useful. The most useful constraints here are the strong upper and lower limits on Kepler-105 c.

The eccentricity joint posteriors show the expected degeneracy and slope described by Equation~\ref{eqn:gradient}, with a slight offset indicating that relative eccentricities must be non-zero.

 \begin{table}[h!]
  \begin{center}
       \begin{tabular}{||c||}
      \hline
 \hspace{1.75 in} Adopted  parameters  with 1$\sigma$ (2$\sigma$) credible intervals     \hspace{1.75 in} \\
      \hline
    \end{tabular} 
    \begin{tabular}{||c|c|c|c|c|c||}
      \hline
 planet  & $P$ (days)  &      $T_0$ (days)    &      $e\cos\omega$   &  $e\sin\omega$     &     $ \frac{ M_p  }{M_{ \oplus} } \frac{M_{\odot} } {M_{\star} }  $($\pm 2\sigma$)  \\   
 \hline 
    .03  & \textbf{ 3.4363 }$\pm 0.0003$ &  \textbf{ 780.3187 }$ \pm 0.0021 $ &  \textbf{ --0.015 }$^{+ 0.059 }_{ - 0.052 }$ &  \textbf{ --0.050 }$ \pm 0.054 $ &  \textbf{  1.28 }$^{+  1.25 }_{ -  0.71 }$  $\left(^{+  3.10 }_{ -  1.00 }\right)$   \\ 
 b & \textbf{ 5.4119 }$\pm 0.0001 $ &  \textbf{ 780.5529 }$^{+ 0.0003 }_{ - 0.0004 }$ &  \textbf{ --0.042 }$^{+ 0.041 }_{ - 0.040 }$ &  \textbf{ --0.020 }$^{+ 0.044 }_{ - 0.043 }$ &  \textbf{  4.06 }$^{+  1.97 }_{ -  1.97 }$  $\left(^{+  3.93 }_{ -  3.02 }\right)$   \\ 
 c  & \textbf{ 7.1262 }$\pm  0.0002 $ &  \textbf{ 784.5992 }$ \pm  0.0006 $ &  \textbf{ --0.036 }$^{+ 0.035 }_{ - 0.035 }$ &  \textbf{ --0.028 }$^{+ 0.038 }_{ - 0.037 }$ &  \textbf{  4.78 }$^{+  0.91 }_{ -  0.89 }$  $\left(^{+  1.84 }_{ -  1.37 }\right)$  \\ 
 \hline
    \end{tabular}
        \begin{tabular}{||c||c||c||c||}
      \hline
  \hspace{0.48 in}  Test 1   \hspace{0.48 in}    &   \hspace{0.48 in} Test 2  \hspace{0.48 in}  &   \hspace{0.48 in} Test 3  \hspace{0.48 in} &   \hspace{0.48 in} Test 4  \hspace{0.48 in}  \\
      \hline
    \end{tabular}   
    \begin{tabular}{||c|c||c|c||c|c||c|c||}
 planet  \hspace{0.1 in}    &    $ \frac{ M_p  }{M_{ \oplus} } \frac{M_{\odot} } {M_{\star} }  $   \hspace{0.1 in}  &  planet  \hspace{0.1 in}  &    $ \frac{ M_p  }{M_{ \oplus} } \frac{M_{\odot} } {M_{\star} }  $ \hspace{0.1 in}  &  planet  \hspace{0.1 in}  &    $ \frac{ M_p  }{M_{ \oplus} } \frac{M_{\odot} } {M_{\star} }  $ \hspace{0.1 in} &  planet  \hspace{0.1 in}  &    $ \frac{ M_p  }{M_{ \oplus} } \frac{M_{\odot} } {M_{\star} }  $ \hspace{0.1 in}  \\
 \hline 
 .03  &   2.09 $^{+  1.74 }_{ -  1.07 }$   &   .03   &   1.59$^{+1.93 }_{-1.02}$      &          &    &          &     \\ 
b  &  3.67 $^{+  1.82 }_{ -  1.81 }$        &    b      &     4.21$^{+ 2.22 }_{ - 2.32  }$    &    b   &    3.51  $^{+  2.06 }_{ -  1.96 }$ &    b   &    6.85  $^{+  2.49 }_{ -  2.56 }$ \\ 
c  &  4.79 $^{+  0.91 }_{ -  0.91 }$        &    c       &    4.93$^{+  1.01 }_{ - 0.97  }$     &    c   &  4.79 $^{+  0.91 }_{ -  0.93 }$    &    c   &  5.96 $^{+  0.96 }_{ -  0.95 }$    \\ 
      \hline
    \end{tabular}  
    \caption{TTV solutions for Kepler-105 .03, b and c (KOI-115) and  the results of various tests on dynamical masses. Test 1: An alternative eccentricity prior. Test 2: Robust fitting.  Test 3: Two-planet model.  Test 4: Holczer catalog }\label{tbl-koi115}
  \end{center}
\end{table}

Our results for Kepler-105, as shown in Table~\ref{tbl-koi115}, indicate agreement at the 1$\sigma$ level between our nominal transit times and the Holczer catalog for the outermost planet only. This is consistent with the detection of TTVs in the middle planet only.  Hence, we include the outermost planet as having a robust mass measurement, but we exclude the inner two planets from the mass-radius diagram.

%Finally, we address whether KOI-115.03 merits a promotion to confirmed status given the constraints in its mass from the analysis above. We seek to measure a Bayes factor to compare two possible models for this candidate, that it has a mass, or that it has zero mass. Given our posterior samples with a three-planet model, the zero-mass model is embedded within the fully marginalized mass posterior. We estimated the Bayes Factor using the Savage-Dickey Density Ratio (SDRR) (\citealt{verd95,jont15}). This accounts for the width and shape of the posterior distribution for the mass of KOI-115.03.

%We assumed a uniform prior on KOI-115.03 from a mass of zero to a planet made of iron at 0.73$R_{\oplus}$, using the theoretical models of \citet{for07}, I.e.; ($M_{.03} \in U[0:0.74] M_{\oplus}$). Comparing the two models for KOI-115.03, we find calculated the Bayes Factor for an optimized bandwidth of 0.1 $M_{\oplus} \frac{M_{\odot}}{M_{\star}}$. We found that the three-massive planet model is favored at a 99.9\% confidence level.

\subsection{Kepler-177 (KOI-523)}
\begin{figure}[h!]
\includegraphics [height = 2.1 in]{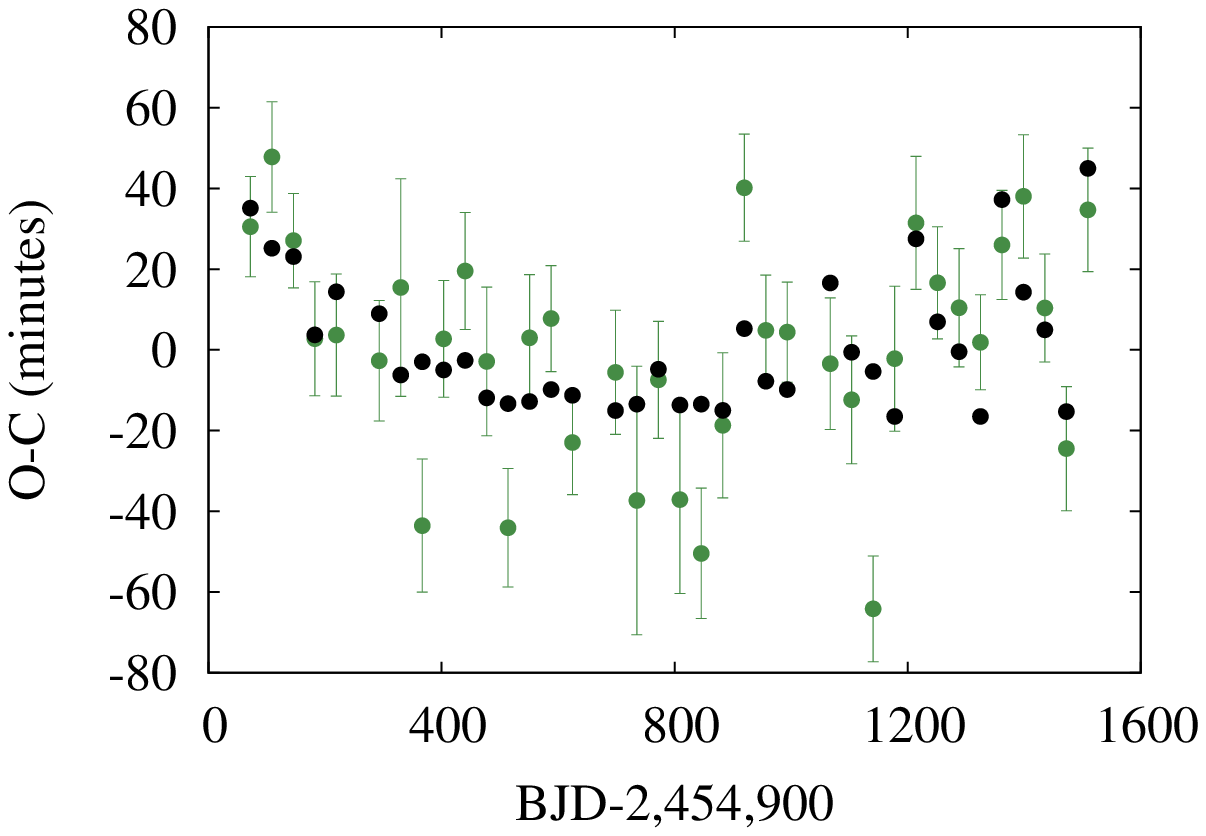}
\includegraphics [height = 2.1 in]{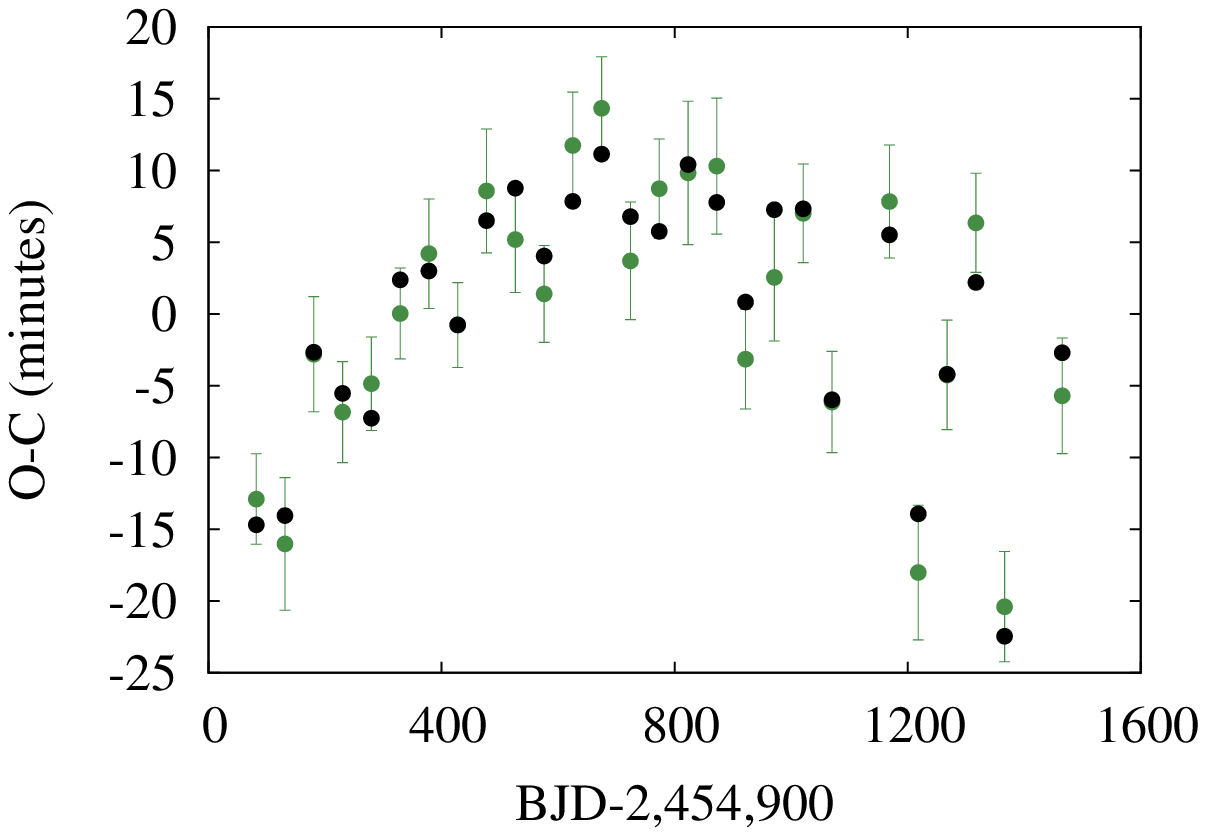}
\caption{Best fit dynamical model for Kepler-177 b (left) and c (right). In black are simulated transit times, and in green are measured transit times with their uncertainties.
}
\label{fig:523-sim} 
\end{figure}

\citet{stef12} identified significant TTVs at Kepler-177. Its planets have the longest orbital periods among the planets of the systems in this study. Figure~\ref{fig:523-sim} shows the TTV signal and our best fit solution for Kepler-177. The TTV cycle is not complete over the four-year \textit{Kepler} baseline. However, strongly detected chopping and a low mass for the inner planet make this an interesting system for TTV analysis.

\begin{figure}[!htb]
\includegraphics [height = 1.5 in]{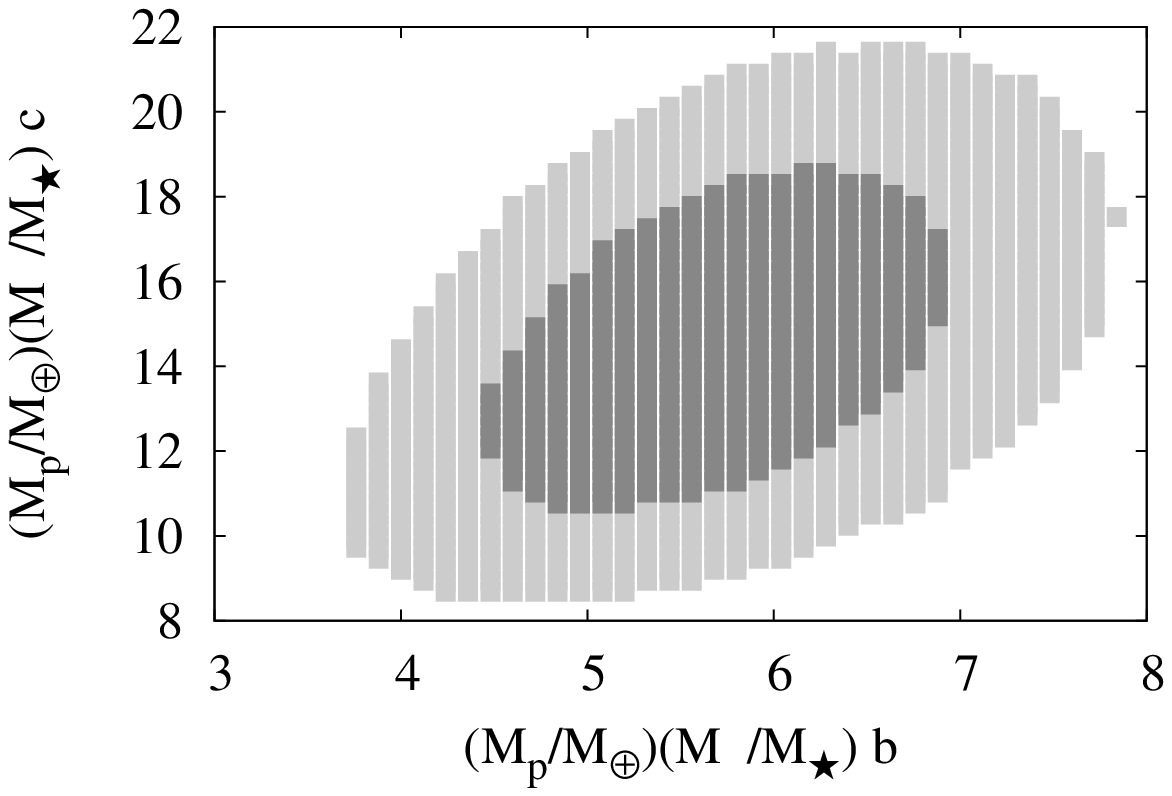}
\includegraphics [height = 1.5 in]{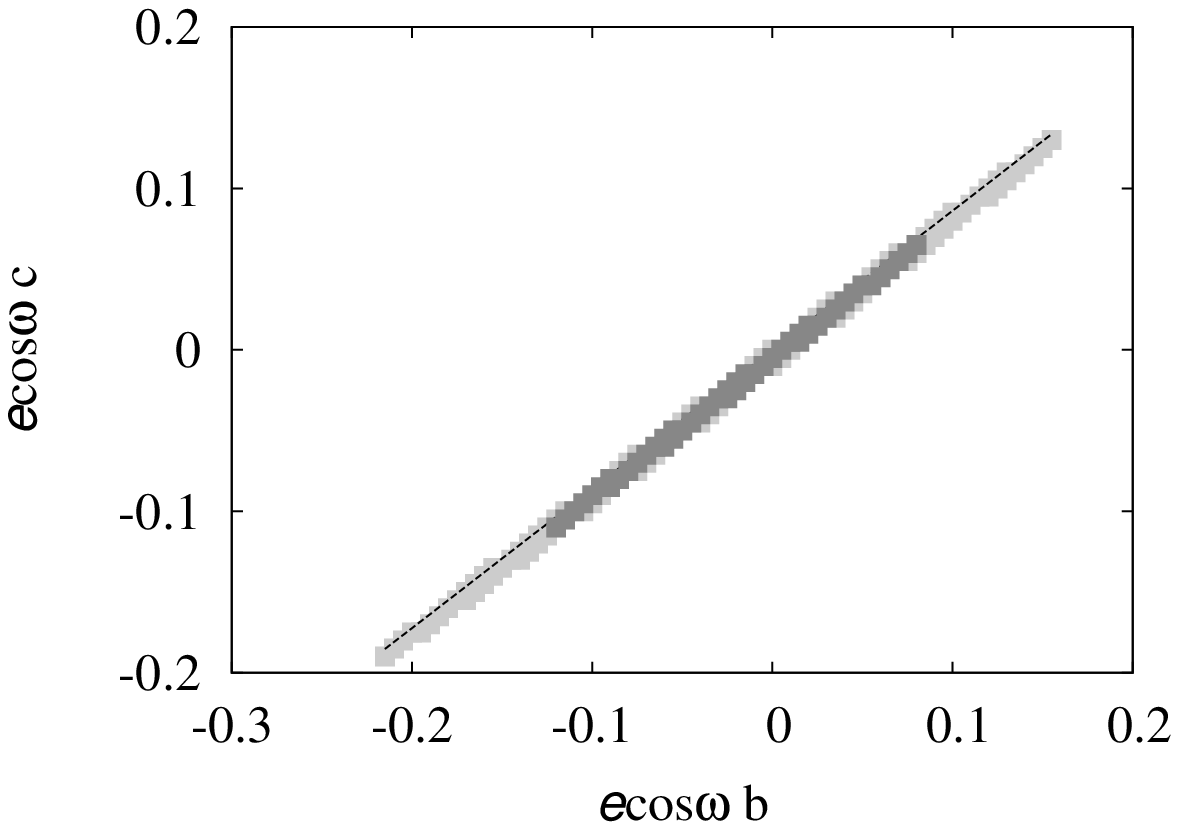}
\includegraphics [height = 1.5 in]{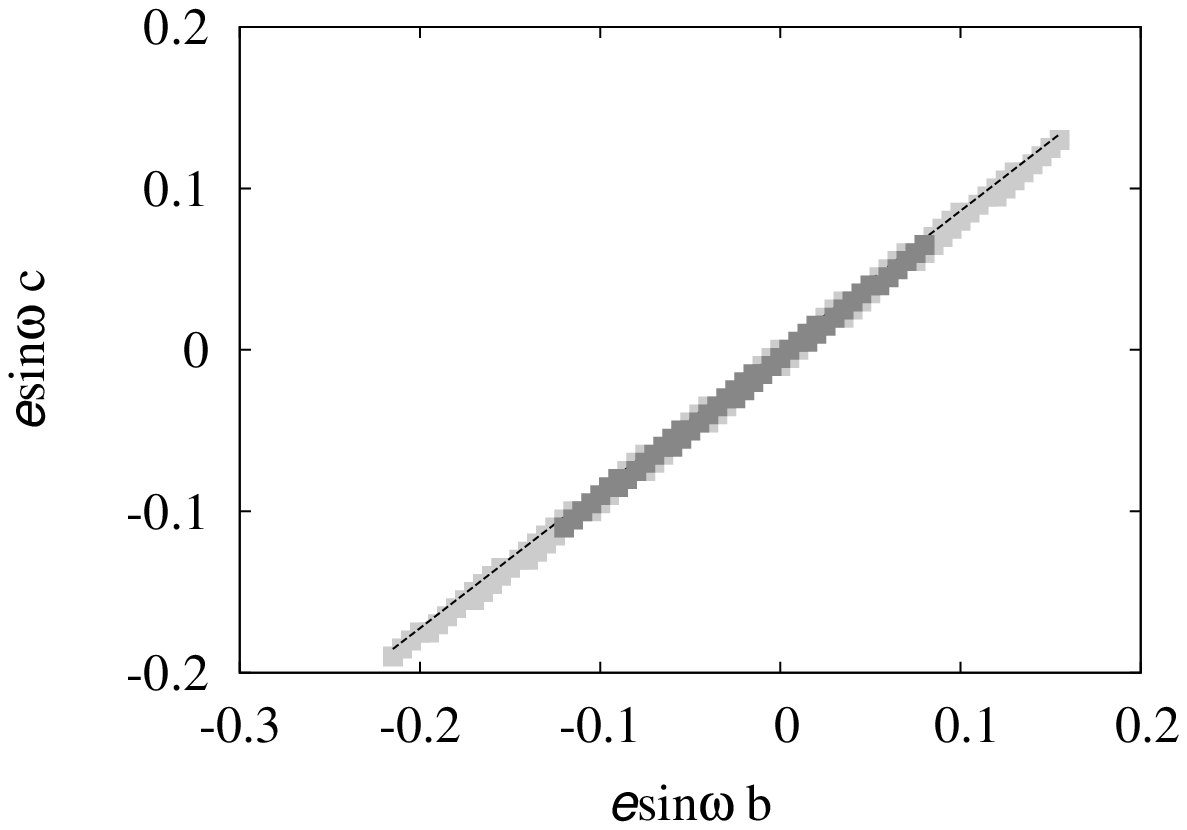}
\caption{Joint Posteriors for planetary masses relative to the host star for Kepler-177 (KOI-523). 95.4\% confidence intervals are in light grey, 68.3\% confidence intervals in dark grey. The dashed lines mark the expected correlation between eccentricity vector components from Equation~\ref{eqn:gradient}.}
\label{fig:523} 
\end{figure}

The joint posteriors for dynamical masses and eccentricity vector components are shown in Figure~\ref{fig:523}. In this case, there are tight constraints on the masses of both planets. To assess the long-term stability of our TTV solutions, we performed long term integrations for this system. All simulations were stable to 1 Myr.

 \begin{table}[h!]
  \begin{center}
       \begin{tabular}{||c||}
      \hline
 \hspace{1.75 in} Adopted  parameters  with 1$\sigma$ (2$\sigma$) credible intervals     \hspace{1.75 in} \\
      \hline
    \end{tabular} 
    \begin{tabular}{||c|c|c|c|c|c||}
      \hline
 planet  & $P$ (days)  &      $T_0$ (days)    &      $e\cos\omega$   &  $e\sin\omega$     &     $ \frac{ M_p  }{M_{ \oplus} } \frac{M_{\odot} } {M_{\star} }  $($\pm 2\sigma$)  \\   
 \hline 
   b  & \textbf{ 36.8590 }$^{+ 0.0019 }_{ - 0.0017 }$ &  \textbf{ 809.0397 }$^{+ 0.0032 }_{ - 0.0030 }$ &  \textbf{ --0.027 }$^{+ 0.074 }_{ - 0.075 }$ &  \textbf{ --0.015 }$^{+ 0.065 }_{ - 0.068 }$ &  \textbf{  5.75 }$^{+  0.84 }_{ -  0.81 }$  $\left(^{+  1.72 }_{ -  1.24 }\right)$   \\ 
 c & \textbf{ 49.4096 }$\pm 0.0010$ &  \textbf{ 822.9979 }$\pm  0.0013 $ &  \textbf{ --0.029 }$^{+ 0.064 }_{ - 0.065 }$ &  \textbf{ --0.015 }$^{+ 0.056 }_{ - 0.059 }$ &  \textbf{ 14.58 }$^{+  2.69 }_{ -  2.53 }$  $\left(^{+   5.50 }_{ -  3.94 }\right)$   \\ 
    \hline
    \end{tabular}
        \begin{tabular}{||c||c||c||c||}
      \hline
  \hspace{0.55 in}  Test 1   \hspace{0.55 in}    &   \hspace{0.55 in} Test 2  \hspace{0.55 in}    &   \hspace{0.55 in} Test 3  \hspace{0.55 in}   &  \hspace{0.21 in}  Analytical Model   \hspace{0.21 in}  \\
      \hline
    \end{tabular}   
    \begin{tabular}{||c|c||c|c||c|c||c|c||}
 planet  \hspace{0.2 in}    &    $ \frac{ M_p  }{M_{ \oplus} } \frac{M_{\odot} } {M_{\star} }  $   \hspace{0.2 in}  &  planet  \hspace{0.2 in}  &    $ \frac{ M_p  }{M_{ \oplus} } \frac{M_{\odot} } {M_{\star} }  $ \hspace{0.2 in} &  planet  \hspace{0.2 in}  &    $ \frac{ M_p  }{M_{ \oplus} } \frac{M_{\odot} } {M_{\star} }  $ \hspace{0.2 in} &  planet  \hspace{0.2 in}  &    $ \frac{ M_p  }{M_{ \oplus} } \frac{M_{\odot} } {M_{\star} }  $ \hspace{0.2 in}  \\
 \hline 
 b  &  5.72 $^{+  0.82 }_{ -  0.79 }$   &  b  &   5.59 $^{+  1.03 }_{ -  0.99 }$     &    b  &  8.68 $^{+  0.83 }_{ -  0.81 }$  &    b  &  5.15 $\pm 0.71$      \\ 
   c  &  14.66 $^{+  2.69 }_{ -  2.56 }$   &  c  &   11.74 $^{+  3.31 }_{ -  2.76 }$      &  c  &   21.91$^{+  4.09 }_{ -  3.92 }$    &  c  &   12.64 $\pm 2.19$   \\ 
      \hline
    \end{tabular}  
    \caption{TTV solutions for Kepler-177 b and c (KOI-523) and  the results of various tests on dynamical masses. Test 1: An alternative eccentricity prior. Test 2: Robust fitting. Test 3. Holczer catalog. The last column in the bottom panel gives an analytical result using the approximation of \citet{agol15} in close agreement with the dynamical fits. }\label{tbl-koi0523}
  \end{center}
\end{table}

Table~\ref{tbl-koi0523} lists our results for Kepler-177, including the tests that we have performed. In this case, our results were robust against the eccentricity prior and the measurement uncertainies. Our analytical fits are accurate to first order in eccentricity (given the proximity of the planets to the first order 4:3 resonance) and show close agreement with our dynamical fits. We obtained higher
dynamical masses with the Holczer transit timing catalog. We found that with fewer transits and an incomplete TTV cycle, this system was more sensitive to the methodology of the light curve analysis than others. We therefore omit these planets from the mass-radius diagram, although we note that fits against both sets of transit times agree that these planets have relatively low mass given their sizes and the upper limits on the mass of Kepler-117 c imply an extremely low bulk density (see Discussion).

\newpage
\subsection{Kepler-307 (KOI-1576)}
\begin{figure}[h!]
\includegraphics [height = 2.1 in]{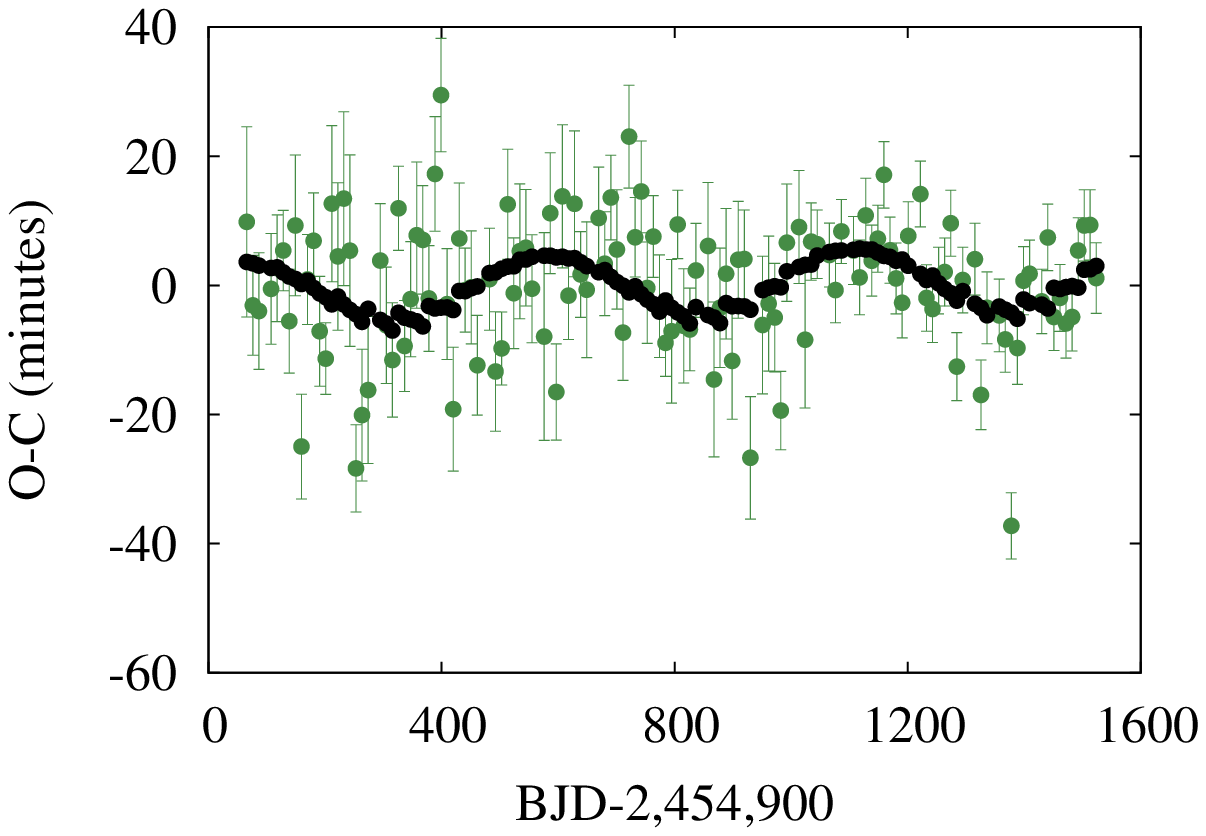}
\includegraphics [height = 2.1 in]{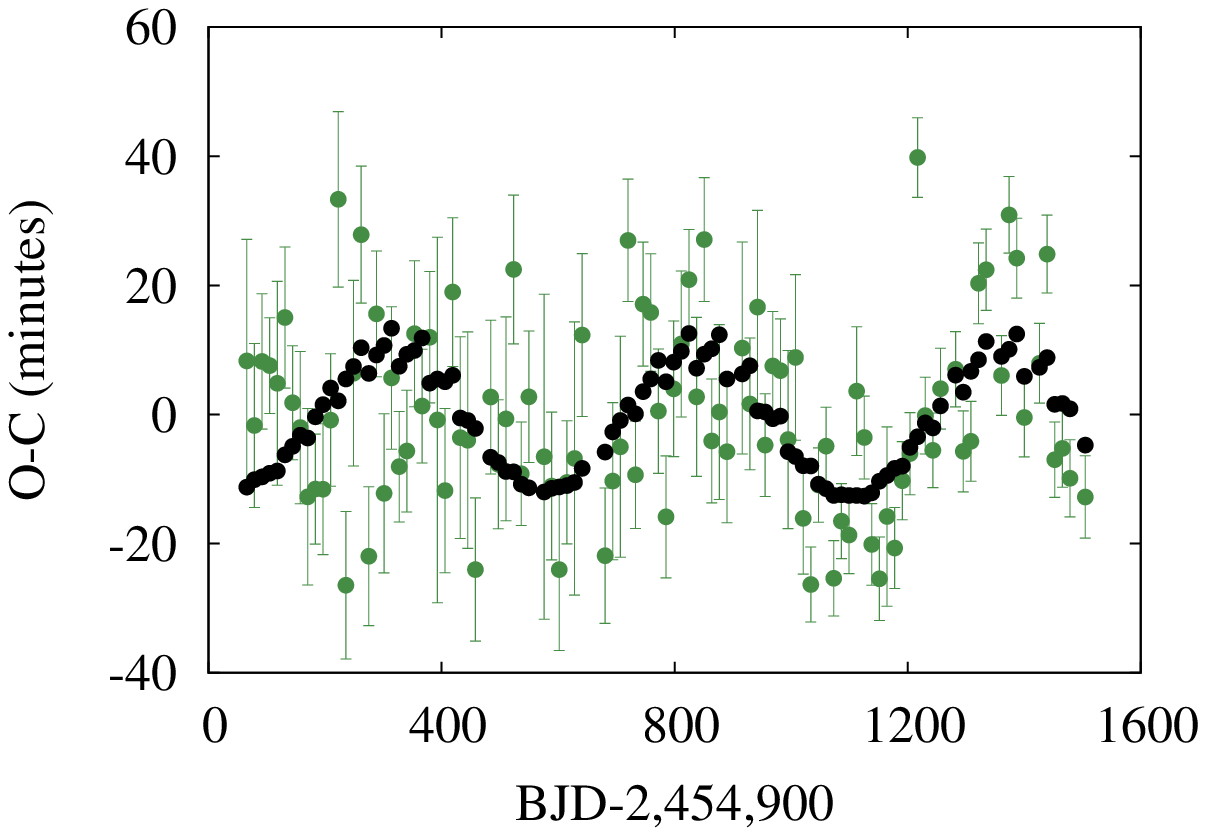}
\caption{Best fit dynamical model for Kepler-307 b (left) and c (right). In black are simulated transit times, and in green are measured transit times with their uncertainties.
}
\label{fig:1576-sim} 
\end{figure}

Two of the planetary candidates that transit KOI-1576 were validated by \citet{rowe14} and \citet{liss14a}. They were also confirmed by virtue of their detected TTVs by \citet{xie14}. The two confirmed planets, Kepler-307 b and Kepler-307 c, orbit every 10.4 days and every 13.1 days respectively. There is a third candidate with an orbital period of 23.34 days that is significantly smaller than the two validated planets with a period 1.78 times that of Kepler-307 c. It is far from any first or second order mean motion resonance. Therefore, we include just the two confirmed planets, near the 5:4 resonance, in our dynamical modeling. Figure~\ref{fig:1576-sim} shows the TTVs of Kepler-307 b and Kepler-307 c. The predicted 727 day TTV period due to their near-resonance is easily discernible in the data.

 \begin{table}[h!]
  \begin{center}
       \begin{tabular}{||c||}
      \hline
 \hspace{1.75 in} Adopted  parameters  with 1$\sigma$ (2$\sigma$) credible intervals     \hspace{1.75 in} \\
      \hline
    \end{tabular} 
    \begin{tabular}{||c|c|c|c|c|c||}
      \hline
 planet  & $P$ (days)  &      $T_0$ (days)    &      $e\cos\omega$   &  $e\sin\omega$     &     $ \frac{ M_p  }{M_{ \oplus} } \frac{M_{\odot} } {M_{\star} }  $($\pm 2\sigma$)  \\   
 \hline 
   b  & \textbf{ 10.4208 }$^{+ 0.0009 }_{ - 0.0008 }$ &  \textbf{ 784.3157 }$\pm 0.0006 $ &  \textbf{ 0.011 }$^{+ 0.038 }_{ - 0.035 }$ &  \textbf{ --0.040 }$^{+ 0.053 }_{ - 0.058 }$ &  \textbf{  8.16 }$^{+  0.97 }_{ -  0.89 }$ $\left(^{+  2.02 }_{ -  1.36 }\right)$   \\ 
  c & \textbf{ 13.0729 }$\pm 0.0012 $ &  \textbf{ 785.2666 }$ \pm 0.0010 $ &  \textbf{ 0.004 }$^{+ 0.034 }_{ - 0.032 }$ &  \textbf{ --0.029 }$^{+ 0.048 }_{ - 0.052 }$ &  \textbf{  4.02 }$^{+  0.68 }_{ -  0.62 }$ $\left(^{+  1.44 }_{ -  0.96 }\right)$   \\ 
    \hline
    \end{tabular}
        \begin{tabular}{||c||c||c||}
      \hline
  \hspace{0.78 in}  Test 1   \hspace{0.78 in}    &   \hspace{0.78 in} Test 2  \hspace{0.78 in}  &   \hspace{0.78 in} Test 3  \hspace{0.78 in}  \\
      \hline
    \end{tabular}   
    \begin{tabular}{||c|c||c|c||c|c||}
 planet  \hspace{0.48 in}    &    $ \frac{ M_p  }{M_{ \oplus} } \frac{M_{\odot} } {M_{\star} }  $   \hspace{0.48 in}  &  planet  \hspace{0.48 in}  &    $ \frac{ M_p  }{M_{ \oplus} } \frac{M_{\odot} } {M_{\star} }  $ \hspace{0.48 in}   &  planet  \hspace{0.48 in}  &    $ \frac{ M_p  }{M_{ \oplus} } \frac{M_{\odot} } {M_{\star} }  $ \hspace{0.48 in}  \\
 \hline 
  b  &  8.63 $^{+  0.65 }_{ -  0.65 }$   &  b  &   8.69 $^{+  1.09 }_{ -  1.02 }$     &   b  &  9.24 $^{+  0.98 }_{ -  0.91 }$       \\ 
   c  &  4.30 $^{+  0.58 }_{ -  0.58 }$   &  c  &   3.52 $^{+  0.80 }_{ -  0.73 }$  &  c  &   4.81 $^{+  0.68 }_{ -  0.62 }$      \\ 
      \hline
    \end{tabular}  
    \caption{TTV solutions for Kepler-307 b and c (KOI-1576) and  the results of various tests on dynamical masses. Test 1: An alternative eccentricity prior. Test 2: Robust fitting.  Test 3 Holczer catalog.}\label{tbl-koi1576}
  \end{center}
\end{table}
We list the results of our TTV analysis in Table~\ref{tbl-koi1576}. We obtained secure solutions for the dynamical masses of Kepler-307 b and Kepler-307 c. 

\begin{figure}[h!]
\includegraphics [height = 1.5 in]{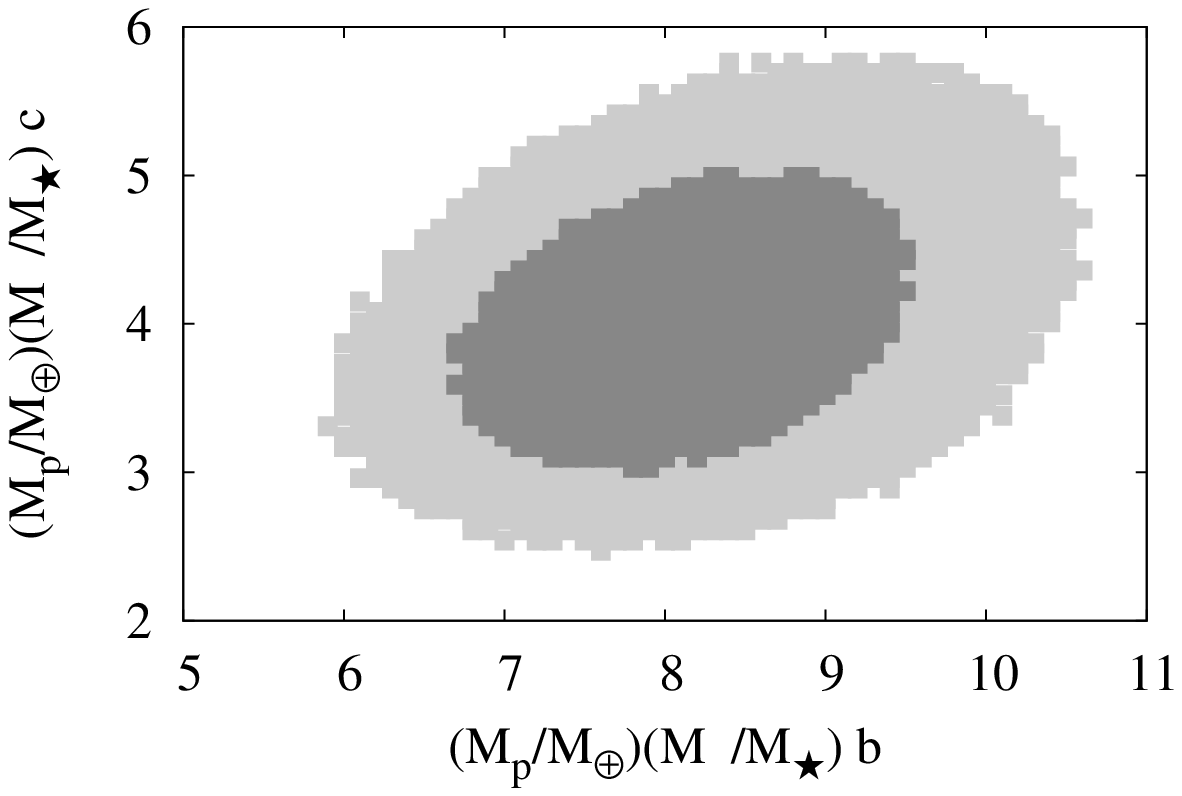}
\includegraphics [height = 1.5 in]{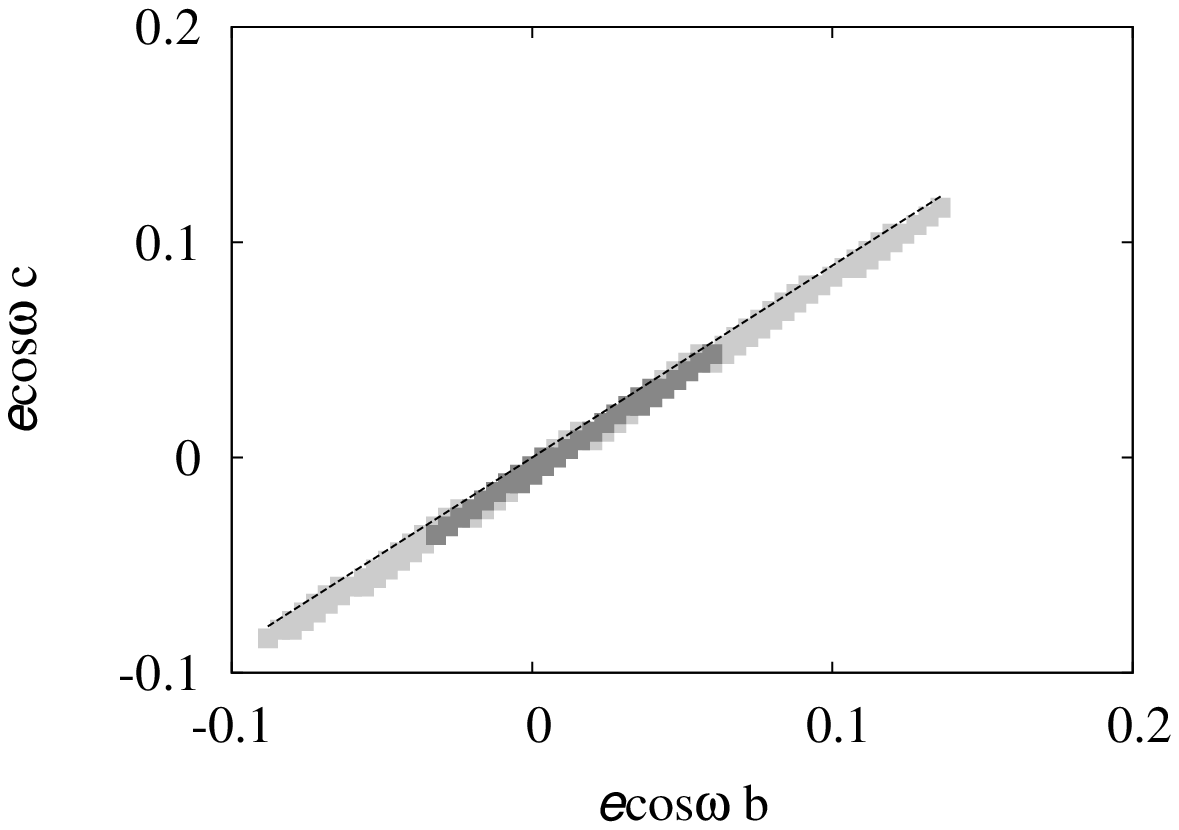}
\includegraphics [height = 1.5 in]{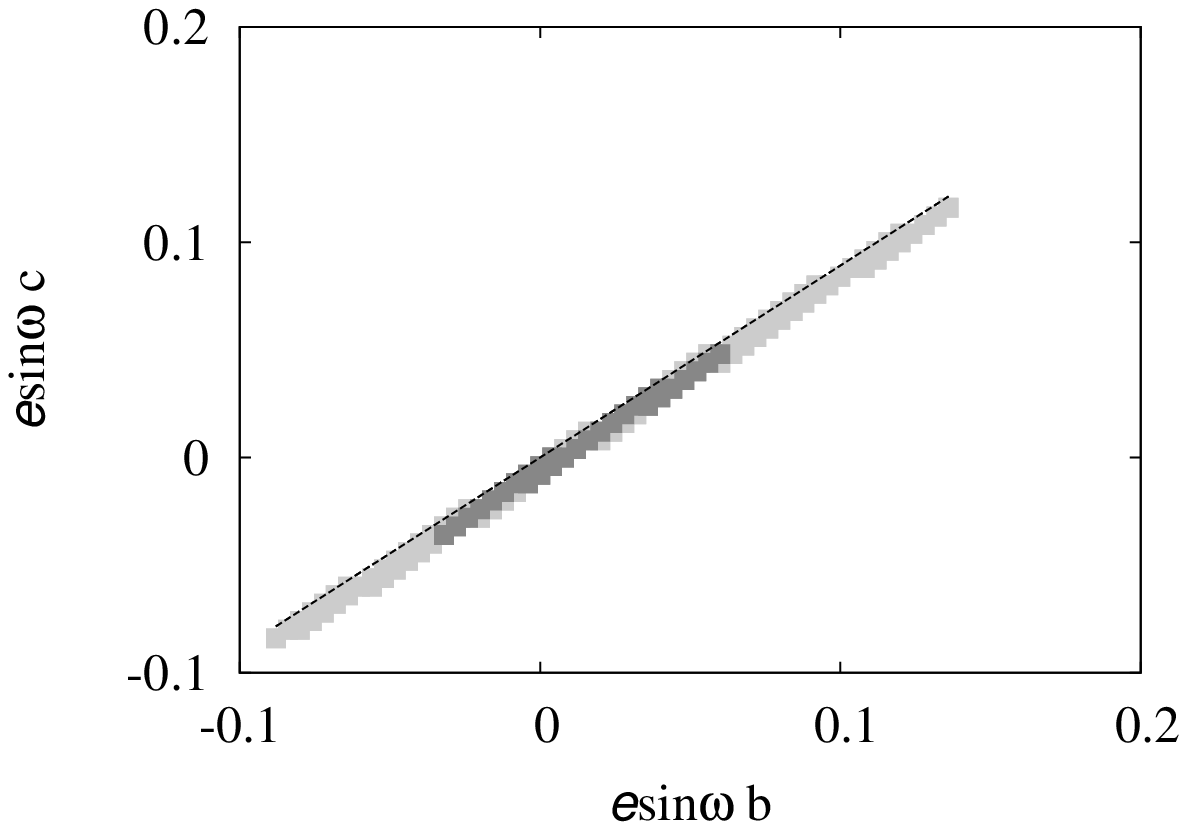}
\caption{Joint Posteriors for planetary masses relative to the host star for Kepler-307 (KOI-1576). 95.4\% confidence intervals are in light grey, 68.3\% confidence intervals in dark grey. The dashed lines mark the expected correlation between eccentricity vector components from Equation~\ref{eqn:gradient}
}
\label{fig:1576} 
\end{figure}

The TTVs tightly constrain the dynamical masses of Kepler-307 b and c. A sample of 50 solutions from our posteriors were integrated for 1 Myr, and all were found to be stable.  \citet{had15} performed an independent analysis of the TTVs of Kepler-307 using transit times measured from long cadence data and their dynamical results are consistent with ours, with mass (and radius) estimates for Kepler-307 b and c at $8.6^{+1.6}_{-1.4} M_{\oplus}$ ($3.2^{+1.2}_{-0.5} R_{\oplus}$) and $3.7^{+1.0}_{-0.8} M_{\oplus}$ ($2.8^{+1.0}_{-0.4} R_{\oplus}$) respectively. However, our precise constraints on the stellar parameters imply slightly lower masses for the planets, smaller radii and higher densities, as shown in Section 5.

\subsection{Summary}
We thus have tight constraints on the dynamical masses of ten transiting planets. We note that the ratio of the 2$\sigma$ lower bounds on the planet masses to 1$\sigma$ values cluster just above 1.5, and none exceed 2; the negative uncertainties are almost always less than the positive uncertainties. Hence, zero mass (or non-detection of TTVs) is much more strongly rejected by the data than would be implied with assumed Gaussian distributions about the median.

Although we have not recovered precise eccentricities, we have tightly constrained the relative eccentricities between neighboring planets in all of these systems. We examine the posteriors of relative apses for planet pairs below. 

\section{Eccentricities and Apses}
\begin{figure}[!h]
\includegraphics [height = 1.5 in]{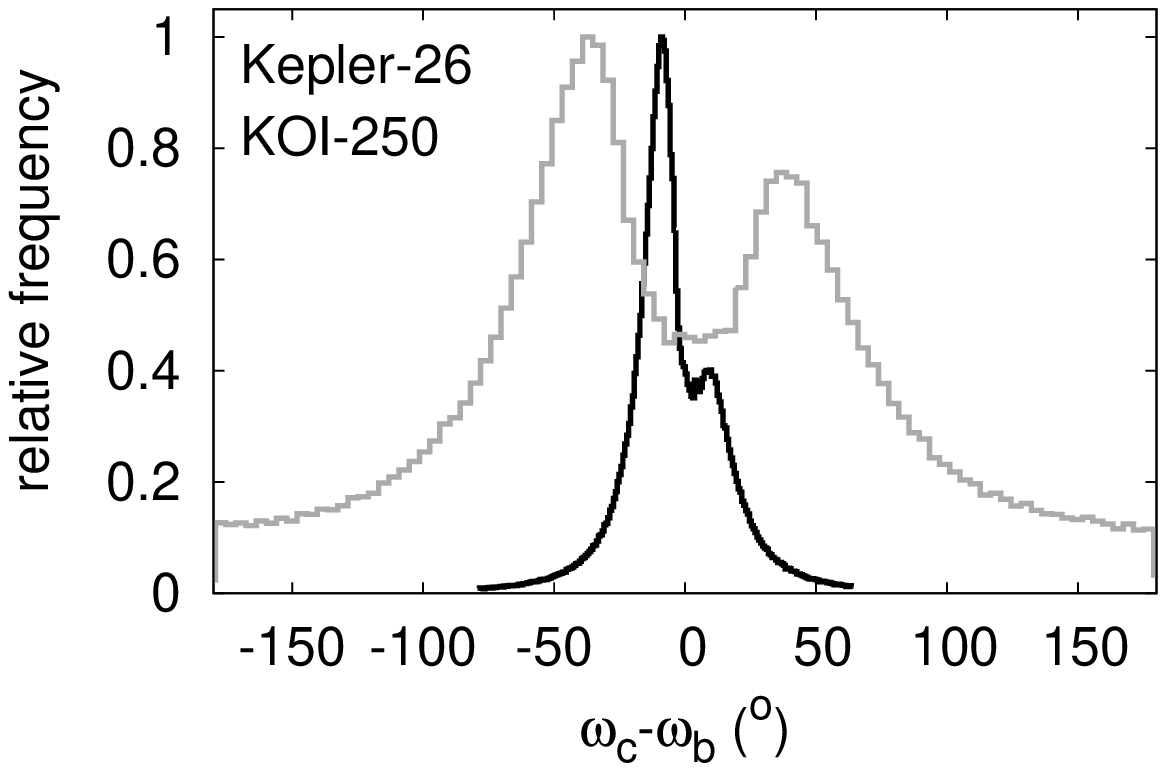}
\includegraphics [height = 1.5 in]{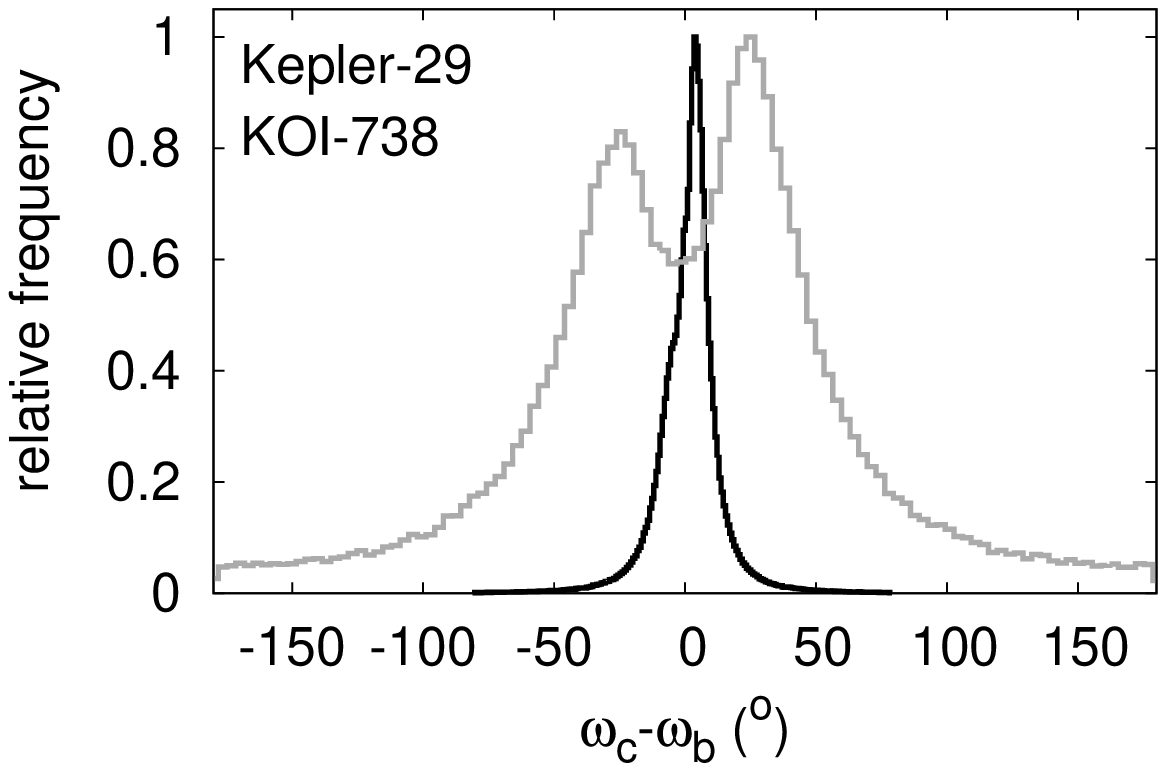}
\includegraphics [height = 1.5 in]{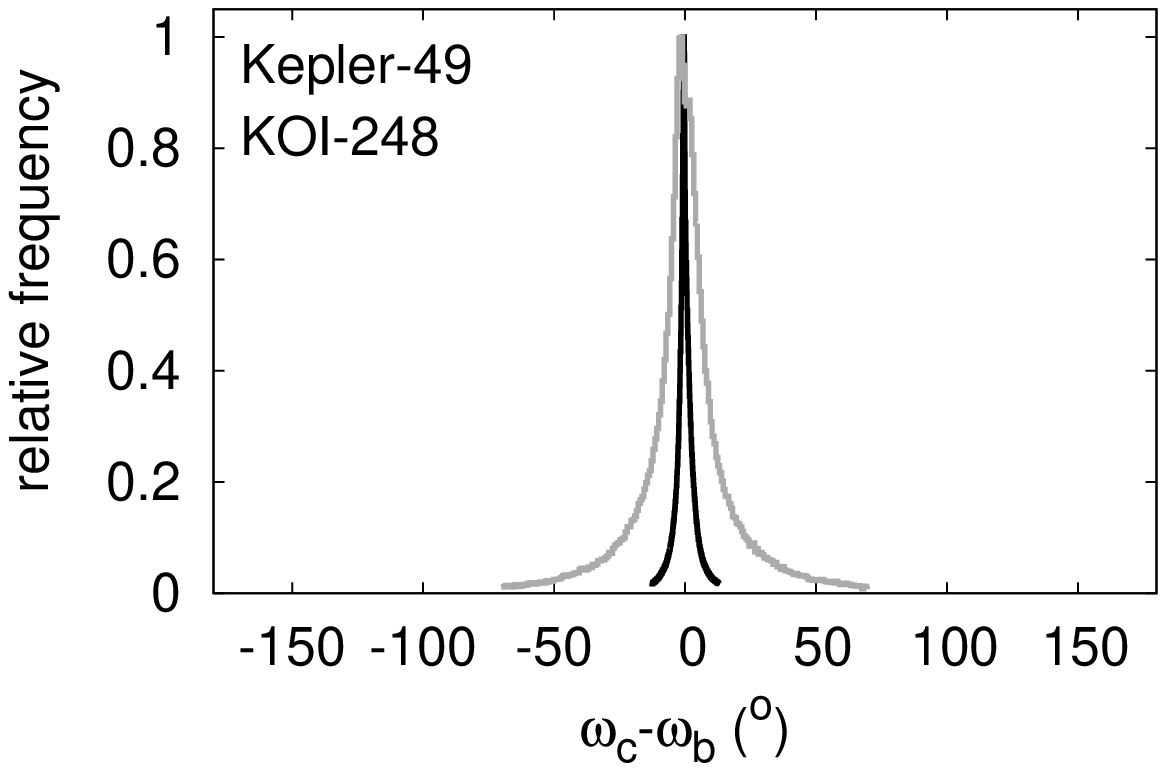}
\includegraphics [height = 1.5 in]{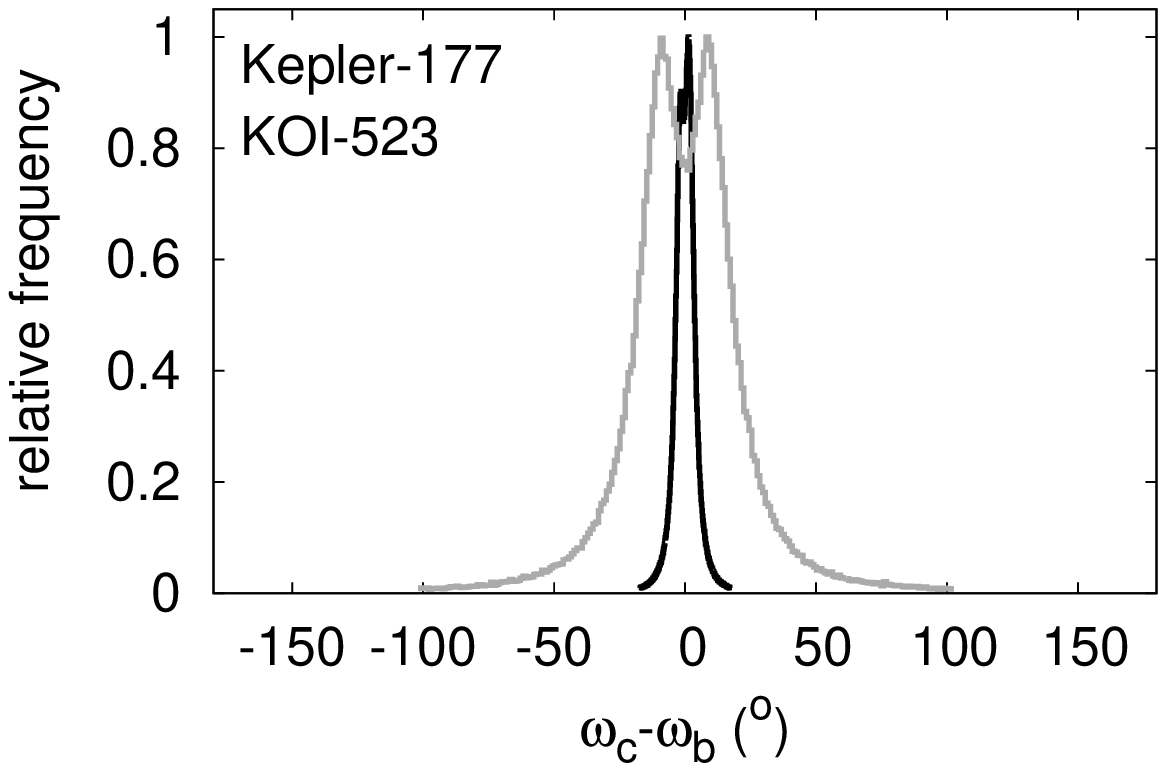}
\includegraphics [height = 1.5 in]{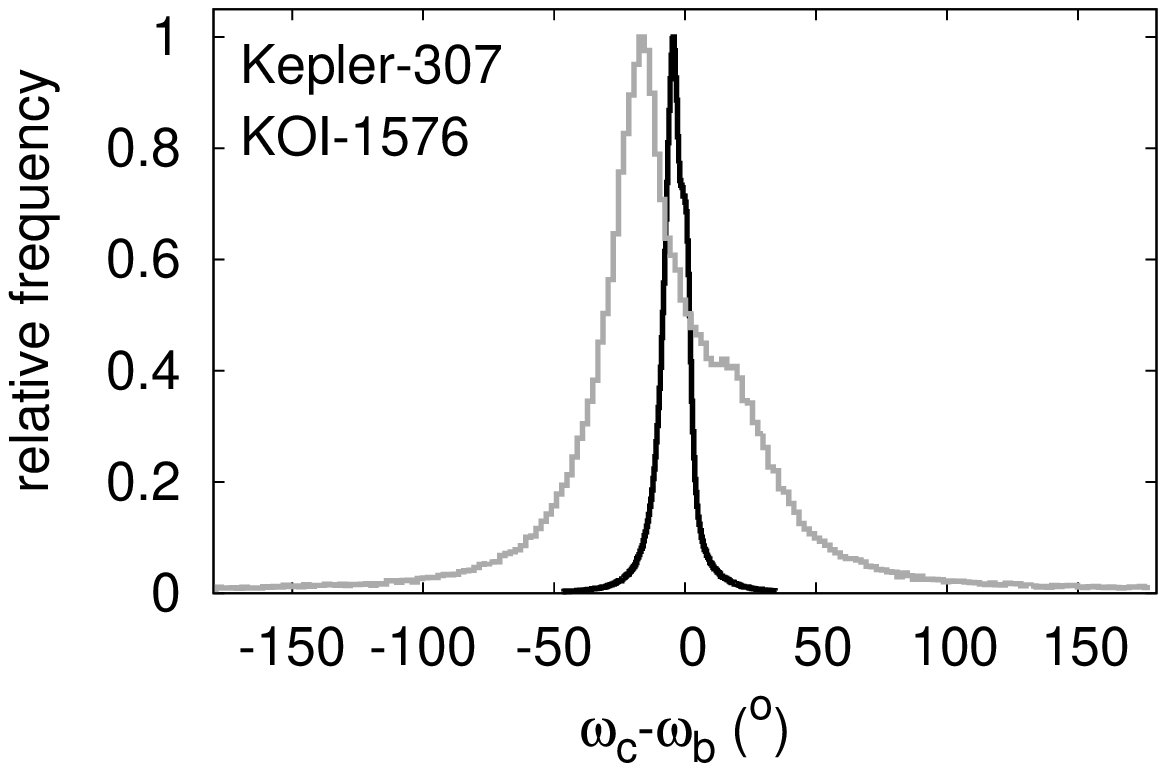}
\caption{Marginal posterior distributions for the relative apses of two-planet systems in this study. In these histograms, the relative frequencies have been normalized to the peak. The black curves mark posteriors using a broad prior on orbital eccentricity, a Rayleigh distribution of Rayleigh parameter 0.1, while the grey curves correspond to a narrow prior a Rayleigh distribution of Rayleigh parameter 0.02 that disfavors high eccentricities. In the cases of Kepler-26 and Kepler-29, only the high eccentricity solutions have closely aligned pericenters. For Kepler-49, Kepler-177 and Kepler-307, marginal posteriors show apses are aligned, either tightly (Kepler-49) or loosely (Kepler-177 and Kepler-307) for both choices of eccentricity prior.}
\label{fig:2planetapses} 
\end{figure}

Here, we compare the distributions of the relative positions of apses for each planet pair ($\omega'-\omega$) given our posterior samples. In Figure~\ref{fig:2planetapses}, the histograms of relative apses from our posterior samples for two choices of prior on orbital eccentricity are compared. For the wide prior on eccentricity (with vector components normally distributed: N($\mu=$0,$\sigma$ =0.1), the relative apses all show a sharp peak near zero, corresponding to apsidally aligned solutions. However, in each case our analysis with a narrower eccentricity prior for vector components N($\mu$=0,$\sigma$=0.02) results in a wider posterior in relative apses. Therefore, our inference about the preference for alignment is extremely sensitive to our choice of prior for eccentricity. In some cases (Kepler-49, Kepler-177 and Kepler-307), we find TTVs provide evidence of apsidal alignment for a wide range of choices for eccentricity priors.

\begin{figure}[!h]
%\includegraphics [height = 1.2 in]{KOI-2086-histo-wdeg-bc.eps}
%\hspace{-0.7 cm}
%\includegraphics [height = 1.2 in]{KOI-2086-histo-wdeg-cd.eps}
%\hspace{-0.4 cm}
%\includegraphics [height = 1.2 in]{KOI-115-histo-wdeg-bc.eps}
%\hspace{-0.7 cm}
%\includegraphics [height = 1.2 in]{KOI-115-histo-wdeg-cd.eps}
\includegraphics [height = 1.2 in]{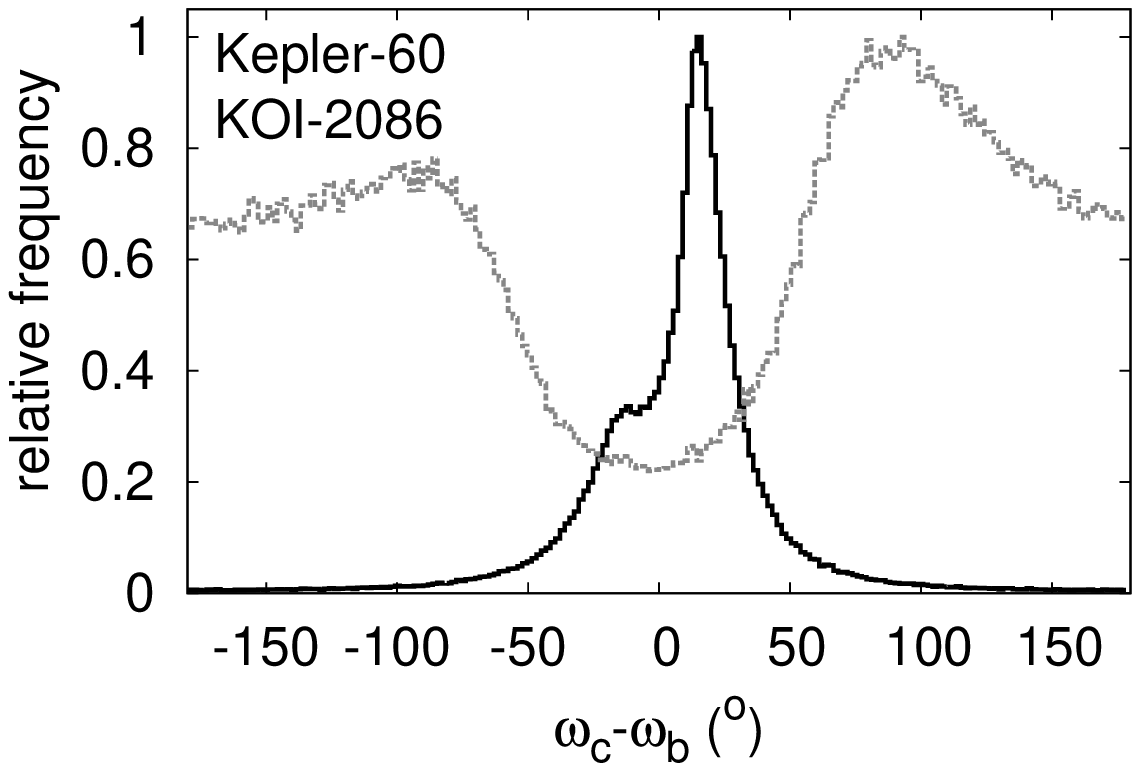}
\hspace{-0.7 cm}
\includegraphics [height = 1.2 in]{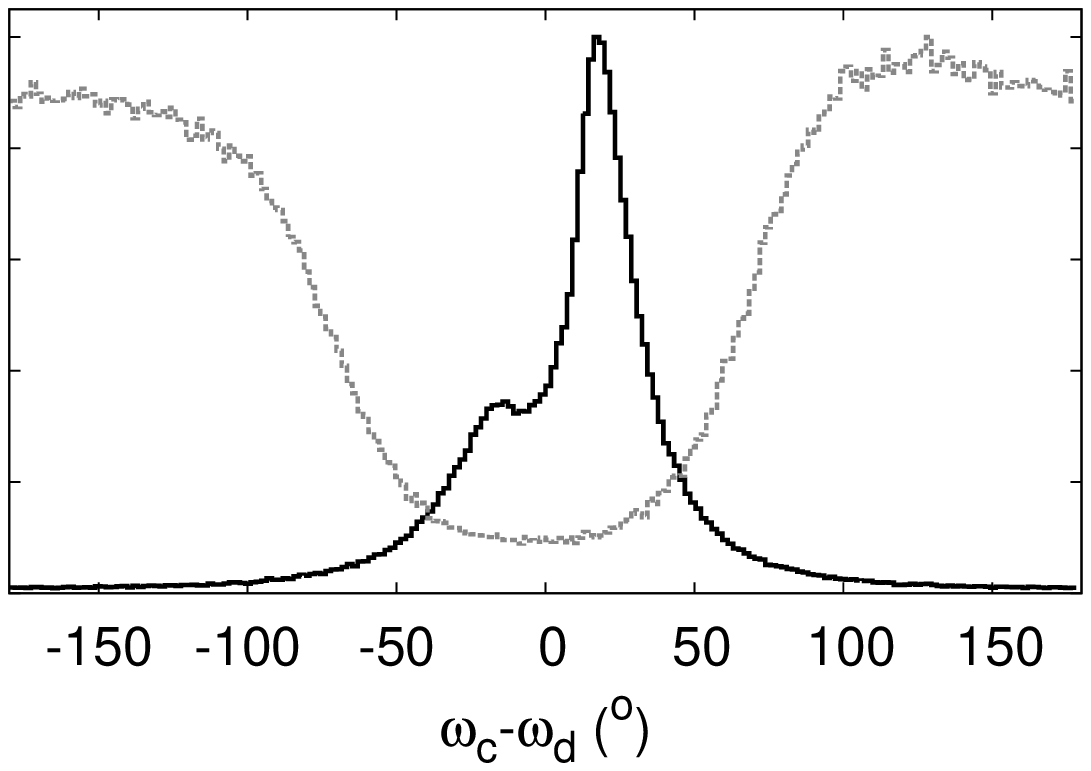}
\hspace{-0.4 cm}
\includegraphics [height = 1.2 in]{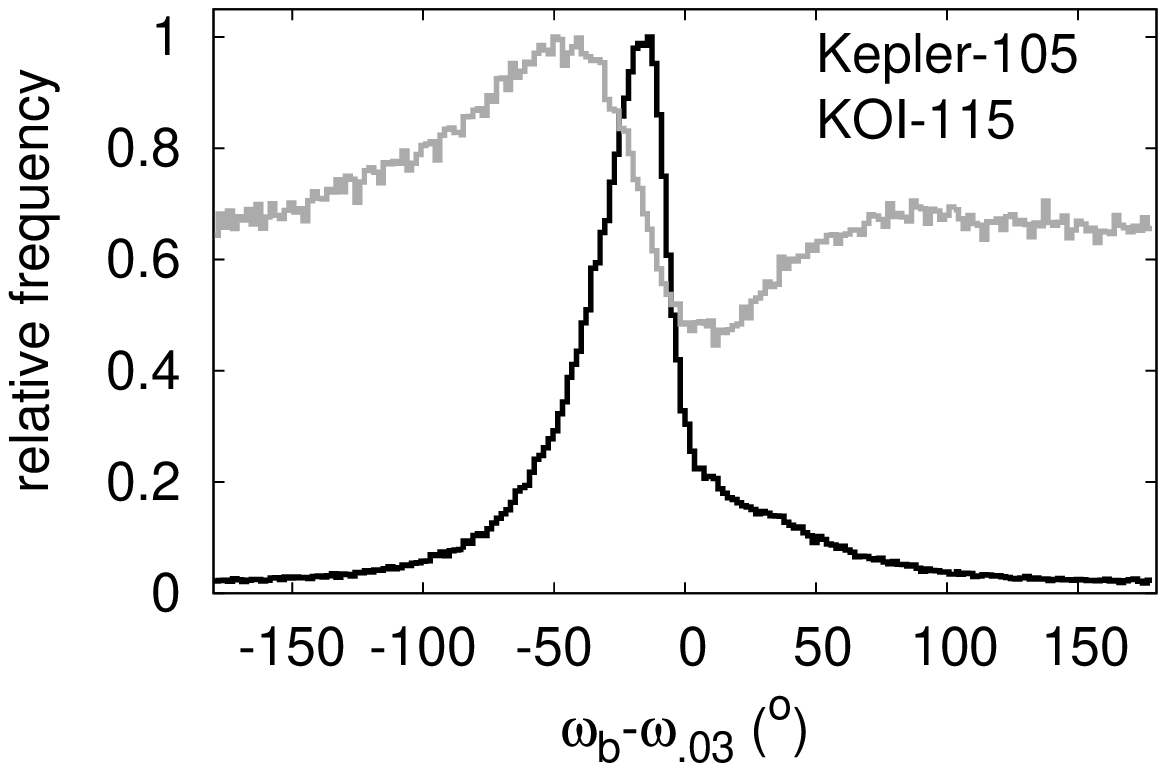}
\hspace{-0.7 cm}
\includegraphics [height = 1.2 in]{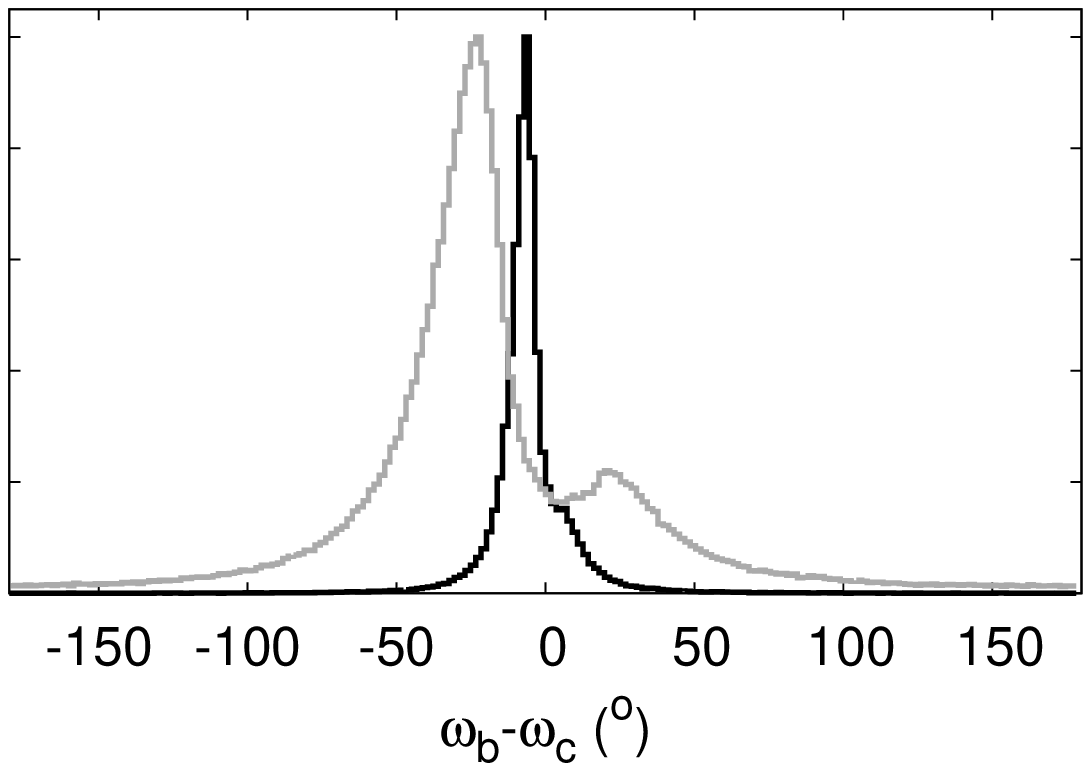}
\caption{Marginal posteriors for the relative apses of three-planet systems in this study; Kepler-60 (left panels) and Kepler-105 (right panels). The black curves mark posteriors using a broad prior on orbital eccentricity, while the grey curves correspond to a narrow prior that disfavors high eccentricities. In all cases, the sharp peak near zero for the wide eccentricity prior reveals that apsidal alignment is favored among high eccentricity solutions. For the low-eccentricity samples, only the outer pair of Kepler-105 has a sharp peak in relative apses indicative of apsidal alignment.}
\label{fig:3planetapses} 
\end{figure}

Figure~\ref{fig:3planetapses} shows histograms of relative apses in the three-planet systems, for both the wide and narrow eccentricity priors. At Kepler-60, we see that while the high eccentricity solutions give a very narrow peak in relative apses near alignment, the low eccentricity solutions disfavor apsidal alignment. At Kepler-105, the outer pair show a narrow range of relative apses for both priors in eccentricity, whereas the inner pair shows evidence of apsidal alignment with high eccentricities only.

Next, we consider the secular evolution of these systems. The planets of all of these systems are secularly highly coupled. In classical second-order secular theory, each planet's eccentricity vector components ($k = e\cos \varpi $, $h = e\sin \varpi$) are the vector sum of contributions from N$_{p}$ eigenmodes, where N$_{p}$ is the number of planets in the system. For all cases here, we have assumed the systems are coplanar and set the ascending node to zero so that the argument of periapse $\omega$ and its longitude of periapse $\varpi$ are the same. We analyze the eigenmodes of systems with interacting pairs of planets, beginning with a detailed explanation of Kepler-26. 

\begin{figure}[h!]
\includegraphics [height = 1.8 in]{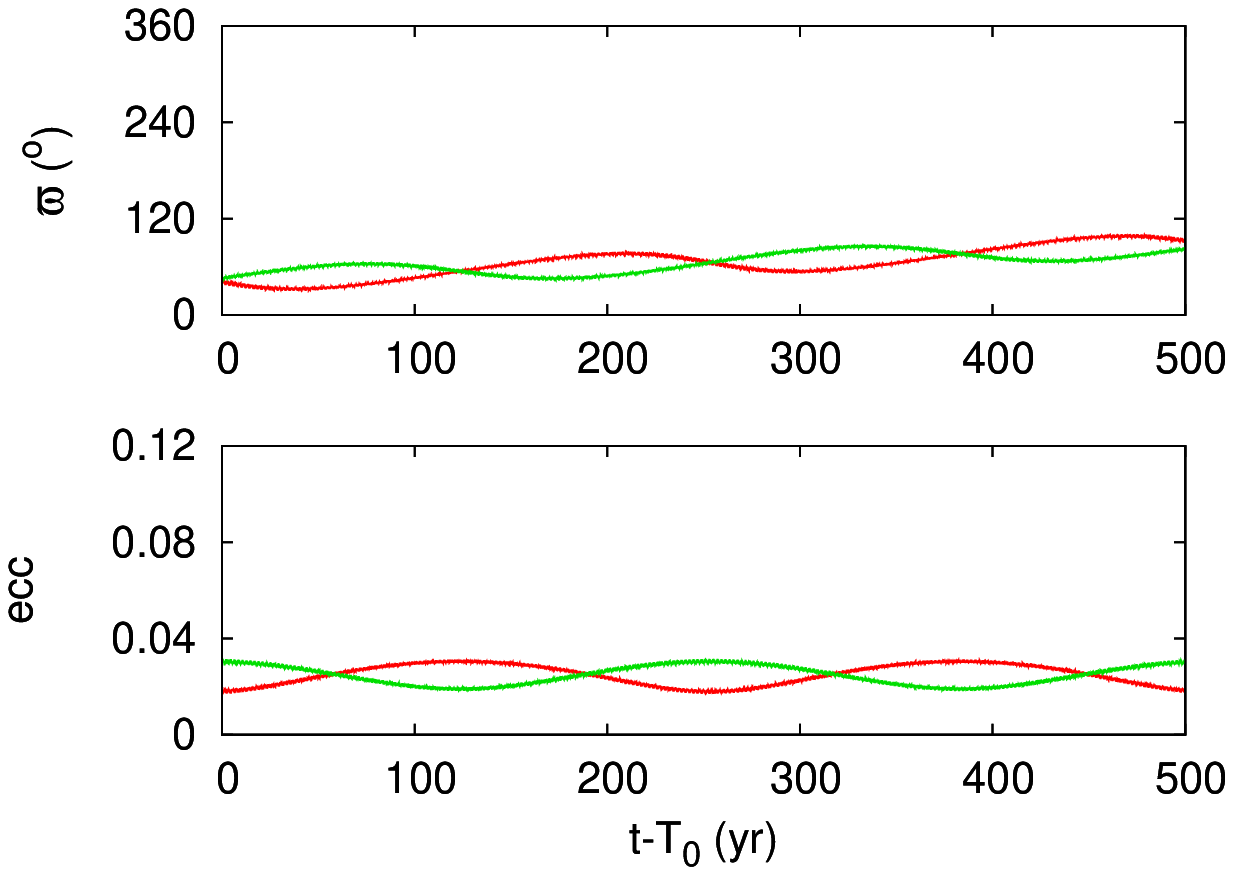}
\includegraphics [height = 1.8 in]{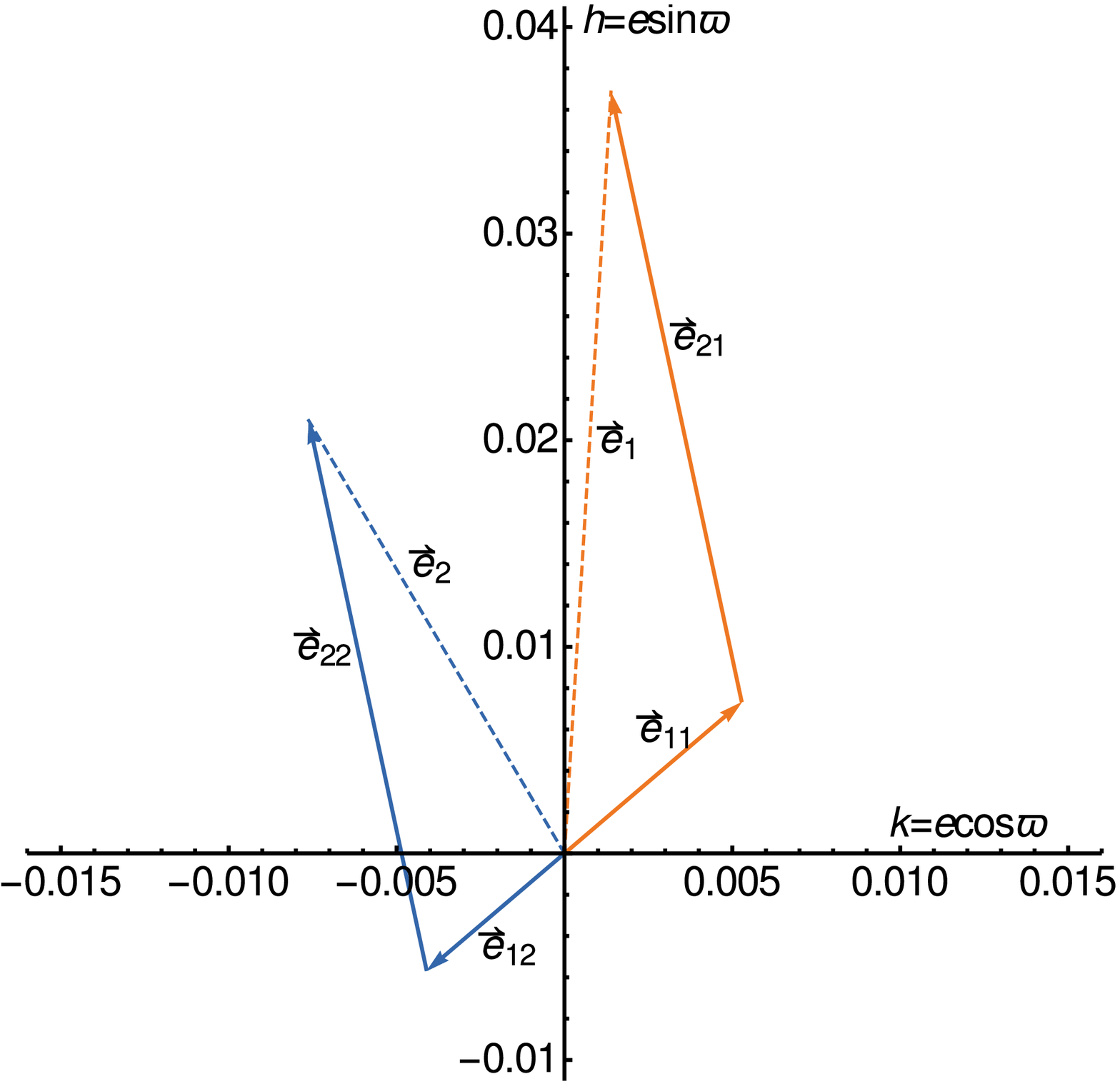}
\includegraphics [height = 1.8 in]{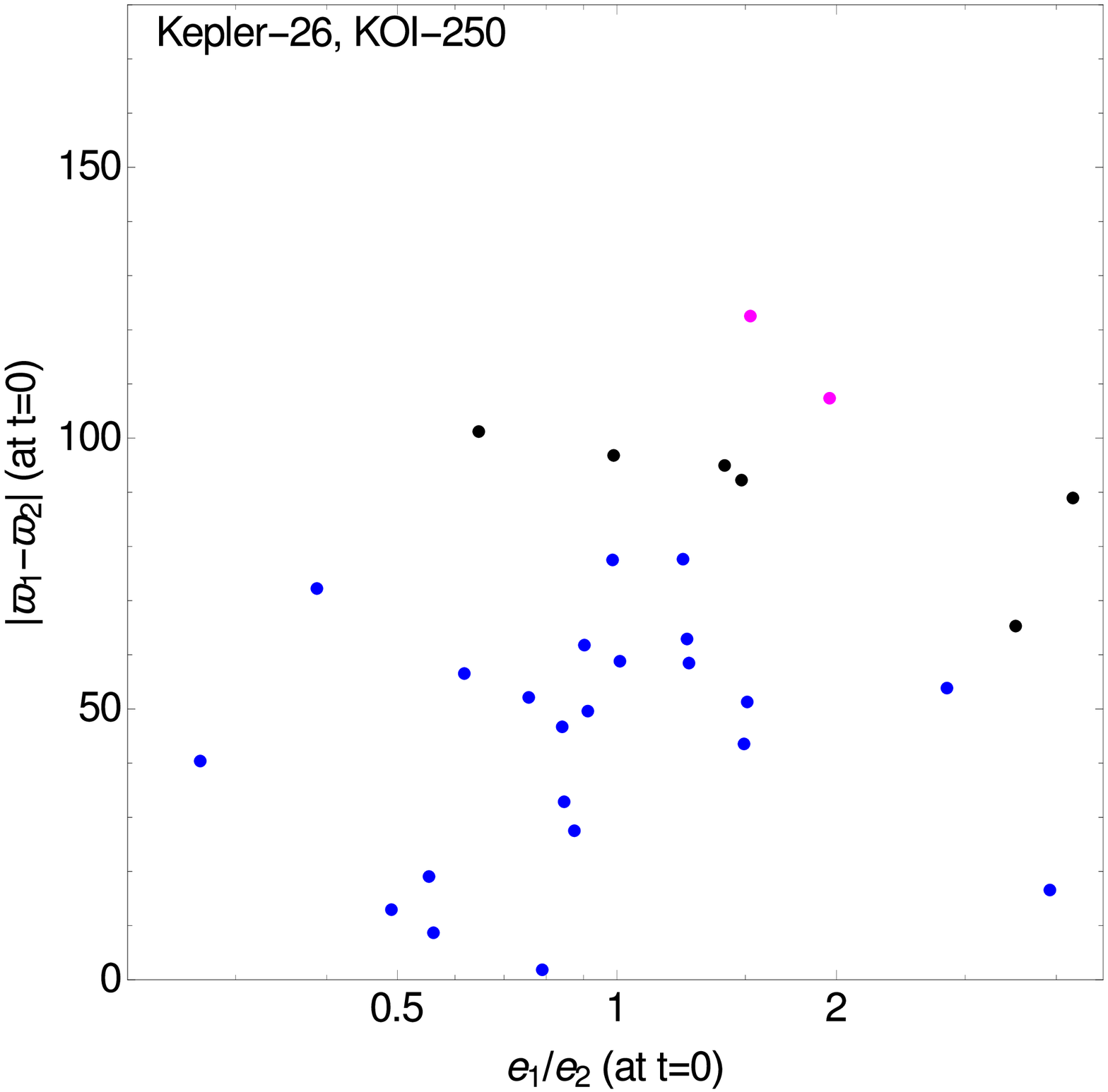}
\caption{On the left is an example of apsidal alignment over a 500-year simulation, with apsides in the upper panel and eccentricities on the lower panel, where Kepler-26 b is in red, and Kepler-26 c is in green. The middle panel illustrates eigenmode decomposition of one example of our posterior samples, with aligned and anti-aligned eigenmodes. On the right we plot the difference in apses for a sample of our posteriors, as a function of eccentricity ratios. The blue points are in libration about apsidal alignment, while the magenta points are solutions that are in libration about anti-alignment. The black points are circulating without any alignment.  }
\label{fig:eigenmodesK26} 
\end{figure}

In the case of Kepler-26, both eigenmodes affect each planet almost equally.  Most of our samples from our nominal prior in eccentricity (Gaussians in (\textit{k,h}) with $\sigma =$ 0.1) were apsidally aligned, as expected due to the eccentricity-eccentricity degeneracy highlighted in Equation~\ref{eqn:Zfree}. To test whether there is any preference for apsidal alignment in our posterior sample based on an alternative eccentricity prior (Gaussians in (\textit{k,h}) with $\sigma =$ 0.02), we took 30 samples from our MCMC chains, and integrated the solutions for 500 years. An example of apsidal alignment is shown in the left panel of Figure~\ref{fig:eigenmodesK26}. 
 
In (\textit{k, h}) space, the vector contributions to the planets' eccentricities from the faster eigenmode ($\vec{e_{11}}$, $\vec{e_{12}}$) initially point in opposite directions. For the slower eigenmode, its vector contributions ($\vec{e_{21}}$, $\vec{e_{22}}$) are initially parallel. If the anti-parallel eigenmodes dominate the overall eccentricity for both planets, the longitude of the orbital apses ($\varpi$) will librate around anti-alignment. However, if the parallel eigenmodes dominate the eccentricity for both planets, their apsides librate around alignment. These parallel and anti-parallel components are illustrated for one sample of the posteriors for Kepler-26 in the middle panel of Figure~\ref{fig:eigenmodesK26}.   %This outcome is favored by the eccentricity-eccentricity degeneracy in TTVs. 

For Kepler-26, we found that majority of our sample were in apsidal alignment, and a small fraction were apsidally anti-aligned. These are shown in the right panel of Figure~\ref{fig:eigenmodesK26}. 

\begin{figure}[h!]
%\includegraphics [height = 2.9 in]{248_cvl.eps}
%\includegraphics [height = 2.9 in]{0738_cvl.eps}
%\includegraphics [height = 2.9 in]{523_cvl.eps}
%\hspace{0.3 in}
%\includegraphics [height = 2.9 in]{1576_cvl.eps}
\includegraphics [height = 2.9 in]{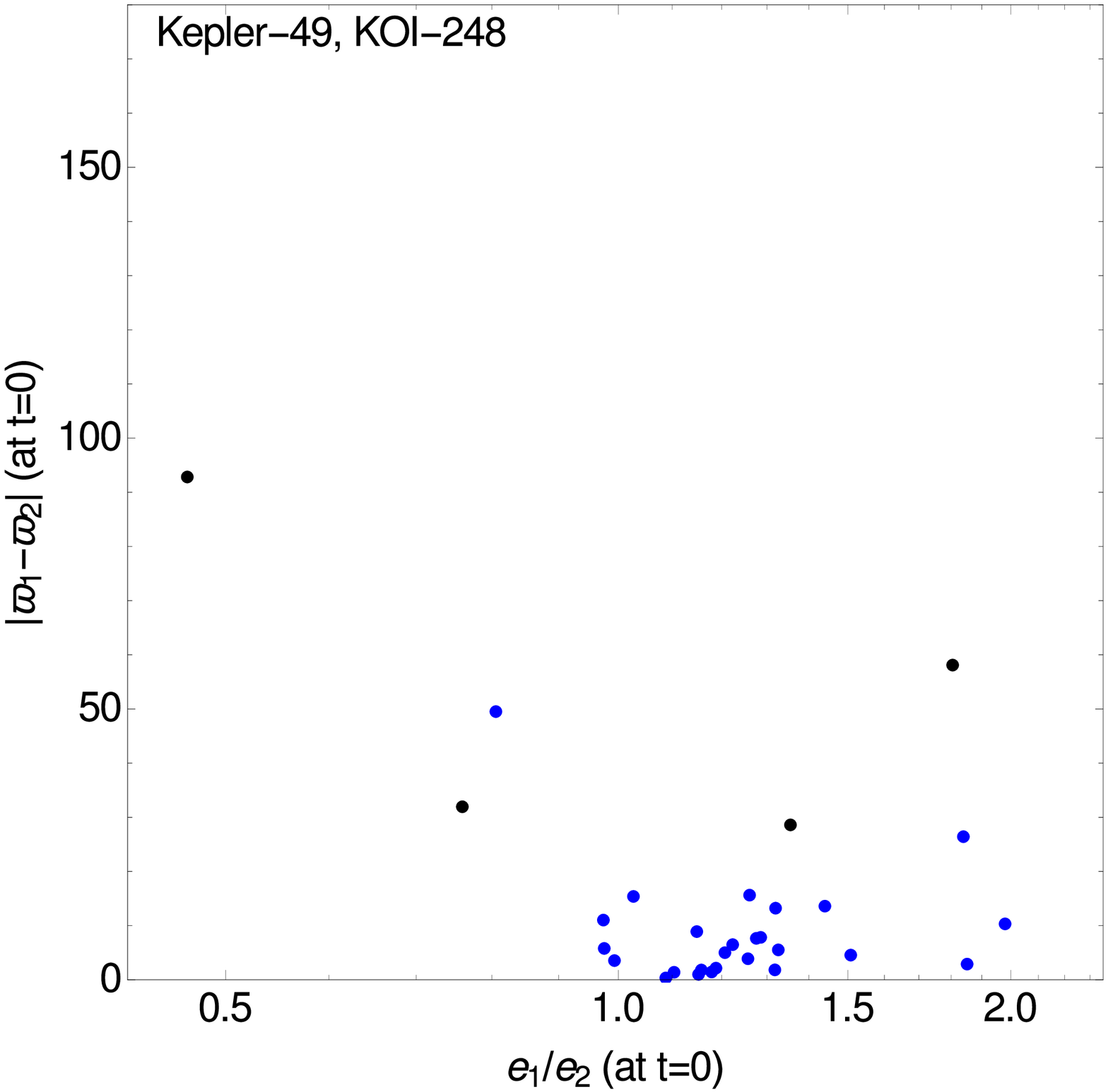}
\includegraphics [height = 2.9 in]{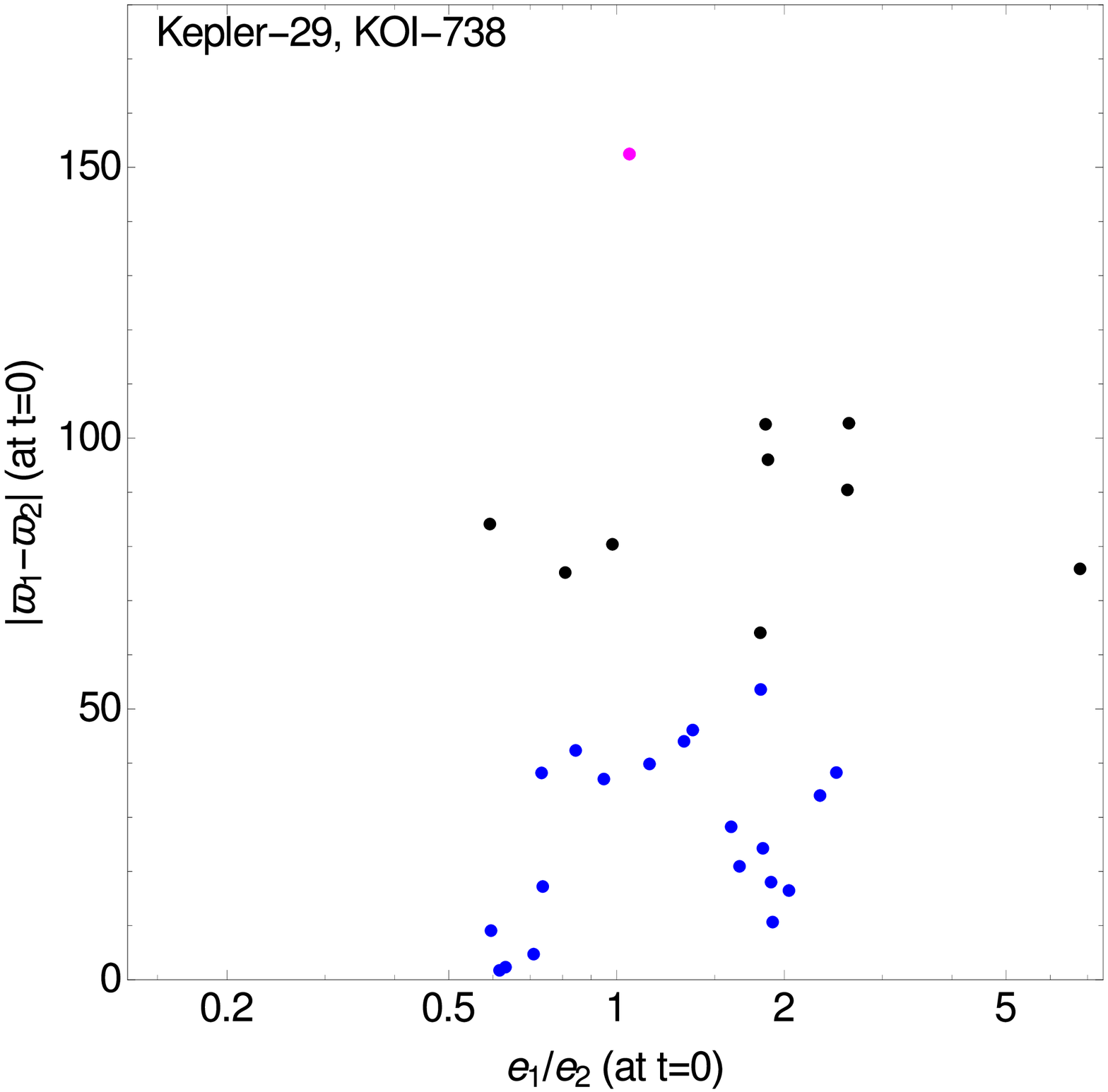}
\includegraphics [height = 2.9 in]{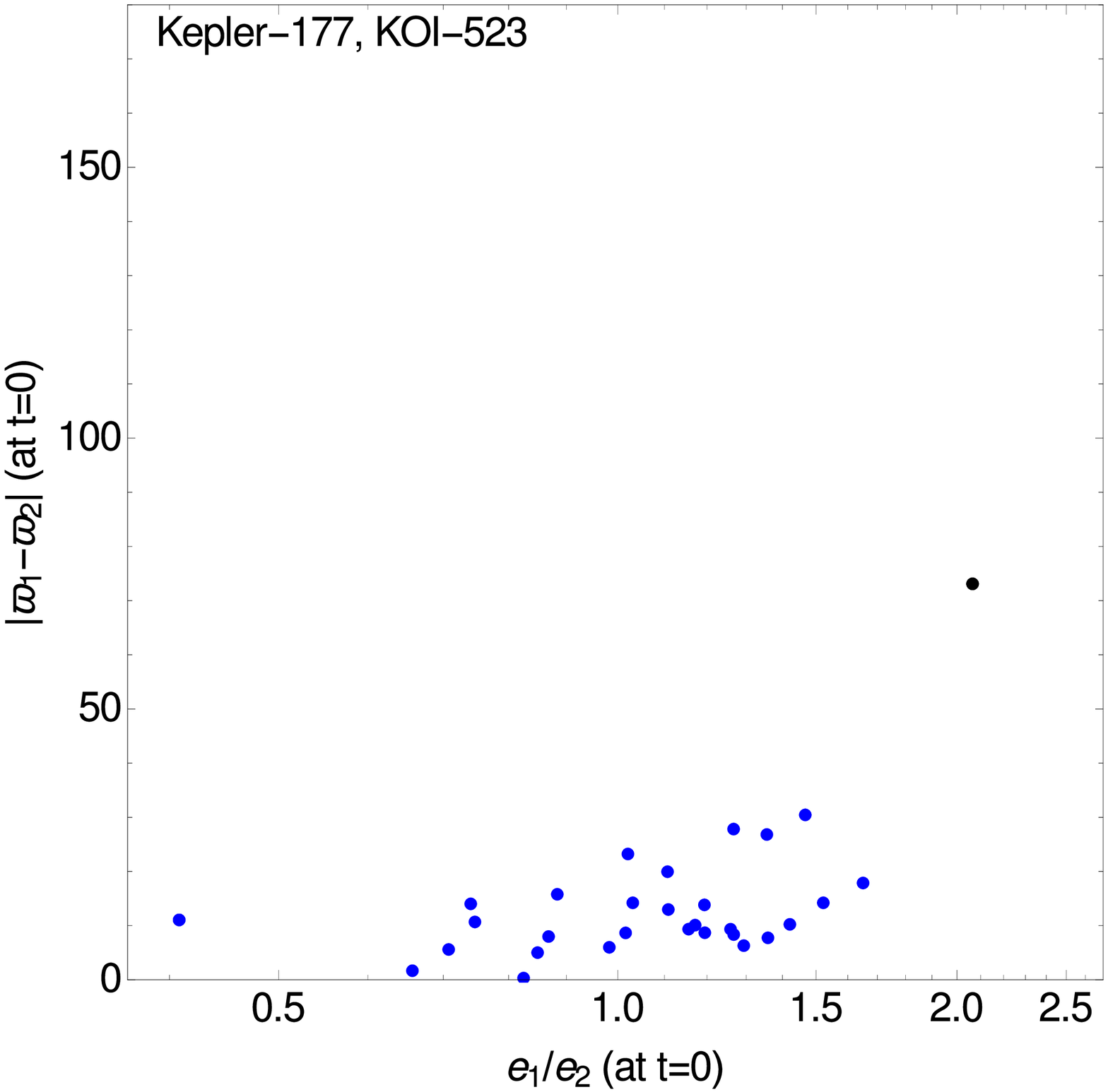}
\hspace{0.3 in}
\includegraphics [height = 2.9 in]{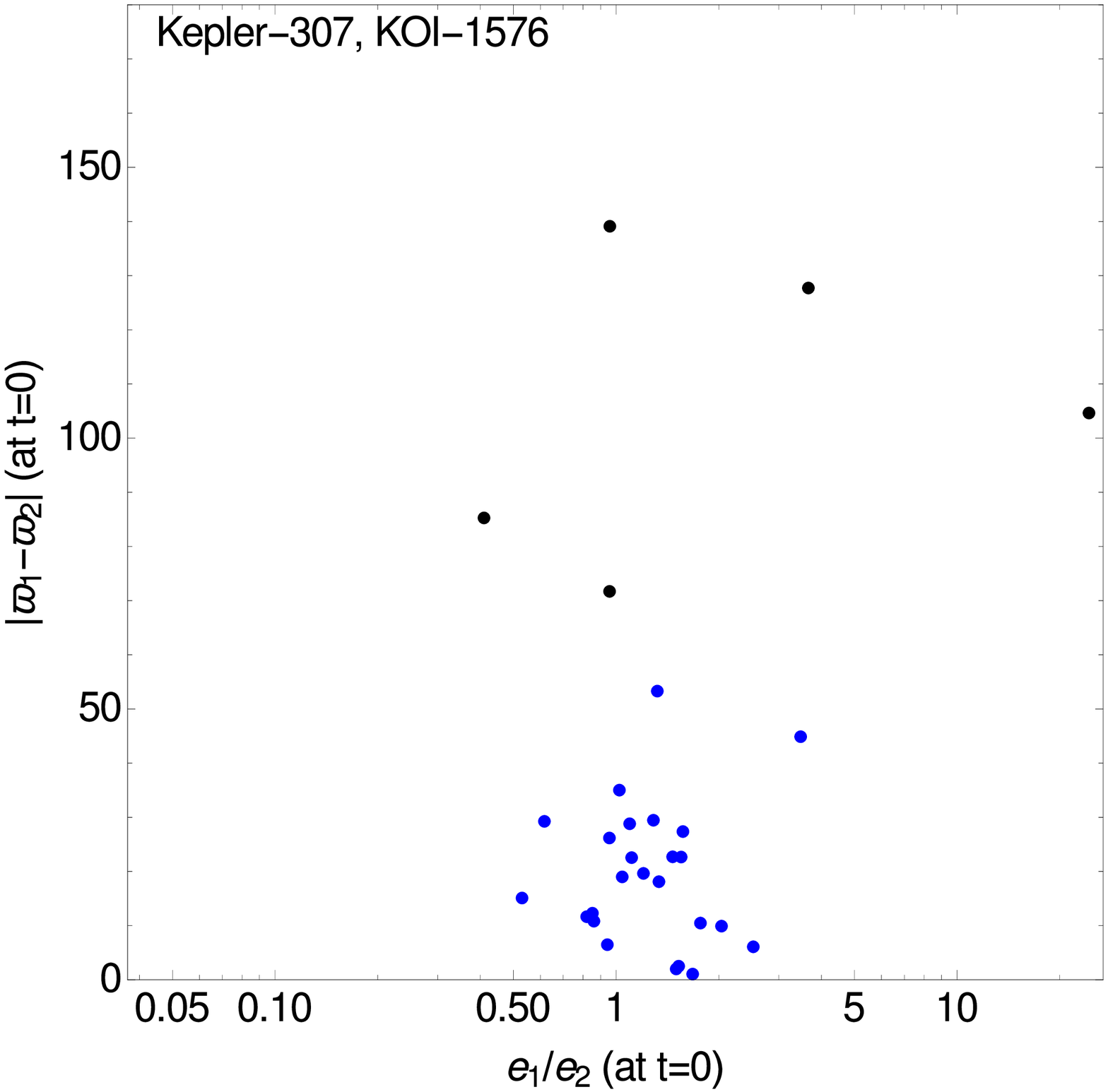}\caption{Initial differences in apses for 30 posterior samples of low eccentricity for each of four systems, plotted as a function of eccentricity ratios. Panels show results for Kepler-49 (KOI-248, top left), Kepler-29 (KOI-738, top right), Kepler-177 (KOI-523, bottom left) and Kepler-307 (KOI-1576, bottom right). The blue points are in libration about apsidal alignment, while the magenta points are solutions that are in libration about anti-alignment. The black points are circulating without any alignment. A high fraction of samples in all four cases show apsidal alignment.}
\label{fig:eigenmodes-all} 
\end{figure}

We repeated this analysis for the remaining systems where the TTVs are due to a pair of interacting planets only. Figure~\ref{fig:eigenmodes-all} compares outcomes in relative apsidal orientations among the interacting planet pairs of four systems. For our nominal choice of eccentricity prior, we find strong evidence of apsidal alignment in all cases, particularly among high eccentricity solutions. We also analyzed the eigenmodes from posterior samples taken with a narrow eccentricity prior. In each of these four cases, and especially Kepler-49 and Kepler-177, while the fraction of samples that were aligned was lower than a sample that includes high eccentricities, the majority of posterior samples were apsidally locked for low eccentricities too.

\section{Stellar Parameters}
For all stellar parameters, we constrain the stellar density, $\rho_{\star}$, from our light curve analysis, after removing the effects of measured TTVs from our transit models \citep{rowe14}. We adopt the analytic model of \citet{man02} for the transit profile, with quadratic limb darkening parameters from \citet{cla11}. Atmospheric parameters come from the most recent version of the Kepler Input Catalog retrieval from the NASA Exoplanet Archive \citep{Huber2014}. These include parameters based on either spectroscopy or photometry depending on what observations are available. We re-iterate the light curve analysis after inferring the eccentricity vector posteriors from the TTV analysis to obtain a transit-model measurement of $\rho_\star$ (\citealt{liss13, jont14, jont15}). Combining this estimate of $\rho_{\star}$ with published stellar atmospheric values of the effective stellar temperature $T_{eff}$, metallicity [Fe/H], and surface gravity $\log g$, we matched stellar evolution models to estimate the stellar mass $M_{\star}$ and radius $R_{\star}$ \citep{dem04}. We checked for consistency between \logg\ and \rhostar\ based on stellar models and find good agreement (within 1$\sigma$) for all systems except for Kepler-49 (KOI-248) and Kepler-57 (KOI-1270).

In the case of Kepler-49, the multi-planet transit models suggests a star with \rhostar\ $\sim$ 1 \gcmc, which is in strong disagreement with stellar parameters reported by Swift et al. (2015). The Kepler-49 system consists of four planets, with planets b and c found to exhibit anti-correlated TTVs and KOI-248.03 and KOI-248.04 not exhibiting TTV behaviour.  We fitted a transit model to each transiting planet independently, thus allowing an estimate of \rhostar\ from each transiting planet.  We find that all transiting planets, except Kepler-49 c have transit models consistent with Swift et al. (2015).  Since planets b and c show anti-correlated TTVs they must be part of the same planetary system, thus we conclude that the transit model for Kepler-49 c must have a systematic error.  An explanation can be found if the observed transit depth is shallower than expected due to spot-crossing events.  \citet{rowe15a} found that a diluted transit depth will result in a systematic error on the transit-derived value of \rhostar.  A further examination of the transit models for the Kepler-49 system shows that all planets, except Kepler-49 c, are consistent with a central crossing transit (where impact parameter $b=0$).  In the case of Kepler-49 c, the models and transit shape are consistent with a near grazing ($b\sim$0.9) model. 

The host is known to be an active star with a likely rotation period $\sim$18.5 days \citep{McQuillan2013}. We speculate that Kepler-49 c was seen to be consistently crossing a polar star spot resulting in a diluted transit depth. This can cause both anomalous TTVs and an underestimate of the planet radius \citep{Oshagh2013}. Thus the true radius of this planet is larger than reported.  The anomalous transit profile was for Kepler-49 c, which is part of our TTV analysis. Hence, we report our TTV results and adopt the published stellar parameters of Swift et al. (2015) here, but demote this system to the less secure class of systems that require further analysis (see Table~\ref{tbl-results}). 

In the case of Kepler-57, a two-planet system, our multi-planet transit model gives \rhostar $\sim$ 15 \gcmc. Fitting models to individual planets, we find consistent values of \rhostar\ from model fits both planet transit curves. However, the inferred stellar parameters disagree with stellar classification based on spectral data listed in \textit{Kepler}'s Community Follow-up Program\footnote{which can be found at https://cfop.ipac.caltech.edu.}. Spectral models infer \rhostar $\sim$ 2.5 g cm$^{-3}$. The discrepant stellar densities from the transit light curve and spectral data may be due to a stellar blend with the planets of the Kepler-57 orbiting a fainter unidentified stellar companion. In this case, the planetary masses would be lower than inferred with a single star model, and the planet sizes would be larger than in the single star model. This would cause the planet densities to be much lower than the nominal results shown in Table~\ref{tbl-results}. For this system, we report our TTV results and adopt the published stellar parameters of {\it CFOP}, but demote this system to the less class of systems that require further analysis (see Table~\ref{tbl-results}).

 \begin{table}[h!]
  \begin{center}
    \begin{tabular}{||c|c|c|c|c|c|c||}
      \hline
 \hline 
 Kepler \#  & KOI  &      $M_{\star}$  ($M_{\odot}$)  &      $R_{\star}$ ($R_{\odot}$)  &  $T_{eff}$   &    [Fe/H]  (dex) &  Reference  \\   
 \hline
 Kepler-26 & 250 &  $0.544\pm 0.025$ &  $0.512\pm0.017$ &  $3914\pm 119$ &  $-0.13 \pm 0.13$ & this work   \\ 
  Kepler-29 & 738 &  $0.979\pm 0.052$ &  $0.932\pm0.060$ &  $5701 \pm 102$ &  $-0.04 \pm 0.12$ & this work   \\  
 Kepler-49 & 248 &  $0.55\pm 0.04$ &  $0.52\pm0.04$ &  $3838^{+111}_{-74}$ &   $-0.02\pm0.14$  &\citet{swift15}   \\
Kepler-57  & 1270  &   $0.884\pm0.066$ &  0.791$\pm 0.179$ & $5324\pm166$ &   $-0.04\pm0.28$  & \textit{CFOP}  \\  
Kepler-60  &  2086  & 1.041$\pm 0.077$  &  $1.257\pm0.094$ &  $5905\pm144$ &   $-0.09\pm0.10$  & this work   \\  
 Kepler-105 & 115  &  0.961$\pm 0.046$ &  $ 0.894\pm0.044$ &  $5827\pm94$ &   $-0.19\pm0.10$  & this work  \\  
  Kepler-177  & 523  & 1.259$\pm 0.108$ &  $1.399\pm0.097$ &  $6189\pm183$ &   $-0.09\pm0.50$  & this work  \\  
 Kepler-307 & 1576  &  0.907$\pm 0.034$&  $0.814\pm0.024$ &  $5367\pm94$ &   $0.19\pm0.10$  & this work   \\ 
%  \textit{ Kepler-326 } & \textit{ 1835 } &  0.84$^{+0.37}_{-0.26}$&  $0.801\pm0.052$ &  $5105\pm100$ &   $0.05\pm0.10$  &\citet{rowe14}   \\ 
        \hline
    \end{tabular}
    \caption{Adopted stellar parameters. In most cases, we have improved constraints on stellar properties with our light curve analysis following TTV analysis. In the case of Kepler-49, we identified an anomalous transit profile which may indicate spot-crossing events. Hence, we adopt the stellar parameters of \citet{swift15}. For Kepler-57, the stellar density from light curve analysis is inconsistent with spectroscopic observations of the star, and we adopt the spectroscopic parameters listed on \textit{Kepler's} Community Follow-up Program (\textit{CFOP}), as explained in the text.}\label{tbl-stars}
  \end{center}
\end{table}

 \begin{table}[h!]
  \begin{center}
    \begin{tabular}{|c|c|c|c|c|c|}
 \hline 
 KOI  \& Kepler \#  \hspace{0.7 in}  & $P$ (days) &   $M_{p}$  ($M_{\oplus}$)  \hspace{0.2 in} &      $R_{p}$ ($R_{\oplus}$)   \hspace{0.2 in} &  $\rho_{p}$ (g cm$^{-3}$) & Flux ($F_{\oplus}$) \hspace{0.1 in}    \\   
 \hline 
 \end{tabular}
  \begin{tabular}{c}
\hspace{1.7 in}  \textbf{Precise Exoplanet Parameters that are insensitive to assumptions} \hspace{1.7 in} \\

\end{tabular}
  \begin{tabular}{|c|c|c|c|c|c|}
  \hline
%KOI-244.02  \hspace{0.1 in}   \textbf{Kepler-25 b}   &  6.23820 & \textbf{  1.824 }$^{+  3.150 }_{ -  1.161 }$ &   \textbf{   2.71 } $ \pm    0.05   $    &   \textbf{  0.505 }$^{+  0.881 }_{ -  0.360 }$   &   \textbf{ 481.40 }$^{+  16.80 }_{ -  15.20 }$   \\ 
%KOI-244.01 \hspace{0.1 in} \textbf{Kepler-25 c} &   12.7206 & \textbf{  6.877 }$^{+  5.260 }_{ -  3.787 }$ &   \textbf{   5.20 } $ \pm    0.09   $    &   \textbf{  0.270 }$^{+  0.206 }_{ -  0.150 }$   &   \textbf{ 186.20 }$^{+   6.40 }_{ -   6.00 }$   \\ 
%\hline
KOI-250.01 \hspace{0.1 in}  \textbf{Kepler-26 b} & 12.2796  & \textbf{  5.12 }$^{+  0.65 }_{ -  0.61 }$ &   \textbf{   2.78 } $ \pm    0.11   $    &   \textbf{  1.26 }$^{+  0.21 }_{ -  0.19 }$   &   \textbf{   7.60 }$\pm 1.1$   \\ 

KOI-250.02\hspace{0.1 in}  \textbf{Kepler-26 c}  & 17.2559  & \textbf{  6.20 }$^{+  0.65 }_{ -  0.65 }$ &   \textbf{   2.72 } $ \pm    0.12   $    &   \textbf{  1.61 }$^{+  0.27 }_{ -  0.22 }$   &   \textbf{   4.82 }$\pm 0.72$   \\ 

\hline
KOI-738.01  \hspace{0.1 in} \textbf{Kepler-29 b}  & 10.3393 & \textbf{  4.51 }$^{+  1.41 }_{ -  1.47 }$ &   \textbf{   3.35 } $ \pm    0.22   $    &   \textbf{  0.65 }$^{+  0.27 }_{ -  0.23 }$   &   \textbf{  96.0 }$\pm 15.0$   \\
KOI-738.02 \hspace{0.1 in} \textbf{Kepler-29 c}  & 13.2869 & \textbf{  4.00 }$^{+  1.23 }_{ -  1.29 }$ &   \textbf{   3.14 } $ \pm    0.20   $    &   \textbf{  0.70 }$^{+  0.29 }_{ -  0.25 }$   &   \textbf{  69.0 }$\pm 10.5$   \\

\hline
KOI-2086.01 \hspace{0.1 in}  \textbf{Kepler-60 b}  & 7.1334 & \textbf{  4.19 }$^{+  0.56 }_{ -  0.52 }$ &   \textbf{   1.71 } $ \pm    0.13   $    &   \textbf{  4.62 }$^{+  1.40 }_{ -  1.10 }$   &   \textbf{ 318 }$\pm 52$   \\ 

KOI-2086.02 \hspace{0.1 in}  \textbf{Kepler-60 c} & 8.9187 & \textbf{  3.85 }$^{+  0.81 }_{ -  0.81 }$ &   \textbf{   1.90 } $ \pm    0.15   $    &   \textbf{  3.06 }$^{+  1.14 }_{ -  0.86 }$   &   \textbf{ 236 }$\pm 39$   \\ 
 
KOI-2086.03 \hspace{0.1 in}  \textbf{Kepler-60 d}  & 11.8981 & \textbf{  4.16 }$^{+  0.84 }_{ -  0.75 }$ &   \textbf{   1.99 } $ \pm    0.16   $    &   \textbf{  2.91 }$^{+  1.03 }_{ -  0.78 }$   &   \textbf{ 161 }$\pm 27$   \\

\hline
KOI-115.02 \hspace{0.1 in} \textbf{Kepler-105 c}  & 7.1262 & \textbf{  4.60 }$^{+  0.92 }_{ -  0.85 }$ &   \textbf{   1.31 } $ \pm    0.07   $    &   \textbf{ 11.20 }$^{+  3.00 }_{ -  2.56 }$   &   \textbf{ 161 }$\pm19$   \\ 
\hline

KOI-1576.01 \hspace{0.1 in}  \textbf{Kepler-307 b} &   10.4208  & \textbf{  7.44 }$^{+  0.91 }_{ -  0.87 }$ &   \textbf{   2.43 } $ \pm    0.09   $    &   \textbf{  2.62 }$^{+  0.38 }_{ -  0.34 }$   &   \textbf{ 59.7 }$\pm 5.0$   \\ 

KOI-1576.02 \hspace{0.1 in}  \textbf{Kepler-307 c} &  13.0729  & \textbf{  3.64 }$^{+  0.65 }_{ -  0.58 }$ &   \textbf{   2.20 } $ \pm    0.07   $    &   \textbf{  1.74 }$^{+  0.30 }_{ -  0.30 }$   &   \textbf{  44.0 }$\pm 3.7$   \\ 
\hline
\end{tabular}
  \begin{tabular}{c}
 \textbf{Less secure TTV solutions} \\
\end{tabular}
  \begin{tabular}{|c|c|c|c|c|c|}
  \hline
KOI-248.01 \hspace{0.1 in}  \textbf{Kepler-49 b} &   7.2040 & \textbf{  5.090 }$^{+  2.112 }_{ -  1.936 }$   & \textbf{   2.35 } $ \pm    0.09   $    &   \textbf{  2.048 }$^{+  0.885 }_{ -  0.799 }$   &   \textbf{  41.3}$\pm 2.0$   \\  
KOI-248.02  \hspace{0.1 in} \textbf{Kepler-49 c}  &   10.9123  &  \textbf{  3.280 }$^{+  1.452 }_{ -  1.320 }$ &  \textbf{   2.06 } $ \pm    0.09   $    &   \textbf{  1.932 }$^{+  0.928 }_{ -  0.765 }$   &   \textbf{  23.8 }$\pm1.2$   \\ 
\hline
% using our rhostar solution
%KOI-1270.01 \hspace{0.1 in}  \textbf{Kepler-57 b} &   5.72947  & \textbf{ 16.855 }$^{+  7.239 }_{ -  6.949 }$ &   \textbf{   1.88 } $ \pm    0.14   $    &   \textbf{ 11.786 }$^{+  5.446 }_{ -  4.883 }$   &   \textbf{  73 }$\pm 12$   \\ 
%KOI-1270.02 \hspace{0.1 in}  \textbf{Kepler-57 c} &  11.60653  & \textbf{  4.123 }$^{+  1.969 }_{ -  1.890 }$ &   \textbf{   1.39 } $ \pm    0.10   $    &   \textbf{  7.033 }$^{+  3.644 }_{ -  3.189 }$   &   \textbf{  28.4 }$\pm 4.7$   \\ 
%\hline
% using CFOP stellar params
KOI-1270.01 \hspace{0.1 in}  \textbf{Kepler-57 b} &  5.7295  & \textbf{ 23.13 }$^{+  9.76 }_{ -  7.64 }$ &   \textbf{   1.88 } $ \pm    0.14   $    &   \textbf{ 16.68 }$^{+  8.83 }_{ -  6.97 }$   &   \textbf{ 137 }$\pm 24$   \\ 
 KOI-1270.02 \hspace{0.1 in}  \textbf{Kepler-57 c} &11.6065  & \textbf{  5.68 }$^{+  2.55 }_{ -  1.96 }$ &   \textbf{   1.39 } $ \pm    0.10   $    &   \textbf{  9.74 }$^{+  5.75 }_{ -  4.31 }$   &   \textbf{  53 }$\pm12$   \\
\hline
KOI-115.03 \hspace{1.0 in} & 3.4363 & \textbf{  1.21 }$^{+  1.28 }_{ -  0.68 }$ &   \textbf{   0.73 } $ \pm    0.04   $    &   \textbf{ 17.11 }$^{+ 16.15 }_{ -  9.48 }$   &   \textbf{ 427 }$\pm 51$   \\ 
KOI-115.01 \hspace{0.1 in}   \textbf{Kepler-105 b}  &  5.4119 & \textbf{  3.88 }$^{+  1.92 }_{ -  1.85 }$ &   \textbf{   2.22 } $ \pm    0.11   $    &   \textbf{  1.94 }$^{+  0.96 }_{ -  0.94 }$   &   \textbf{ 233}$\pm 28$   \\ 
\hline
KOI-523.02 \hspace{0.1 in}  \textbf{Kepler-177 b} & 36.8590  & \textbf{  7.24 }$^{+  1.26 }_{ -  1.16 }$ &   \textbf{   4.04 } $ \pm    0.29   $    &   \textbf{  0.51 }$^{+  0.11 }_{ -  0.10 }$   &   \textbf{  47.2 }$\pm 8.0$   \\ 
KOI-523.01 \hspace{0.1 in}  \textbf{Kepler-177 c} & 49.4096  & \textbf{ 18.36 }$^{+  3.93 }_{ -  3.48 }$ &   \textbf{   9.77 } $ \pm    0.68   $    &   \textbf{  0.09 }$^{+  0.02 }_{ -  0.02 }$   &   \textbf{  31.9 }$\pm 5.4$   \\ 

\hline

  %\textit{ Kepler-57 b} & & & & \\ 
     %   \hline
    \end{tabular}
    \caption{Inferred planetary parameters. The top panel lists the precise results of this study that are secure against outliers, insensitive to our priors in eccentricity and consistent between independently measured transit times. The lower panel lists results which are less secure or are flagged for any reason. These include disagreement between independently measured transit times for Kepler-177 and Kepler-57. Additionally, the size of the planets orbiting Kepler-57 may be underestimated, causing the measured planetary densities to be too high (see Section 5).  Kepler-49 is deemed 'less secure' because of an anomalous transit profile for one of the planets of the TTV analysis. For these systems, accurate planetary parameters await further analysis or future data. The inner two planets of Kepler-105 (KOI-115.03 and Kepler-105 b) have upper limits only inferred by the TTVs, whereas we infer useful upper and lower limits on the mass of Kepler-105 c.}\label{tbl-results}
  \end{center}
\end{table}

\section{Discussion}
\begin{figure}[h!]
%\hspace{0.5 in}
%\includegraphics [height = 3.3 in]{TenSystems-M-R-limitedrange.eps}
\includegraphics [height = 3.3 in]{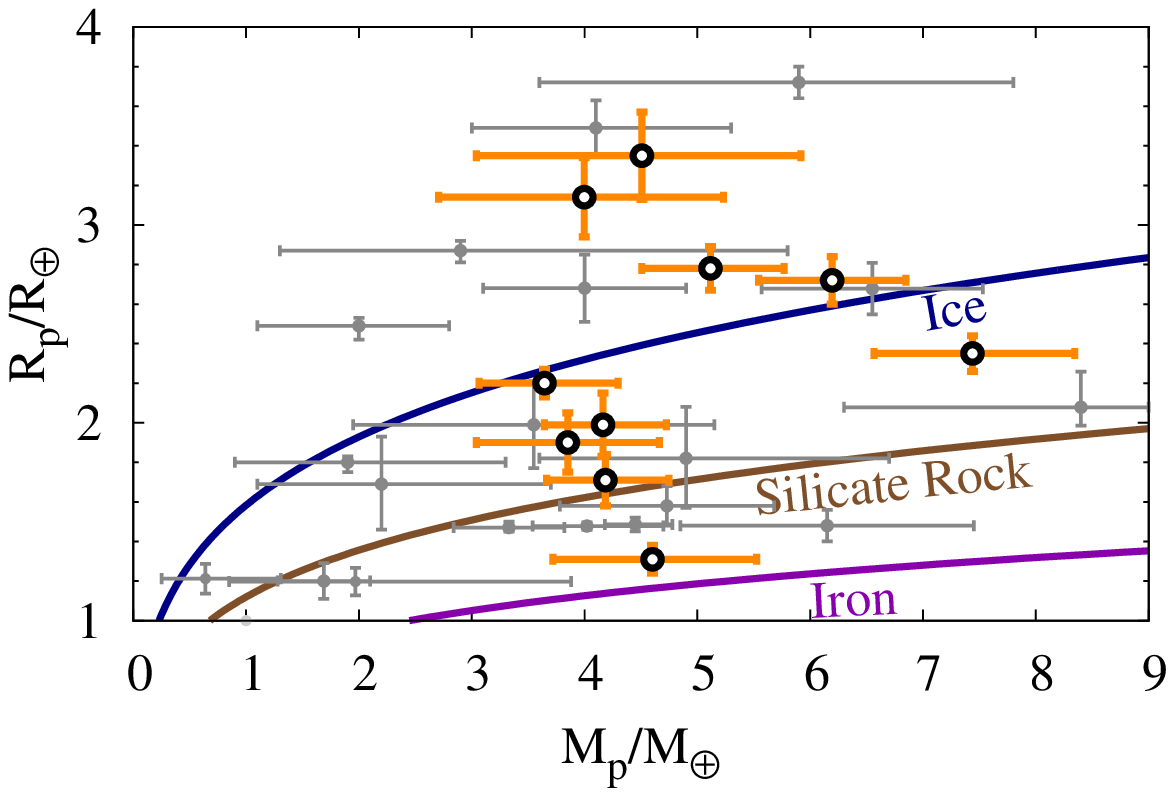}
\caption{Mass-Radius diagram of exoplanets of mass $M_{p} < 9$ $M_{\oplus}$ and from 1 to 4 $R_{\oplus}$ in size, compared to theoretical curves of pure water ice, silicate rock and iron \citep{for07}. Previously characterized exoplanets in this range are shown in grey; the ten that we add have open circles and orange error bars. Despite a small range in mass, they span a wide range in density and hence bulk composition.}
\label{fig:MRlimitedrange} 
\end{figure}

The ten well-characterized planets in this study have a narrow range of masses, from $\sim$3--8 $M_{\oplus}$. Their sizes range from 1.31 $R_{\oplus}$ to 3.35 $R_{\oplus}$, and hence they span over an order of magnitude in density. Our results substantially raise the number of exoplanets in this size and mass range on the mass-radius diagram, as shown in Figure~\ref{fig:MRlimitedrange}. The stellar hosts of our well-characterized sample range in mass from 0.54 $M_{\odot}$ (Kepler-26) to 1.04 $M_{\odot}$ (Kepler-60), representing a significant variety of stellar properties. Figure~\ref{fig:MRlimitedrange} highlights the well-characterized exoplanets that trace the transition from rocky planets to those that retain deep atmospheres in a limited mass range. The densest planets we have characterized include Kepler-105 c, which has a density consistent with an Earth-like rocky composition. It joins other super-earth-size planets likely denser than pure silicate rock including CoRoT-7 b, HD219134 b, Kepler-10 b, Kepler-20 b, Kepler-36 b, Kepler-78 b, Kepler-89 b, Kepler-93 b,  Kepler-99 b, Kepler-138 c, Kepler-406 b, and WASP-47 c (see Appendix A for citations). 
 
A larger sample of well-characterized low-mass planets in this study and in others have densities consistent with mixtures of either rock and ice, or rock and gas. In this study we add three planets orbiting Kepler-60 and two planets orbiting Kepler-307. The remainder of our sample, however, are all less dense than water ice given their masses and likely retain deep atmospheres. These include the four well-characterized planets of Kepler-26 and Kepler-29. Upper limits on the mass of Kepler-105 b imply a density less than that of rock, but consistent with a mixture of either rock and water or rock and gases. 

Although we have not included several planets that did not pass all of our tests for robustness in Figure~\ref{fig:MRlimitedrange}, we note that Kepler-177 c has very strong mass upper limits that indicate an extreme low bulk density, and a remarkably thick atmosphere given its low mass.

\begin{figure}[h!]
\vspace{0.5 in}
\includegraphics [height = 2.3 in]{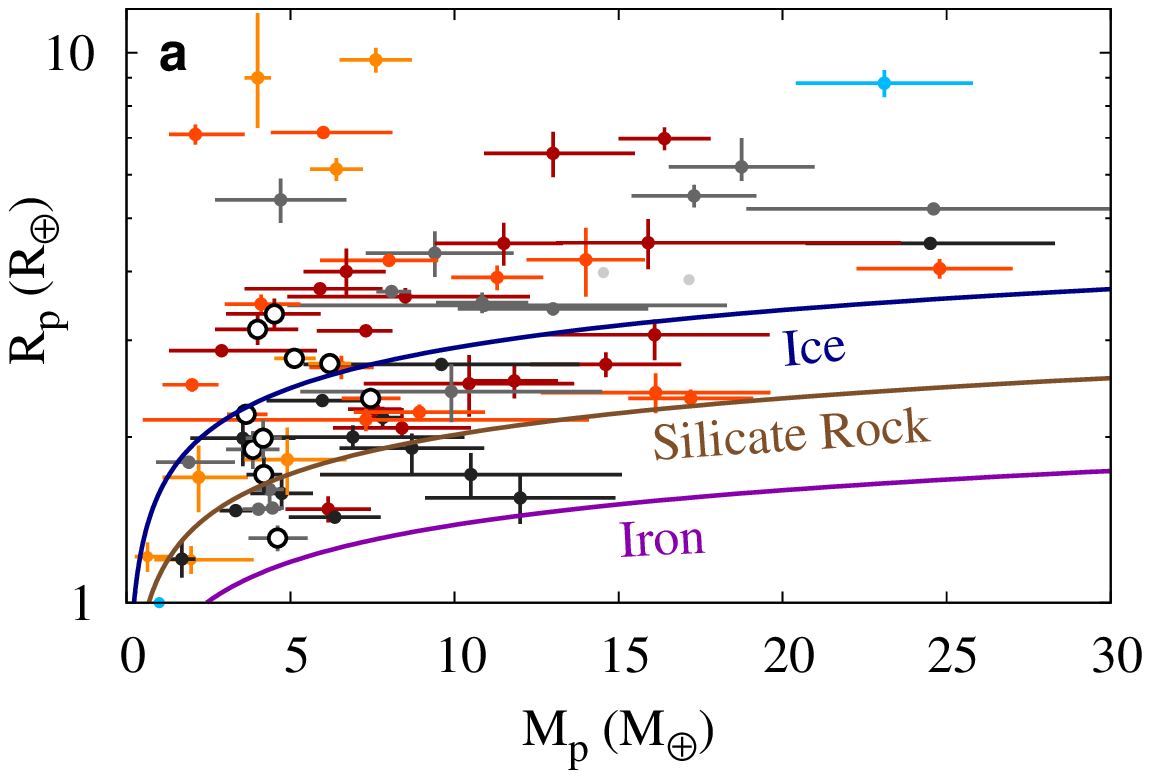}
\includegraphics [height = 2.3 in]{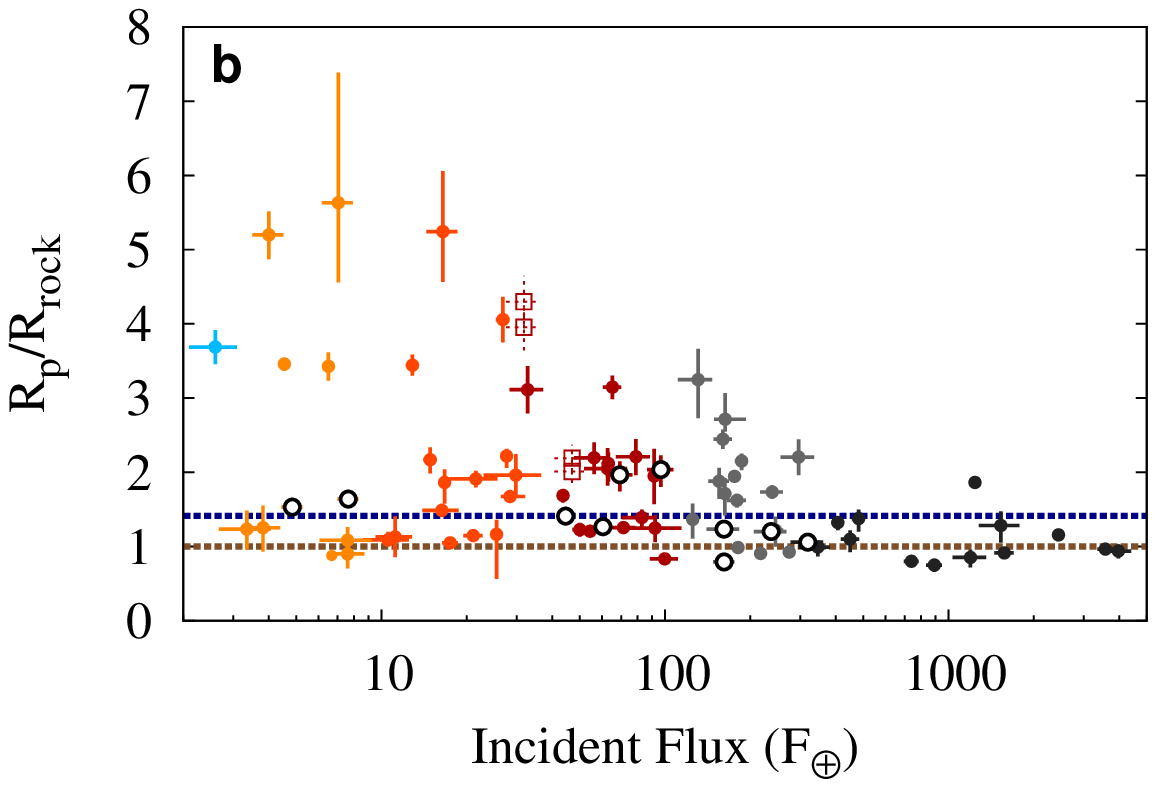}
\caption{Mass-Radius diagram for planets less than 30 $M_{\oplus}$ and larger than 1 $R_{\oplus}$ (a) and planetary sizes relative to pure silicate rock as a function of incident flux (b). In the mass-radius diagram, the curves mark \citet{for07} solutions for planets made of pure iron (purple), silicate rock (brown) or water ice (navy blue). In the right panel, planet sizes are compared to pure rock. Below the brown line, planets are denser than rock and are unlikely to retain appreciable amounts of volatiles. Above the navy blue line, planets are less dense than water ice and must retain deep atmospheres. The colors correspond to the amount of incident flux compared to Earth ($F_{\oplus}$), with exoplanets $<$3 $F_{\oplus}$ in light blue, planets $>$300$F_{\oplus}$ in black, and orange, red, maroon and dark grey respectively in-between, as shown in the right panel. Well outside this regime in flux, Neptune and Uranus are marked as light grey points on the mass-radius diagram. Incident fluxes were compared to Earth assuming low eccentricities ($e^2 << 1$) and using the following equation:
$F (F_{\oplus}) = 4.62 \times 10^{4} \left(\frac{T_{eff}}{T_{\odot}}\right)^4   \left( \frac{R_{\star}}{a} \right)^2$ with uncertainties added in quadrature. Stellar parameters were taken from the references in Table~\ref{tbl-mrdiagram}. The open circles mark the secure planet characterizations of this study. The open squares mark our two possible solutions for Kepler-177 b, using independently measured transit times. The range in observed planet sizes given their measured mass appears strongly anti-correlated with incident flux.}
\label{fig:MR} 
\end{figure}

To place these planet characterizations in context, we plot in Figure~\ref{fig:MR} the mass-radius-flux diagram for planets less than 25 $M_{\oplus}$. Apart from a few extremely low density low-mass planets in the upper left of the mass-radius diagram, the majority of well-characterized planets below 8 $M_{\oplus}$ are smaller than Neptune. The ten planets that we add in this study span a wide range in radii from super-Earth size to Neptune size. The wide scatter in Figure~\ref{fig:MR}a shows a lack of rocky planets more massive than 10 $M_{\oplus}$, consistent with \citet{rog15}.  All the planets more massive than Neptune in Figure~\ref{fig:MR}a are less dense than water ice and must retain deep atmospheres. Both of these constraints on the mass-radius relation are based on the upper limits of well-characterized masses and hence are unlikely to be affected by detection biases. I.e., if there were rocky planets more massive than Neptune, detection biases would have made their detection more likely than low density planets of the same size. Nevertheless, the wide range in planetary radii among planets less massive than Neptune implies that a simple mass-radius relation will not provide a good fit to the dataset of low-mass exoplanets \citep{wolf15}.

Figure~\ref{fig:MR}b compares planet radii normalized to planets made of pure rock, as a function of incident flux. Planets that are larger than a rocky world of the same mass (with uncertainties in mass and radius added in quadrature) are volatile rich, and must retain significant amounts of ices or gases to explain their volumes. However, planets that are larger and less dense than worlds made of pure water ice likely retain deep atmospheres of H/He. %

Despite the uncertainties on the mass of the planets of Kepler-177, the strong upper limits on the mass imply very low densities for both planets. Kepler-177 c joins a class of extreme low density super-Earth mass planets including Kepler-51 b, c and d; Kepler-79 d, Kepler-87 c and Kepler-33 d. Kepler-177 c however, stands out as being the most massive of this group, and is the largest known planet in this mass range.

The upper bounds on size relative to a planet made of pure rock at any given mass show a clear trend as a function of incident flux, as shown in Figure~\ref{fig:MR}b. Below 20 $F_{\oplus}$ (where $F_{\oplus}$ is the flux received by Earth), bulk densities range from 6.3 g cm$^{-3}$ (rocky) to less than 0.05 g cm$^{-3}$. Low-mass planets that have incident fluxes ($F$) greater than $\sim300$ times that of Earth are predominantly rocky-- in this mass range only Kepler-4 b receives more than 300 $F_{\oplus}$ and must retain a deep atmosphere, although this case is complicated since the host has likely evolved off the main sequence, increasing the incident flux \citep{silv15}. Hence, the range of planet densities seen in low-mass exoplanets appears strongly anti-correlated with incident flux \citep{jont14}. In Figure~\ref{fig:MR}b we include our posteriors for Kepler-177 b and c given both sets of independently measured transit times for this system. The larger of the two, Kepler-177 c, has a similar low density and incident flux to Kepler-79 d, and traces the upper bound on the range of sizes relative to planets made of pure rock around 30 $F_{\oplus}$. 

Atmospheric mass-loss may contribute to the trend that low-mass planets with higher incident flux have a narrow range of densities as shown in Figure~\ref{fig:MR}b. \citet{owe13} found that an increase in the range of detected transiting planet radii in the \textit{Kepler} sample was consistent with the evaporation of planetary atmospheres over time at high incident flux. Most of the mass loss is assumed to occur in the first 100 Myr while the star is young and bright in X-ray and EUV (\citealt{ribas05,jackson12,lop13a}). During this time, the intense high energy flux heats the outer planetary atmosphere causing light elements to escape and drag heavy elements to escape.

Explaining the wide range in densities among low-mass planets remains an active problem for atmospheric mass loss models. \citet{lop13a} found that the mass of the planet's core is an additional factor in the lifetime of a planetary envelope. They invoked a higher core mass for Kepler-36 c to slow its mass loss enough to leave a large difference in bulk density with its inner neighbor. Other possible causes of atmospheric mass loss include giant impacts, invoked by \citet{liu15} to explain the large density ratio between Kepler-36 b and c. In their model, the stochastic nature of giant impacts causes the wide range in density of low-mass planets. 

It is unclear whether or not the extremely low density exoplanets are still undergoing significant mass loss. The lowest density planets, all orbiting Kepler-51, are likely very young ($\sim$300 Myr), and may still be cooling or losing mass \citep{mas13}. Kepler-79 d on the other hand, likely has a much deeper atmosphere than its two inner neighbors and is $\sim$3.4 Gyr in age. Hence, cooling time alone cannot explain the different densities at Kepler-79. Similarly, Kepler-177 is 1--4 Gyr in age, and the inner planet is significantly denser than the extreme low-density planet Kepler-177 c. 

\section{Conclusion}
We have inferred dynamical mass posteriors on 18 planets in \textit{Kepler}'s multiplanet systems that show strong evidence of inducing TTVs. Our dynamical masses confirm the accuracy of analytical expressions for TTV inversion. We find moderate evidence of apsidal alignment over a wide range of possible eccentricities for the planets orbiting Kepler-26, Kepler-49, Kepler-29, Kepler-177 and Kepler-307. All but two of our sample systems (Kepler-49 and Kepler-57) have benefitted from revised stellar parameters following our transit light curve fitting and dynamical models. Following tests for robustness and stability we add ten well-characterized exoplanets below 8 $M_{\oplus}$ to the planetary mass-radius-flux diagram (orbiting Kepler-26, Kepler-29, Kepler-60, Kepler-307 and a likely rocky planet orbiting Kepler-105). Additionally we find strong upper limits on mass on two planets orbiting Kepler-177, and therefore an extreme low density for Kepler-177 c, the largest characterized planet below 25 $M_{\oplus}$. Finally, we find that characterized low-mass exoplanets show a clear decrease in the range of observed bulk densities with increasing incident flux.

We wish to thank our anonymous referee for a helpful review that improved this paper. DJ, EF  and JL acknowledge support from NASA Exoplanets Research Program award \#NNX15AE21G. The results reported herein benefitted from collaborations and/or information exchange within NASA's Nexus for Exoplanet System Science (NExSS) research coordination network sponsored by NASA's Science Mission Directorate. DJ and EF gratefully acknowledge that this work was partially supported by funding from the Center for Exoplanets and Habitable Worlds.  The Center for Exoplanets and Habitable Worlds is supported by the Pennsylvania State University, the Eberly College of Science, and the Pennsylvania Space Grant Consortium. EA acknowledges support from NASA grants NNX13AF20G, NNX13AF62G, and NASA Astrobiology Institutes Virtual Planetary Laboratory, supported by NASA under cooperative agreement NNH05ZDA001C. KD acknowledges support from the Joint Center for Planetary Astronomy fellowship at Caltech.

\section*{Appendix A: Planetary Mass and Radius Data}
Table note: Tables~\ref{tbl-mrdiagram} and~\ref{tbl-mrdiagram2} list the planetary masses and radii used in Figure~\ref{fig:MR}. Note that the \citet{mar14} sample includes many non-detections and upper limits, which we do not plot. %For any multi-planet model in \citet{mar14} that included negative nominal masses, we list only the 3$\sigma$ detections in the relevant mass range. Otherwise, we list all 2$\sigma$ detections. 

%Incident fluxes plotted in Figure~\ref{fig:MR} were compared to Earth assuming low eccentricities ($e^2 << 1$) and using the following equation:
%\begin{equation}
%F (F_{\oplus}) = 4.62 \times 10^{4} \left(\frac{T_{eff}}{T_{\odot}}\right)^4   \left( \frac{R_{\star}}{a} \right)^2,
%\label{eqn:flux}
%\end{equation}
%with uncertainties added in quadrature. Stellar parameters were taken from the references in Table~\ref{tbl-mrdiagram}. 

 \begin{table}[h!]
 %\tiny
  \begin{center}
    \begin{tabular}{|c|c|c|c|r|}
 \hline 
 Name  & Mass ($M_{\oplus}$) &      Radius ($R_{\oplus}$) &     References    \\   
 \hline  
55 Cnc e  & 7.81$^{+0.58}_{-0.53}$ &     2.17$\pm 0.10$  &   \citet{win11}, \citet{gill12}   \\   

EPIC204221263 b  &  12.0$\pm 2.9$ &     1.55$\pm 0.16$  &   \citet{Sinukoff2015}   \\   
EPIC204221263 c  & 9.9$\pm 4.6$ &     2.42$\pm 0.29$  &   \citet{Sinukoff2015}   \\   

GJ1214 b  & 6.55$\pm 0.98$ &     2.68$\pm 0.13$  &   \citet{cha09}   \\   
GJ3470 b  & 14.00$\pm 1.80$ &     4.20$\pm 0.60$  &   \citet{bon12}   \\   
GJ436 b  & 25.79$^{+2.22}_{-2.54}$ &     4.05$\pm 0.17$  &   \citet{von12}   \\   
CoRoT-7 b  & 4.73$\pm 0.95$ &     1.58$\pm 0.10$  &   \citet{hay14}   \\   
%HAT-P-11 b  & 26.70$\pm 2.22$ &     4.36$\pm 0.10$  &   \citet{sou11}   \\   
HAT-P-26 b  & 18.75$\pm 2.22$ &     6.20$^{+0.79}_{-0.35}$  &   \citet{har11}   \\   
HD97658 b  & 7.55$^{+0.83}_{-0.79}$ &     2.25$\pm 0.10$  &   \citet{vangrootel2014}   \\   
HD219134 b  & 4.46$\pm 0.47$ &     1.61$\pm 0.09$  &   \citet{motalebi2015}   \\   

HIP116454 b  & 11.82$\pm 1.33$ &     2.53$\pm 0.18$  &   \citet{vanderburg2015}   \\   

Kepler-4 b  & 24.5$\pm 3.8$ &     4.50$\pm 0.12$  &   \citet{bor10} \citet{silv15}   \\   

Kepler-10 b  &  3.33$\pm$0.49 &     1.47$^{+0.03}_{-0.02}$  &   \citet{bat11}   \\   
Kepler-10 c  &  17.20$\pm1.90$ &     2.35$^{+0.09}_{-0.04}$  &   \citet{dum14}   \\   

Kepler-11 b  &1.9$^{+1.4}_{-1.0}$ &     1.80$^{+0.03}_{-0.05}$  &   \citet{liss13}   \\   
Kepler-11 c  & 2.9$^{+2.9}_{-1.6}$ &     2.87$^{+0.05}_{-0.06}$  &   \citet{liss13}   \\   
Kepler-11 d  &7.3$^{+0.8}_{-1.5}$ &     3.12$^{+0.06}_{-0.07}$  &   \citet{liss13}   \\   
Kepler-11 e  &8.0$^{+1.5}_{-2.1}$ &     4.19$^{+0.07}_{-0.09}$  &   \citet{liss13}   \\   
Kepler-11 f  &2.0$^{+0.8}_{-0.9}$ &     2.49$^{+0.04}_{-0.07}$  &   \citet{liss13}   \\   
Kepler-18 b  &6.9$\pm 3.4$ &     2.0$\pm 0.1$  &   \citet{coch11}   \\   
Kepler-18 c  & 17.3$\pm 3.4$ &     5.49$\pm 0.26$  &   \citet{coch11}   \\   
Kepler-18 d  & 16.40$\pm 1.4$ &     6.98$\pm 0.33$  &   \citet{coch11}   \\   
Kepler-20 b  & 8.7$\pm 2.2$ &     1.91$^{+0.12}_{-0.21}$  &   \citet{fres12}, \citet{gau12}   \\   
Kepler-20 c  & 16.1$\pm 3.5$ &     3.07$^{+0.20}_{-0.31}$  &   \citet{fres12}, \citet{gau12}   \\   

Kepler-25 b  & 9.6$\pm 4.2$ &     2.71$\pm 0.05$  &   \citet{mar14}   \\   
Kepler-25 c  & 24.6$\pm 5.7$ &     5.2$\pm 0.09$  &   \citet{mar14}   \\   

Kepler-26 b  & 5.1$\pm 0.7$ &     2.78$\pm 0.11$  &   This work   \\   
Kepler-26 c  & 6.2$\pm 0.7$ &     2.72$\pm 0.12$  &   This work   \\   

Kepler-29 b  & 4.5$\pm 1.5$ &     3.35$\pm 0.22$  &   This work   \\   
Kepler-29 c  & 4.0$\pm  1.3$ &     3.14$\pm 0.20$  &   This work   \\

Kepler-30 b  & 11.3$\pm 1.4$ &     3.9$\pm 0.2$  &   \citet{san12}    \\   
Kepler-30 d  & 23.1$\pm 2.7$ &     8.8$\pm 0.5$  &   \citet{san12}    \\   

%Kepler-33c  & 0.8$^{+2.5}_{-0.7}$ &     3.2$\pm 0.3$  &   \citet{had15} \citet{liss12}   \\   
Kepler-33 d  & 4.7$\pm 2.0$ &     5.4$\pm 0.5$  &   \citet{had15}, \citet{liss12}   \\   
Kepler-33 e  & 6.7$^{+1.2}_{-1.3}$ &     4.0$\pm 0.4$  &   \citet{had15}, \citet{liss12}   \\   
Kepler-33 f  & 11.5$^{+1.8}_{-2.1}$ &     4.5$\pm 0.4$  &   \citet{had15}, \citet{liss12}   \\   

Kepler-36 b  & 4.45$^{+0.33}_{-0.27}$ &     1.49$^{+0.03}_{-0.04}$  &   \citet{car12}    \\   
Kepler-36 c  & 8.08$^{+0.60}_{-0.46}$ &     3.68$^{+0.05}_{-0.05}$  &   \citet{car12}    \\

Kepler-48 c  & 14.61$\pm2.30$ &     2.71$\pm0.14$  &   \citet{mar14}    \\

 \hline
    \end{tabular}
    \caption{Mass-radius data for planets less than 25 M$_{\oplus}$ (Part 1 of 2).  The last column lists references for both planetary and stellar parameters used in Figure~\ref{fig:MR}.}\label{tbl-mrdiagram}
  \end{center}
\end{table}

 \begin{table}[h!]
% \tiny
  \begin{center}
    \begin{tabular}{|c|c|c|c|r|}
 \hline 
 Name  & Mass ($M_{\oplus}$) &      Radius ($R_{\oplus}$) &      References    \\   
 \hline

Kepler-51 b  & 2.1$^{+1.5}_{-0.8}$ &     7.1$\pm 0.3$  &   \citet{mas14}    \\   
Kepler-51 c  & 4.0$\pm 0.4$ &     9.0$^{+2.8}_{-1.7}$  &   \citet{mas14}    \\   
Kepler-51 d  & 7.6$\pm 1.1$ &     9.7$\pm 0.5$  &   \citet{mas14}    \\   

Kepler-60 b  & 4.2$\pm 0.6$ &     1.71$\pm0.13$  &  This work    \\   
Kepler-60 c  & 3.9$\pm 0.8$ &     1.90$\pm0.15$  &  This work    \\   
Kepler-60 d  & 4.2$\pm 0.8$ &     1.99$\pm0.16$  &  This work    \\

Kepler-68 b  & 5.97$\pm 1.70$ &     2.33$\pm0.02$  &   \citet{mar14}    \\   
Kepler-78 b  & 1.69$\pm 0.41$ &     1.20$\pm0.09$  &   \citet{pepe13}, \citet{how13}    \\   

Kepler-79 b  & 10.9$^{+7.4}_{-6.0}$ &     3.47$\pm0.07$  &  \citet{jont14}    \\   
Kepler-79 c  & 5.9$^{+1.9}_{-2.3}$ &     3.72$\pm0.08$  &  \citet{jont14}    \\   
Kepler-79 d  & 6.0$^{+2.1}_{-1.6}$ &     7.16$^{+0.13}_{-0.16}$  &  \citet{jont14}    \\   
Kepler-79 e  & 4.1$^{+1.2}_{-1.1}$ &     3.49$\pm0.14$  &  \citet{jont14}    \\   

Kepler-87 c  & 6.4$\pm 0.8$ &     6.14$\pm0.29$  &   \citet{ofir14}    \\   

Kepler-89 b  & 10.5$\pm 4.6$ &     1.71$\pm0.16$  &   \citet{weis13}    \\   
Kepler-89 c  & 9.40$^{+2.4}_{-2.1}$ &     4.32$\pm0.41$  &   \citet{weis13}, \citet{mas13}    \\   
Kepler-89 e  & 13.0$^{+2.5}_{-2.1}$ &     6.56$\pm0.62$  &   \citet{weis13}, \citet{mas13}   \\   

Kepler-93 b  & 4.02$^{+0.68}_{-0.48}$ &     1.48$\pm0.02$  &   \citet{dres15}   \\   

Kepler-94 b  & 10.84$\pm1.40$ &     3.51$\pm0.15$  &   \citet{mar14}   \\   
Kepler-95 b  & 13.0$\pm2.9$ &     3.42$\pm0.09$  &   \citet{mar14}   \\   
Kepler-98 b  & 3.55$\pm1.60$ &     1.99$\pm0.22$  &   \citet{mar14}   \\   

Kepler-99 b  & 6.15$\pm1.30$ &     1.48$\pm0.08$  &   \citet{mar14}   \\   

Kepler-102 e  & 8.93$\pm2.00$ &     2.22$\pm0.07$  &   \citet{mar14}   \\   
Kepler-105 c  & 4.60$\pm0.9$ &     1.31$\pm0.07$  &   This work  \\   

Kepler-106 c  & 10.44$\pm3.20$ &     2.50$\pm0.32$  &   \citet{mar14}   \\   

Kepler-131 b  & 16.13$\pm3.50$ &     2.41$\pm0.20$  &   \citet{mar14}   \\   

Kepler-138 b  & 0.07$^{+0.06}_{-0.04}$ &     0.52$\pm0.03$  &   \citet{jont15}   \\   
Kepler-138 c  & 1.97$^{+1.91}_{-1.12}$ &     1.20$\pm0.07$  &   \citet{jont15}   \\   
Kepler-138 d  & 0.64$^{+0.67}_{-0.39}$ &     1.21$\pm0.07$  &   \citet{jont15}   \\   
Kepler-231 b  & 4.9$^{+1.8}_{-1.7}$ &     1.82$^{+0.26}_{-0.25}$  &   \citet{kip14}   \\   
Kepler-231 c  & 2.2$^{+1.5}_{-1.1}$ &     1.69$^{+0.24}_{-0.23}$  &   \citet{kip14}   \\   
Kepler-289 b  & 7.3$\pm 6.8$ &     2.15$\pm 0.10$  &   \citet{sch14}   \\   
Kepler-289 c  & 4.0$\pm 0.9$ &     2.68$\pm 0.17$  &   \citet{sch14}   \\ 

Kepler-307 b  & 7.4$\pm 0.9$ &     2.43$\pm0.09$  &  This work    \\   
Kepler-307 c  & 3.6$\pm 0.7$ &     2.20$\pm0.07$  &  This work    \\   

Kepler-406 b  & 6.35$\pm 1.40$ &     1.43$\pm 0.03$  &   \citet{mar14}   \\ 

K2-3 b & 8.4$\pm 2.1$ &     2.08$^{+0.18}_{-0.09}$  &   \citet{almenara2015}   \\

%K2-3 c & 2.1$^{+2.1}_{-1.3}$ &     1.66$^{+0.16}_{-0.07}$  &   \citet{almenara2015}   \\ 
%K2-3 d & 11.1$\pm 3.5$ &     1.53$\pm 0.11$  &   \citet{almenara2015}   \\ 
K2-19 c  & 15.9$^{+7.7}_{-2.8}$ &     4.51$\pm 0.47$  &   \citet{Barros2015}   \\ 

WASP-47 d & $8.5^{+3.8}_{-3.6}$ &     3.6$\pm 0.13$  &   \citet{beck15}   \\ 

 \hline
    \end{tabular}
    \caption{Mass-radius data for planets less than 25 M$_{\oplus}$ (Part 2 of 2).  The last column lists references for both planetary and stellar parameters used in Figure~\ref{fig:MR}.}\label{tbl-mrdiagram2}
  \end{center}
\end{table}

\end{document}